\documentclass[a4paper,11pt]{article}
\pdfoutput=1 
\usepackage{jheppub} 
\usepackage{subfig}
\usepackage[T1]{fontenc} 
\usepackage{graphicx}
\usepackage{epsfig}
\usepackage{axodraw}
\usepackage{amsmath}
\usepackage{amssymb}
\usepackage{float}
\usepackage{bm}
\usepackage{placeins}
\usepackage{breqn}

\newcommand{\be}{\begin{equation}}
\newcommand{\ee}{\end{equation}}

\newcommand{\bea}{\begin{eqnarray}}
\newcommand{\eea}{\end{eqnarray}}
\newcommand{\bfig}{\begin{figure}}
\newcommand{\efig}{\end{figure}}
\newcommand{\bc}{\begin{center}}
\newcommand{\ec}{\end{center}}

\newcommand{\img}[1]{\includegraphics[scale=0.29]{Figs/Plot_#1.pdf}}

\allowdisplaybreaks[4]

\title{\boldmath Generalised power series expansions for the elliptic planar families of Higgs + jet production at two loops}

\author[a]{F.~Moriello}

\affiliation[a]{ETH Zurich, Institut fur theoretische Physik, Wolfgang-Paulistr. 27, 8093, Zurich, Switzerland}
\emailAdd{fmoriell@phys.ethz.ch}

\abstract{
We obtain generalised power series expansions for a family of planar two-loop master integrals relevant for the QCD corrections to Higgs + jet production, with physical heavy-quark mass. This is achieved by defining differential equations along contours connecting two fixed points, and by solving them in terms of one-dimensional generalised power series. The procedure is efficient, and can be repeated in order to reach any point of the kinematic regions. The analytic continuation of the series is straightforward, and we present new results below and above the physical thresholds.  The method we use allows to compute the integrals in all kinematic regions with high precision. For example, performing a series expansion on a typical contour above the heavy-quark threshold takes on average $\mathcal{O}(1 \text{ second})$ per integral with worst relative error of $\mathcal{O}(10^{-32})$, on a single CPU core. After the series is found, the numerical evaluation of the integrals in any point of the contour is virtually instant. Our approach is general, and can be applied to Feynman integrals provided that a set of differential equations is available.}

\begin{document} 
\maketitle
\flushbottom

\section{Introduction}
\label{sec:intro}

The computation of Feynman integrals is a central ingredient for the prediction of collider experiments. In the past decades, we have seen an enormous progress in our capabilities to efficiently compute Feynman integrals in closed analytic form or in a purely numerical way. From the analytic side, several techniques are available. Some of the most effective techniques are the differential equations method ~\cite{Kotikov:1990kg,Kotikov:1991pm,Bern:1993kr,Remiddi:1997ny,Gehrmann:1999as}  and the direct integration methods~\cite{Brown:2009ta, Panzer:2015ida}. In dimensional regularisation, one is able to reduce a given (generally large) set of scalar Feynman integrals to a minimal set of linearly independent integrals, called master integrals (MIs), by using integration-by-parts identities (IBP)~\cite{Tkachov:1981wb,Chetyrkin:1981qh,Laporta:1996mq,Laporta:2001dd}. Once a basis is identified, it is possible to define a closed system of first order linear differential equations, that can be solved in terms of iterated integrals \cite{Chen-iterated}. Our understanding of differential equations for Feynman integrals has been further refined by the identification of canonical bases of integrals~\cite{Henn:2013pwa}, which make the solution of the equations in terms of iterated integrals completely algorithmic. In some cases Feynman integrals can be computed in terms of special functions known as multiple polylogarithms (MPLs)~\cite{Goncharov:1998kja} or, more recently, in terms of their elliptic generalisation, elliptic multiple polylogarithms (eMPLs)~\cite{BrownLevin,Broedel:2014vla,Adams:2015ydq,Broedel:2017kkb} (for analytic results involving functions of elliptic type see e.g.~\cite{Laporta:2004rb,Kniehl:2005bc,Adams:2013kgc,Bloch:2013tra,Bloch:2014qca,Adams:2014vja,Adams:2015gva,Adams:2015ydq,Remiddi:2016gno,Primo:2016ebd,Bonciani:2016qxi,Adams:2016xah,Passarino:2016zcd,Harley:2017qut,vonManteuffel:2017hms,Ablinger:2017bjx,Chen:2017pyi,Hidding:2017jkk,Bogner:2017vim,Bourjaily:2017bsb,Broedel:2017siw,Laporta:2017okg,Broedel:2018iwv,Mistlberger:2018etf,Lee:2018jsw,Broedel:2018qkq,Adams:2018bsn,Adams:2018kez,Broedel:2019hyg,Bogner:2019lfa,Kniehl:2019vwr,Broedel:2019kmn}). Even though having a representation in terms of known functions is important from the conceptual and practical side, in recent years this approach has become challenging. For state-of-the-art computations one usually encounters multi-loop integrals depending on several mass scales. In this case the differential equations exhibit complicated analytic structures, and their solution in terms of known special functions is not well understood yet. Similar challenges are encountered when solving Feynman integrals by direct integration, e.g., in Feynman parameter space. From the purely numerical side, several methods are available to compute Feynman integrals by using Monte Carlo integration techniques~\cite{Borowka:2015mxa,Smirnov:2015mct}. As opposed to the analytic approach, these methods are fully algorithmic. Nonetheless multi-loop multi-scale integrals generally present numerical instabilities that make their numerical integration challenging. A different numerical approach has been used in \cite{Boughezal:2007ny,Czakon:2007qi,Czakon:2008zk,Mandal:2018cdj}, where the solution of differential equations for Feynman integrals is obtained by using Runge-Kutta algorithms.  

A third route of exploration has been methods based on series expansions. When it is difficult to obtain a closed form solution for a given integral, it is usually possible to obtain a (generalised) power series expansion of the solution. Series representations have a number of useful features. It is usually possible to compute several orders of the expansion and obtain an arbitrarily good approximation of the full solution. Moreover their numerical evaluation is virtually instant, since each term of the expansion is an elementary or a relatively simple function. All these features make series expansions a natural candidate to solve large classes of complicated Feynman integrals. Series expansion methods have been mostly applied to single scale problems in e.g. \cite{Pozzorini:2005ff,Aglietti:2007as,Mueller:2015lrx,Lee:2017qql,Lee:2018ojn,Bonciani:2018uvv,Mistlberger:2018etf}. On the other hand, for integrals depending on several scales, series expansions have been performed with respect to one variable, parametrising special kinematic configurations, while keeping the dependence on the remaining variables exact (see e.g.~\cite{Caffo:2002ch, Caffo:2008aw,Melnikov:2016qoc,Melnikov:2017pgf,Bonciani:2018omm, Bruser:2018jnc,Davies:2018ood,Davies:2018qvx}), or to transport analytic boundary conditions in various regions \cite{Heller:2019gkq}. Therefore, in the multivariate case, it is desirable to study a systematic approach to obtain results in all points of the kinematic regions.

In this paper we reduce the computation of a set of multivariate Feynman integrals \cite{Bonciani:2016qxi} to a single scale problem, by defining differential equations along contours connecting two generic points of the kinematic regions. We then find generalised power series solutions by solving the (single-scale) differential equations with respect to the contour parameter (while replacing all the other variables with numbers). In this way the solution can be transported from a base point, where the integrals are assumed to be known, to a generic target point. We show that this approach is efficient, and can be repeated to compute the integrals in any point of the kinematic regions, with high numerical precision. More specifically, we apply this method to a family of planar (elliptic) Feynman integrals relevant for the two-loop QCD corrections to Higgs + jet production, below and above the heavy-quark threshold. Previously \cite{Bonciani:2016qxi} these integrals were computed in the Euclidean region by using integral representations. Our results are new, and provide at the same time the analytic continuation of these integrals to the physical region, and an efficient method for their numerical evaluation. Further applications of these methods to the non-planar integral topologies are presented in a companion paper \cite{non-planars}.

The paper is organised as follows. In Section \ref{sec:DE review} we review general properties of the differential equations for dimensionally regulated scalar Feynman integrals, and their solution in terms of iterated integrals. In Section \ref{sec:series along countour} we describe the series expansion strategy used in this paper. We show that after defining the (multi-scale) differential equations along a (one-dimensional) contour, the series solution can be obtained by series expanding the differential equations and iteratively integrating them up to the desired order of the dimensional regulator. In Section \ref{sec:HJ application} we apply this strategy to a family of planar integrals relevant for Higgs + jet production in the physical region. We show how, once a series expansion is found, the analytic continuation is performed in a straightforward manner. We finally show high precision numerical results and timings, with comparisons to sector decomposition programs. In Section \ref{sec:conclusion} we draw our conclusions.

\section{Differential equations for dimensionally regulated Feynman integrals}
\label{sec:DE review}
By using standard IBP reduction techniques, it is possible to identify a basis $\vec{f}(\vec{x},\epsilon)$ for a set of dimensionally regulated scalar Feynman integrals, where $\vec{f}(\vec{x},\epsilon)=\left\{f_1(\vec{x},\epsilon),\dots,f_n(\vec{x},\epsilon)\right\}$ and $\vec{x}=\left\{x_1,\dots,x_m\right\}$ is the set of kinematic invariants. Given a basis, one is also able to define a system of first order linear differential equations satisfied by the basis, that in full generality takes the form
\begin{equation}
\label{eq:DE FI General}
\frac{\partial}{\partial x_i} \vec{f}(\vec{x},\epsilon)=\mathbf{A}_{x_i}(\vec{x},\epsilon)\vec{f}(\vec{x},\epsilon),
\end{equation}    
where $\mathbf{A}_{x_i}$ is an $n$-by-$n$ matrix that depends rationally on its variables. If the set of basis integrals is minimal the differential equations satisfy the integrability condition,
\begin{equation}
\frac{\partial\mathbf{A}_{x_j}}{\partial x_i}-\frac{\partial\mathbf{A}_{x_i}}{\partial x_j}+ [\mathbf{A}_{x_i},\mathbf{A}_{x_j}]=0,
\end{equation}
where the last term is a commutator.  Nonetheless the applicability of our method does not rely on this condition, and it can be applied also to an overcomplete set of integrals. 

The basis $\vec{f}$ is not unique. In \cite{Henn:2013pwa} it was conjectured that, with a basis change, it is possible to cast differential equations for Feynman integrals in a canonical form, where the dependence on the dimensional regulator is factorised. In differential form the canonical equations have the following form
\begin{equation}
\label{eq:canonicalDEdiff}
d \vec{f}(\vec{x},\epsilon)=\epsilon\; d \tilde{\mathbf{A}}(\vec{x})\vec{f}(\vec{x},\epsilon).
\end{equation}
In dimensional regularisation we are interested in a solution around $\epsilon=0$. By series expanding $\vec{f}(\vec{x},\epsilon)$,
\begin{equation}
\label{eq:Series_eps_f}
\vec{f}(\vec{x},\epsilon)=\sum_{k=0}^\infty \epsilon^k\vec{f}^{(k)}(\vec{x}),
\end{equation}
the solution of Eq.~(\ref{eq:canonicalDEdiff}) can be written in terms of iterated integrals~\cite{Chen:1977oja}:
\begin{align}
\label{eq:SolutionIteInt}
\vec{f}(\vec{x},\epsilon) = \vec{f}(\vec{x}_0,\epsilon) + \sum_{k\geq 1}\epsilon^k \sum_{j=1}^k \int_0^1 \gamma^*(d\tilde{\mathbf{A}})(t_1) \int_0^{t_1} \gamma^*(d\tilde{\mathbf{A}})(t_2) \ldots \int_0^{t_{j-1}} \gamma^*(d\tilde{\mathbf{A}})(t_j) \,\vec{f}^{(k-j)}(\vec{x}_0)\,,
\end{align}
where $\gamma$ is a path with domain $[0,1]$ in the space of external invariants and $\vec{f}(\vec{x}_0,\epsilon)$ is a vector of boundary conditions. 
If $\vec{f}^{(i)}(x)$ admits a representation in terms of multiple polylogarithms \cite{Goncharov:1998kja},
\begin{equation}
G(a_1,a_1\dots a_w,t)=\int_0^t dt \frac{1}{t-a_1}G(a_2,\dots,a_w,t),\quad G(\vec{0}_w,t)\equiv\frac{\log(t)^{w}}{w!},
\end{equation}
with $G(,t)\equiv 1$, the transformation matrix to the canonical basis is algebraic, and the matrix $\tilde{\mathbf{A}}(x)$ is a $\mathbb{Q}$-linear combination of logarithms
\begin{equation}
\tilde{\mathbf{A}}(\vec{x})=\sum_{i=1}^{d_{\alpha}} \mathbf{C}_i \log(\alpha_i(\vec{x})),
\end{equation}
where $d_\alpha$ is the number of linearly independent logarithms, $\mathbf{C}_i$ are constant rational matrices, and $\alpha_i(\vec{x}), i\in \{1,\dots,d_{\alpha}\}$  are called letters of the differential equations. The set of letters is referred to as the alphabet. If the letters admit a representation in terms of rational functions, then the iterated integrals of (\ref{eq:SolutionIteInt}) can be directly expressed in terms of multiple polylogarithms. On the other hand, if a rational representation is not found, it is often possible to find a polylogarithmic representation by using the knowledge of the symbol of iterated integrals~\cite{Goncharov:1998kja,Goncharov:2010jf}. The symbol of the $i$-th basis element of (\ref{eq:Series_eps_f}) at order $\epsilon^k$ is obtained by the following recursive formula
\begin{align}
    \mathcal{S}\left(f_i^{(k)}(\vec{x})\right) = \sum_j \mathcal{S}\left(f_j^{(k-1)}(\vec{x})\right) \otimes \tilde{\mathbf{A}}_{ij}(\vec{x})\,.
\end{align}
Once the symbol of the solution is known, it is possible to find a corresponding polylogarithmic expression by using an ansatz for the set of polylogarithmic functions, and by imposing boundary conditions (see e.g, \cite{Duhr:2011zq,Bonciani:2016qxi}). 

When considering integral sectors that do not admit a polylogarithmic representation, the properties of the transformation matrix to the canonical form are not yet well understood (but see e.g.~\cite{Adams:2018yfj} for recent progress in this direction). For this reason, when a canonical basis is not available, we will only assume that the differential equations are non-singular in the $\epsilon\rightarrow 0$ limit (note that it is always possible to find a basis of Feynman integrals satisfying such differential equations, see e.g. \cite{Henn:2014qga}).
  
\section{Series expansion along a contour}
\label{sec:series along countour}

Given a base point where the integrals are known, we show how the integrals are computed in a new point by series expanding the integrals along a contour connecting these points. We first define the differential equations along the contour, and then we show how a series expansion of the integrals is found by solving the differential equations around a set of points of the contour. The procedure can be repeated to reach any point of the kinematic regions. In this section we describe the general framework needed to perform series expansions along contours. In appendix \ref{app:1loopexample} we discuss in detail a simple one-loop example.

Provided that a set of differential equations with respect to a complete set of kinematic invariants is available, we can define the differential equations along a contour $\gamma(t)$ connecting two  fixed points $\vec{a}=\{a_1,\dots,a_m\},\; \vec{b}=\{b_1,\dots,b_m\}$. This is achieved by parametrising the contour with a parameter $t$:
\begin{equation}
\gamma(t) : \; t\mapsto \{x_1(t),\dots,x_m(t)\},\qquad \vec{x}(t_a)=\vec{a}, \; \; \vec{x}(t_b)=\vec{b},
\end{equation}
and by considering the differential equations with respect to $t$:
\begin{equation}
\label{eq:DE along contour}
\frac{\partial}{\partial t} \vec{f}(t,\epsilon)=\mathbf{A}_{t}(t,\epsilon)\vec{f}(t,\epsilon),
\end{equation}
where the new differential equations matrix can be readily obtained by using the chain rule,
\begin{equation}
\mathbf{A}_{t}(t,\epsilon)=\sum_{i=1}^{m}\mathbf{A}_{x_i}(t,\epsilon)\frac{\partial x_i(t)}{\partial t}.
\end{equation}
It is known that a set of master integrals, at a given order of the dimensional regulator, admit a solution in the vicinity of a singular point $\tau$ of the form (see e.g. \cite{Smirnov:1990rz})
 \begin{equation}
\label{eq:Series FI Singular}
\vec{f}_{\text{sing}}^{(i)}(t)=\sum_{j_1\in S_i }  \sum_{j_2=0}^\infty \sum_{j_3=0}^{N_{i}}\vec{c}\;^{(i,j_1,j_2,j_3)}(t-\tau)^{w_{j_1}+j_2}\log{(t-\tau)}^{j_3},
\end{equation} 
where $\vec{c}\;^{(i,j_1,j_2,j_3)} $ are vectors of dimension $n$ (the number of master integrals), $S_i$ is a finite set of integers, $w_k$ is a complex constant (typically a rational number that accounts for the algebraic dependence of the matrix elements), and $N_i$ is the maximal power of the logarithms at order $\epsilon^i$.  On the other hand, in the vicinity of a regular point $\tau$, the integrals admit a standard Taylor series representation
\begin{equation}
\label{eq:Series FI Regular}
\vec{f}_{\text{reg}}^{(i)}(t)=\sum_{j=0}^\infty \vec{c}\;^{(i,j)}(t-\tau)^{j}.
\end{equation} 
More simply, each Feynman integral, in the vicinity of a singular point $\tau$, is expressed as a finite combination of terms of the form
\begin{equation}
(t-\tau)^w\log(t-\tau)^k \rho(t),\quad w\in\mathbb{Q},\;k\in \mathbb{N},
\end{equation}
where $\rho(t)$ is a Taylor series. On the other hand, in the vicinity of a regular point, Feynman integrals admit a Taylor series representation. 

In the remainder of this section we show how the series solutions (\ref{eq:Series FI Singular}) and (\ref{eq:Series FI Regular})  can be found by series expanding the differential equations matrices, and by iteratively integrating them until the desired order of $\epsilon$. 

We know that, for many phenomenologically relevant processes, most integrals admit a polylogarithmic canonical basis. In the next subsection we discuss how we find a power series solution of a canonical set of differential equations. This form of the equations is usually much more compact then the differential equations for a generic basis of master integrals, and working with simplified differential equations renders their series expansion more efficient (for a discussion about timings see Section~\ref{sec:Numerics}). In section \ref{sec:Series Elliptic} we discuss how we obtain series solutions for coupled sectors. We remark that the applicability of our method does not rely on (when it exists) a canonical basis of integrals, and a series solution for an arbitrary basis of integrals can be found  by using the methods of section \ref{sec:Series Elliptic}.
\subsection{Canonical differential equations}
\label{sec:Series Can}
When dealing with single scale problems, generalised power series solutions are usually obtained by defining generic power series, and by fixing the corresponding free coefficients by solving recurrence relations~\cite{Wasow,Lee:2017qql,Mueller:2015lrx}. Here we proceed in a more direct way, which is particularly suited for differential equations in canonical form. Specifically, we consider a canonical system of differential equations of the form
\begin{equation}
\frac{\partial}{\partial t} \vec{f}(t,\epsilon)=\epsilon \mathbf{A}_{t}(t)\vec{f}(t,\epsilon),
\end{equation}
and we assume that the solution is known (analytically or numerically with very high precision, e.g. from a previous expansion) for some $t=t_0$. From Eq.~(\ref{eq:SolutionIteInt}) it is easy to see that the solution is

\begin{equation}
    \label{SolutionIteInt t}
    \vec{f}(t,\epsilon) = \vec{f}(t_0,\epsilon) + \sum_{k=1}^\infty\epsilon^k \sum_{j=1}^k \int_{t_0}^t dt_1 \mathbf{A}_{t}(t_1) \int_{t_0}^{t_1} dt_2 \mathbf{A}_{t}(t_2) \ldots \int_{t_0}^{t_{j-1}} dt_j \mathbf{A}_{t}(t_j) \,\vec{f}^{(k-j)}(t_0)\,.
\end{equation}
Since we are interested in the solution in the vicinity of a point $\tau$, we series expand the differential equations matrix around $\tau$,
\begin{equation}
\mathbf{A}_{t}(t)=\sum_{i=0}^\infty \mathbf{A}^{(i)}_{t} (t-\tau)^{w_i},\quad w_i\in \mathbb{Q} ,
\end{equation}
where $\mathbf{A}^{(i)}_{t}$ are constant matrices \footnote{The series expansion for the matrix  $\mathbf{A}^{(i)}_{t}$ can be obtained for example by using the built-in Mathematica function Series.}. For a canonical basis, $w_i, i\in \mathbb{N}$ is expected to be the set of all half integer numbers with $w_i\geq -1$. However the following discussion holds for generic complex numbers  $w_i$.  By plugging the previous expansion into (\ref{SolutionIteInt t}), we get
\begin{align}
    \label{SolutionIteInt t series}
    &\vec{f}(t,\epsilon) = \vec{f}(t_0,\epsilon) \nonumber\\
    & + \sum_{k=1}^\infty\epsilon^k \sum_{j=1}^k \sum_{i_1,\dots ,i_j=0}^\infty\mathbf{A}^{(i_1)}_{t}\dots \mathbf{A}^{(i_j)}_{t} \int_{t_0}^t dt_1 (t_1-\tau)^{w_{i_1}}   \ldots \int_{t_0}^{t_{j-1}} dt_j   (t_j-\tau)^{w_{i_j}} \,\vec{f}^{(k-j)}(t_0)\,.
\end{align}
At each order $\epsilon^k$, we have to compute integrals of the form,
\begin{equation}
\int (t-\tau)^{w}\log(t-\tau)^k dt,\quad w\in \mathbb{Q}, \quad k\in \mathbb{N}.
\end{equation}
These integrals can be computed analytically in terms of integer powers of $\log(t-\tau)$ and (rational) powers of $(t-\tau)$, by using recursively integration-by-parts identities\footnote{In practice, these integrals can be computed by using the built-in Mathematica function Integrate}.

\subsection{Coupled sectors}
\label{sec:Series Elliptic}
The problem of finding a canonical basis for integrals that do not admit a polylogarithmic representation is still poorly studied in the literature. In practice, we often deal with differential equations where only a subset of the equations is in canonical form, while the other sectors admit a generic rational dependence on the dimensional regulator and an algebraic dependence on the kinematic invariants. In this case,  
by iteratively taking derivatives $\partial/\partial_t$ of equation (\ref{eq:DE along contour}), it is possible to obtain a $k-$th order differential equation for a single integral, say $f_j^{(i)}(t)$, of the form,
\begin{equation}
\label{eq:ODE order N}
\frac{\partial^{k}g^{(i)}(t)}{\partial t^{k}}+a_{1}(t)\frac{\partial^{k-1}g^{(i)}(t)}{\partial t^{k-1}}+\cdots+a_k(t)g^{(i)}(t)=\beta^{(i)}(t),
\end{equation}
where, for ease of notation, we denoted the generic master integral at order $\epsilon^i$ as $g^{(i)}\equiv f_j^{(i)}(t)$. The inhomogeneous term $\beta^{(i)}(t)$ is a linear combination of $f_{\neq j}^{(\leq i)}(t)$, and it is known at every iteration $i$. The homogeneous equation associated to (\ref{eq:ODE order N}) is
\begin{equation}
\label{eq:ODE order N Homo}
\frac{\partial^{k} h(t)}{\partial t^{k}}+a_{1}(t)\frac{\partial^{k-1} h(t)}{\partial t^{k-1}}+\cdots+a_k(t) h(t)=0,
\end{equation}
and it admits $k$ linearly independent solutions that we denote as $ h_{1}(t),\dots, h_{k}(t)$. We note that the homogeneous equation does not depend on the order $\epsilon$ under consideration.

By defining the following fundamental matrix,
\begin{equation}
\hat{\mathbf{H}}(t)=\left(\begin{matrix}
 h_{1}(t) && \cdots  &&  h_{k}(t)\\
\frac{\partial}{\partial t} h_{1}(t) && \cdots  && \frac{\partial}{\partial t} h_{k}(t)\\
\vdots && \ddots && \vdots\\
\frac{\partial^{k-1}}{\partial t^{k}} h_{1}(t) && \cdots  && \frac{\partial^{k-1}}{\partial t^{k-1}} h_{k}(t)
\end{matrix}\right),
\end{equation}
the particular solution of (\ref{eq:ODE order N}) is
\begin{equation}
\label{eq:ODE N particular}
g^{(i)}(t)=\sum_{j=1}^k \left( h_{j}(t) \chi^{(i)}_{j}+ h_{j}(t)\int_{t_0}^{t}\frac{W_j(\mathbf{\hat{ H}}( {}s{}))}{W(\mathbf{\hat{ H}}( {}s{}))}\beta^{(i)}( {}s{})d {}s{}\right),
\end{equation}
\label{eq:SOL Higher order}
where $\chi^{(i)}_{j},\; j\in\{1,\dots,k\}$ is a set of boundary constants, the Wronskian is defined by $W(\mathbf{\hat{ H}}( {}s{}))=\text{det}(\mathbf{\hat{H}}(t))$, while $W_j(\mathbf{\hat{ H}}( {}s{}))$ is the determinant obtained by replacing the $j$-th column of $ \mathbf{\hat{ H}}(t)$ by $(0,\dots,0,1)$. 
 It can be shown that, for functions admitting a series solution of the form (\ref{eq:Series FI Singular}), the equation in the vicinity of a point $\tau$ can be written as,
\begin{equation}
\frac{\partial^{k}g^{(i)}(t)}{\partial t^{k}}+\frac{b_{1}(t)}{t-\tau}\frac{\partial^{k-1}g^{(i)}(t)}{\partial t^{k-1}}+\cdots+\frac{b_k(t)}{(t-\tau)^k}g^{(i)}(t)=\beta^{(i)}(t),
\end{equation}
where the functions $b_i(t)$ are analytic in $t=\tau$. In this case a series solution in the vicinity of $t=\tau$ can be found by applying the Frobenius method, which is discussed in detail in Appendix \ref{App:Frobenious} (see also e.g. \cite{Coddington}). More specifically, the Frobenius method allows to find a complete set of $k$ homogeneous series solutions in the vicinity of $\tau$.  These solutions have the form
\begin{equation}
\label{eq:Series Frob}
h_{i}(t)=\sum_{j_1\in S_{i} }  \sum_{j_2=0}^\infty \sum_{j_3=0}^{K_{i}}   c_{i}^{(j_1,j_2,j_3)}(t-\tau)^{\lambda_{j_1}+j_2}\log{(t-\tau)}^{j_3},
\end{equation} 
where $c_{i}^{(j_1,j_2,j_3)}$ are complex constants, $K_i$ is the maximal power of the logarithm, $S_i\subseteq \{1,\dots,k\}$ and $\lambda_1,\dots,\lambda_k$ are the roots of the indicial equation, defined as
\begin{equation}
\lambda (\lambda-1)\cdots(\lambda-k+1)+b_{1}(\tau) \lambda (\lambda-1)\cdots (\lambda-k+2)+b_{k}(\tau)=0.
\end{equation}
From the form of the homogeneous series solutions, it is clear that the full solution (\ref{eq:SOL Higher order}) will require, at each order $i$, the computation of integrals of the form (\ref{SolutionIteInt t series}) which, as explained in the previous section, can be done analytically in terms of integer powers of logarithms and complex powers of $t$. This shows that, also in the case of coupled differential equations, we can explicitly find a series representation for Feynman integrals in the vicinity of regular or singular points of the form of Eqs.~(\ref{eq:Series FI Regular}) and~(\ref{eq:Series FI Singular}) respectively. 

The Frobenius method is rather standard, and we review its general formulation for linear differential equations of generic order  in Appendix \ref{App:Frobenious}. Here we briefly show its application to second order equations, since this is the order encountered in this paper. Let us consider the case $k=2$ of (\ref{eq:ODE order N Homo}) and, without loss of generality, let us assume that $t=0$ is a (regular) singular point,
\begin{equation}
t^2 \frac{\partial^{2} h(t)}{\partial t^{k}}+t \;b_{1}(t)\frac{\partial h(t)}{\partial t}+b_2(t) h(t)=0.
\end{equation}
We have to distinguish two cases in order to proceed. Let us first assume that the two roots of the indicial equation $\lambda_1,\lambda_2$ do not differ by an integer number, with $\lambda_1>\lambda_2$. In this case two linearly independent solutions are
\begin{equation}
\label{eq:frobsol1}
h_{1}(t)=t^{\lambda_1}\sum_{i=0}^{\infty} c_{1,i} \; t^i,\quad h_{2}(t)=t^{\lambda_2}\sum_{i=0}^{\infty} c_{2,i} \; t^i,\quad (c_{1,0},c_{2,0}=1)\\
\end{equation}
and the coefficients are fixed by requiring that the differential equation is satisfied order-by-order in $t$. If the two roots differ by an integer, the solution associated to $\lambda_1$ is obtained by (\ref{eq:frobsol1}) while the second solution is obtained by 
\begin{equation}
h_2(t)= h_1(t)\int \frac{1}{h_1^2(t)}\exp{\left(-\int^t \frac{b_1(t')}{t'}\; dt'\right)} dt,
\end{equation}    
which can be expressed as a series by expanding all the integrands around $t=0$ and performing the integrations term-by-term.

\subsection{Matching}
\label{sec:matching}
Given a base point where the integrals are assumed to be known, we want to transport the solution to a new point, by using series expansions along a contour connecting these points.

In the previous sections we have seen that, given a singular or regular point of the differential equations, we can find a series solution valid in the vicinity of these points. By construction, these solutions converge only in a region that does not contain any singularity other then the expansion point. In the following we will consider truncated series that, within a given accuracy, provide a good approximation of the full series. When considering a generic contour $\gamma(t)$ for a range of values of $t$, we are interested in finding series expansions along the entire contour. The contour will in general contain multiple singular points of the differential equations, and it is necessary to find multiple series expansions and match them together in order to cover the entire contour. A good criterion for determining the domain of definition of a truncated series is that the series will converge fast enough only when considering it in a region such that the maximum distance from the expansion point is half the distance from the nearest singularity. 

Let us assume that we have defined a contour  $\gamma(t)$, for $t\in [0,1]$ and we want to find truncated series expansions that approximate the full solution within a given accuracy. $\gamma(0)$ is the known boundary point and $\gamma(1)$ is the target point we want to compute. There might be real and complex singularities. Let us denote the real singularities as $R=\{\tau_i| i\in N_s\}$ and the complex singularities by $C=\{\lambda_i=\lambda^{re}_i+i \lambda^{im}_i|i\in N_c\}$ where $N_s,N_c\in \mathbb{N}$. Moreover we define the following set of real regular points $C_r=\bigcup_{i=1}^{N_c}\{\lambda^{re}_i- \lambda^{im}_i,\lambda^{re}_i,\lambda^{re}_i+ \lambda^{im}_i\}$. We now consider a set of points $t_i\in R\bigcup C_r$ such that $t_{-1}<0<t_1<\dots<t_{N_{e}}<1< t_{N_{e}+1}<t_{N_{e}+2}$ with $N_e\in \mathbb{N}$. We then find truncated power series around $t_0=0$ and $t_i$ for $i\in\{1,\dots,N_e+1\}$. We denote these series as $\vec{f}_{[\gamma]}(t,\epsilon)_{[t_i-r_i,t_i,t_i+r_i]}$, where $t_i$ is the expansion point, and $r_i$ is the radius of the series defined as the distance between $t_i$ and the closest point $t_{j\neq i}$. Moreover we define $\vec{f}_{[\gamma]}(t,\epsilon)_{[t_i-r_i,t_i,t_i+r_i]}$ to be equal to the truncated series for $t_i-r_i\leq t\leq t_i+r_i$, while it is zero otherwise.
Since the $t_i$ are in general not equally spaced it can happen that some segments of the contour are not cover by any series. In this case we iteratively introduce new (regular) expansion points $\kappa_i, i\in \mathbb{N}$ , and corresponding series $\vec{f}_{[\gamma]}(t,\epsilon)_{[\kappa_i-\rho_i,\kappa_i,\kappa_i+\rho_i]}$, such that $\kappa_i$ is in the middle of an uncovered region and $\rho_i$ is the distance between $\kappa_i$ and the closest point among $\kappa_{j\neq i},t_{j}$. The procedure is repeated until the entire contour is covered. 

There are two special cases that need some care. If there are no points $t_{i}$, we just expand around $t_0=0$ and the entire contour will be covered. If there are no points $t_{N_e+1},t_{N_e+2}$ we iteratively add new (regular) expansion points such that $0<k_i<t_{N_e}$. At the end of this procedure it can happen that the point $t=1$ in not covered. We then set $t_{N_e+1}=1$ and we add new (regular) expansion points as described above until the entire contour is covered. 

We finally define the solution along the entire contour to be,
\begin{equation}
\vec{f}_{[\gamma]}(t,\epsilon)=\sum_{i\in S_e}\vec{f}_{[\gamma]}(t,\epsilon)_{[t_i-r_i,t_i,t_i+r'_i]}+\sum_{i\in S_r}\vec{f}_{[\gamma]}(t,\epsilon)_{[\kappa_i-\rho_i,\kappa_i,\kappa_i+\rho'_i]},
\end{equation}
$r'_i\leq r_i,\rho'_i\leq \rho_i$ are such that there are no overlapping series, and $S_e\subseteq \{1,\dots,N_e\},S_r \subseteq \{1,\dots,N_r\}$ ($N_r$ is the total number of regular expansion points) are such that no series outside the unit interval $t\in [0,1]$ is considered in the sum.

In Sections \ref{sec:Series Can} and \ref{sec:Series Elliptic} we have seen that the series solutions depend on a set of integration constants that have to be fixed by imposing boundary conditions. When considering a set of series along a contour, the integration constants of one series are fixed by knowing the boundary conditions at a given point (e.g. when the boundary conditions are known analytically by other means, or they are known because the contour under consideration intersects another contour along which the series expansion is already known), while the other series are fixed by imposing that two consecutive series have the same value in the contact point, i.e. the point where they are both defined.

\section{A planar elliptic family for Higgs+jet production}
\label{sec:HJ application}

\begin{figure}[!t]
\centering
\vspace{3mm}
\includegraphics[width=3.2cm]{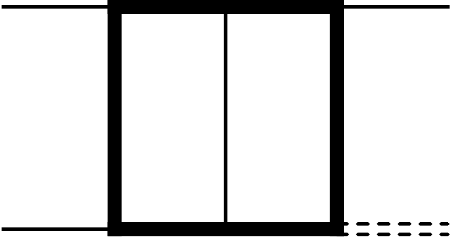} \;
\includegraphics[width=3.2cm]{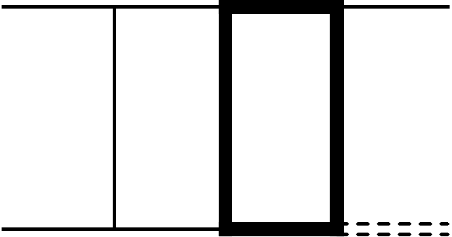} \;
\includegraphics[width=3.2cm]{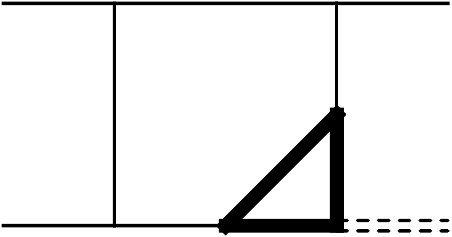} \;
\includegraphics[width=3.2cm]{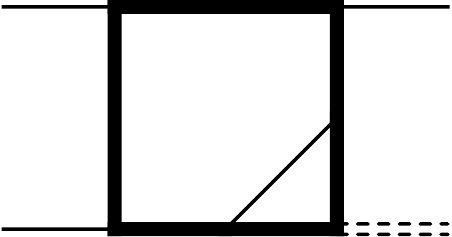} \\
\begin{large}
\vspace{3mm}
A $\qquad\qquad\qquad\,$ B $\qquad\quad\qquad\qquad\;\;\,$ C $\qquad\qquad\qquad\quad$ D $\!\!\!\!\!\!$ \\
\end{large}
\caption{The four planar integral families contributing to two-loop $H{+}j$-production in QCD.}
\label{fig:families}
\end{figure}

The planar two-loop QCD corrections to Higgs+jet productions are mediated by the four integral families depicted in Fig. \ref{fig:families}. These integrals were computed in \cite{Bonciani:2016qxi} in the non-physical region, where all the Mandelstam invariants are negative reals. For the polylogarithmic sectors, the solution was expressed in terms of logarithms and dilogarithms up to weight two, while the weight-three and four solutions were expressed in terms of one-fold integrals over the lower weight solutions. Family A contains two elliptic sectors. The first elliptic sector is $I^{A}_{1,1,0,1,1,1,1,0,0}$; the homogeneous solutions are elliptic integrals, and the solution in dimensional regularisation requires iterated integrations over elliptic kernels. The second elliptic sector is $I^{A}_{1,1,1,1,1,1,1,0,0}$, and it admits a canonical form for the homogeneous differential equations, but it is coupled to the first elliptic sector via inhomogeneous terms. In \cite{Bonciani:2016qxi} the solution for the elliptic sectors was expressed in terms of iterated integrals over elliptic kernels. 

In this section we obtain generalised power series expansions for family A in the  $p_2\leftrightarrow p_3$  channel, which exhibits the most complicated (spurious, see section \ref{sec:Compact phys}) singularity structure of the channels needed for the scattering amplitude.  All the other families are simpler, as they do not involve elliptic structures and can be solved with the same methods.

The family under consideration is defined by the following set of integrals
\begin{align}
\label{eq:fintdef}
I^{A_{p_2 \leftrightarrow p_3}}_{a_1, a_2, a_3 ,a_4, a_5, a_6 ,a_7 ,a_8, a_9} &= \int \!\! \int \frac{d^d k_1 d^d k_2}{(i \pi^{d/2})^2} \frac{d_8^{-a_8} \, d_9^{-a_9}}{d_1^{a_1} d_2^{a_2} d_3^{a_3} d_4^{a_4} d_5^{a_5} d_6^{a_6} d_7^{a_7}}
\end{align}
with
\begin{align}
d_1& =m^2-k_1^2,&d_4& =m^2-\left(k_2+p_1+p_3\right){}^2,&d_7&=m^2-\left(k_2-p_2\right){}^2,\nonumber\\
d_2& =m^2-\left(k_1+p_1+p_3\right){}^2,&d_5 & =m^2-\left(k_1+p_1\right){}^2,&d_8&=-\left(k_2+p_1\right){}^2,\nonumber\\
d_3& =m^2-k_2^2,&d_6& =-\left(k_1-k_2\right){}^2,&d_9&=-\left(k_1-p_2\right){}^2.
\end{align}
$d_1$-$d_7$ are propagators while $d_8$ and $d_9$ are numerators, therefore $a_1,\dots,a_7$ are non-negative integers while $a_9,a_8$ are non-positive integers. The kinematics is such that $p_1^2 = p_2^2 = p_3^2 = 0$ with
\begin{align}
s_{12}&=(p_1{+}p_2)^2, & s_{13} &= (p_1{+}p_3)^2, & s_{23} &= (p_2{+}p_3)^2, & p_4^2 &= (p_1{+}p_2{+}p_3)^2 = s_{12}+s_{13}+s_{23},
\end{align}
where $p_4^2$ is the squared Higgs-mass, $p_4^2=m_H^2$, and $m^2$ is the squared mass of the propagators. The relevant physical region is defined by
\begin{equation}
\label{eq:physical region}
s_{12}>0 \quad \& \quad s_{13}<0 \quad \& \quad s_{12}>-s_{13}+p_4^2.
\end{equation}
We solve the kinematic constraints for $s_{12},s_{13},p_4^2$ and we define the following scaleless variables 
\begin{equation}
x_1=\frac{s_{12}}{m^2},\quad x_2=\frac{p_4^2}{m^2},\quad x_3=\frac{s_{13}}{m^2}.
\end{equation}

The family contains 73 master integrals, and our choice for the integral basis is provided in the ancillary files of the arXiv submission. The first 65 master integrals satisfy a set of differential equations in canonical form,
\begin{equation}
\label{eq:DE A poly}
\frac{\partial}{\partial x_i}\vec{f}_{1-65}(\vec{x},\epsilon)=\epsilon \mathbf{A}_{x_i}(\vec{x})  \vec{f}_{1-65}(\vec{x},\epsilon).
\end{equation}
The alphabet of the differential equations consists of 42 letters, depending on the following set of 6 square roots\footnote{For ease of notation we joined products of square roots into one square root. However, as discussed in Section \ref{sec:Analytic continuation}, in order to perform the analytic continuation, it is convenient to split them in square-root factors. The actual factorisation of the square roots used in this paper is the one of the integral basis provided in the ancillary files of the arXiv submission. We also note that, since the square roots appear as factors in the basis choice, one has the freedom to choose their factorisation provided that, once one choice is made, all the subsequent operations are carried out consistently with this choice.},
\begin{equation}
\begin{array}{ll}
\left[x_1 \left(x_1-4  \right)\right]^{\frac{1}{2}}, &\quad\;\;\left[x_1 x_3 \left(4   x_2-4   x_3+x_1 \left(x_3-4  \right)\right)\right]^{\frac{1}{2}},\\
\left[x_2\left(x_2-4  \right)\right]^{\frac{1}{2}}, & \quad\;\;\left[x_1 \left(4   x_3 \left(x_2-x_3\right)+x_1 \left( -x_3\right){}^2\right)\right]^{\frac{1}{2}},\\
 \left[x_3 \left(x_3-4  \right)\right]^{\frac{1}{2}}, &\quad\;\;\left[ \left(x_2-x_1\right){}^2-2   x_1 x_3\left(-x_2+x_1+2 x_3\right)+x_1^2 x_3^2\right]^{\frac{1}{2}}.
\end{array}
\end{equation}
We remark that our approach does not rely on the rational parametrisation of the set of square roots, and it works for general algebraic dependence of the differential equations. Integrals 66-73 are elliptic and satisfy coupled differential equations of the form
\begin{equation}
\label{eq:DE A elliptic}
\frac{\partial}{\partial x_i}\vec{f}_{66-73}(\vec{x},\epsilon)=\sum_{j=0}^\infty \epsilon^j \mathbf{B}_{x_i}^{(j)}(\vec{x}) \vec{f}_{66-73}(\vec{x},\epsilon)+\vec{g}_{66-73}(\vec{x},\epsilon) ,
\end{equation}
where the vector $\vec{g}_{66-73}(\vec{x},\epsilon)$ depends on the polylogarithmic integrals $\vec{f}_{1-65}(\vec{x},\epsilon)$. Order-by-order in $\epsilon$ the polylogarithmic integrals $\vec{f}_{1-65}(\vec{x},\epsilon)$ satisfy completely decoupled differential equations. On the other hand, the elliptic sectors are coupled and, in general, one needs to solve higher order differential equations for single integrals. Specifically, the homogeneous matrix $\mathbf{B}^{(0)}_{x_i}$ has the following schematic form
\begin{equation}
\mathbf{B}_{x_i}^{(0)}=\left(
\begin{array}{cccc|cccc}
 \;0\; & \bm{*} & \;0\; & \;\bm{*}\; & \;0\; & \;0\; & \;0\; & \;0\; \\
 0 & \bm{*} & 0 & \bm{*} & 0 & 0 & 0 & 0 \\
 0 & \bm{*} & \bm{*} & \bm{*} & 0 & 0 & 0 & 0 \\
 0 & \bm{*} & 0 & \bm{*} & 0 & 0 & 0 & 0 \\
 \hline
 0 & 0 & 0 & 0 & 0 & 0 & 0 & 0 \\
 0 & \bm{*} & 0 & \bm{*} & 0 & 0 & 0 & 0 \\
 0 & 0 & 0 & 0 & 0 & 0 & 0 & 0 \\
 0 & \bm{*} & 0 & \bm{*} & 0 & 0 & 0 & 0 \\
\end{array}
\right),
\end{equation}
where the lines separate sector $I^{A}_{1,1,0,1,1,1,1,0,0}$ (first four rows) from sector $I^{A}_{1,1,1,1,1,1,1,0,0}$ (last four rows). We see that integrals 67 and 69 are coupled. For each of these integrals we can define a second order differential equation. Once integral 67 or 69 is known, all the other integrals are decoupled and can be solved by considering first order equations only.
\subsection{Series solution of the differential equations}
\label{sec:series sol DE A}
In order to solve the differential equations we need a set of boundary conditions. We use the point $x_1=x_2=x_3=0$, and the boundary conditions are~\cite{Bonciani:2016qxi}
\begin{equation}
f_i(\vec{0},\epsilon)=\left\{\begin{array}{cl}
1+\frac{\pi^2}{6}\epsilon^2-\frac{2 \zeta_3}{3}\epsilon^3+\frac{7\pi^4}{360}\epsilon^4+\mathcal{O}(\epsilon^5) & \qquad  \text{if $i=1$},\\
0 & \qquad \text{otherwise}.
\end{array}\right.
\end{equation}
We are interested in a solution in the relevant physical region~(\ref{eq:physical region}). Therefore we transport the boundary condition to the point $\left\{ 2,\;\frac{13}{25} ,\; -1\right\}$ by series expanding the integrals along the contour\footnote{In this paper we assume $p_4^2=13/25\approx (125/172)^2$, which is a good approximation of the Higgs-boson mass, when normalised to the top mass. However, the applicability of the method does not rely on the choice of this numerical value. For example, we could equally well consider values of the dimensionless ratios related to the bottom mass. More generally, we could apply the expansion method by considering the Higgs mass as an additional variable.},
\begin{equation}
\gamma_0(t)=\left\{ x_1(t),\;x_2(t),\; x_3(t) \right\}=\left\{2 t,\;\frac{13}{25} t,-t \right\}, \quad t\in[0,1].
\end{equation}
Here we discuss in detail how we expand along the contour $\gamma_{\text{thr}}(t)$ defined as
\begin{equation}
\gamma_{\text{thr}}(t)= \left\{x_1(t),\;x_2(t),\; x_3(t)\right\}=\left\{2 + 4 t,\;\frac{13}{25},\; -1\right\}, \quad t\in[0,1].
\end{equation}
This contour is interesting because it allows to analytically continue the integrals above the physical threshold $x_1=4$. In order to achieve the decomposition of the contour described in Section \ref{sec:matching}, we need to know where the singularities lie. When considering differential equations with algebraic dependence (square roots in the present case), the singularities are all the non-analytic points of the differential equations, i.e. points where the differential equations diverge but also the zeros of the square roots (branching points). Once the path is fixed, the problem is one dimensional and we can find the position of the singularities by solving for the zeros of the denominators of the differential equations and for the zeros of the square roots (see Section \ref{sec:Analytic continuation} for a detailed discussion about the different classes of singular points of the differential equations). 

For the path $\gamma_{\text{thr}}(t)$, the relevant singularities are 
\begin{equation}
\tau_1=\frac{1}{2},\quad \tau_2 =-\frac{3}{25},
\end{equation}
where $\tau_1$ is the physical threshold and $\tau_2$ is a spurious singularity outside the $[0,1]$ interval. We now partition the contour as described in Section~\ref{sec:matching}. Specifically, we add the regular expansion points $\kappa_1=0,\;\kappa_2=1\; \kappa_3=\frac{1}{8}$ which results in the following partitioning of the contour,
\begin{align}
\vec{f}_{[\gamma_{\text{thr}}]}(t,\epsilon)=&\; \vec{f}_{[\gamma_{\text{thr}}]}(t,\epsilon)_{[0,\;0,\;0.06]}+ \vec{f}_{[\gamma_{\text{thr}}]}(t,\epsilon)_{[0.06,\;0.125,\;0.19]}\nonumber\\
+&\; \vec{f}_{[\gamma_{\text{thr}}]}(t,\epsilon)_{[0.19,\;0.5,\;0.81]}+\vec{f}_{[\gamma_{\text{thr}}]}(t,\epsilon)_{[0.81,\;1,\;1]},
\end{align}
where we replaced rational numbers with (exact) real numbers. All the boundary constants are fixed by imposing the following chain of boundary conditions,
\begin{align}
\label{eq:BC Chain}
\vec{f}_{[\gamma_{\text{thr}}]}(0,\epsilon)_{[0,\;0,\;0.06]}\equiv &\;\vec{f}_{[\gamma_0]}(1,\epsilon),\nonumber\\
\vec{f}_{[\gamma_{\text{thr}}]}(0.06,\epsilon)_{[0.06,\;0.125,\;0.19]}\equiv &\;\vec{f}_{[\gamma_{\text{thr}}]}(0.06,\epsilon)_{[0,\;0,\;0.06]},\nonumber\\
\vec{f}_{[\gamma_{\text{thr}}]}(0.19,\epsilon)_{[0.19,\;0.5,\;0.81]}\equiv &\; \vec{f}_{[\gamma_{\text{thr}}]}(0.19,\epsilon)_{[0.06,\;0.125,\;0.19]},\nonumber\\
\vec{f}_{[\gamma_{\text{thr}}]}(0.81,\epsilon)_{[0.81,\;1,\;1]}\equiv & \;\vec{f}_{[\gamma_{\text{thr}}]}(0.81,\epsilon)_{[0.19,\;0.5,\;0.81]}.
\end{align}

\subsection{Analytic continuation}
\label{sec:Analytic continuation}

In the previous sections we showed that in order to obtain converging power series expansions along a contour it is necessary to expand around the singular points of the differential equations (and regular points in order to ensure fast converging series, as described in Sec.~\ref{sec:matching}). As already mentioned, by singular point we mean any non-analytic point of the differential equations and of the solution, i.e. power and logarithmic divergences, or branching points of the square roots. The singularities of the solution are a subset of the singularities of the differential equations. In this section we discuss the different classes of singularities encountered in the (series) solution of the differential equations and how to perform the analytic continuation across them. Moreover, for the sake of the following discussion, we remark that we solve the differential equations along contours entirely contained in the physical sheet.

\begin{figure}[t]
\minipage{0.5\textwidth}
\begin{center}
  \includegraphics[trim={0cm 0cm 0cm 0cm},width=5.5cm]{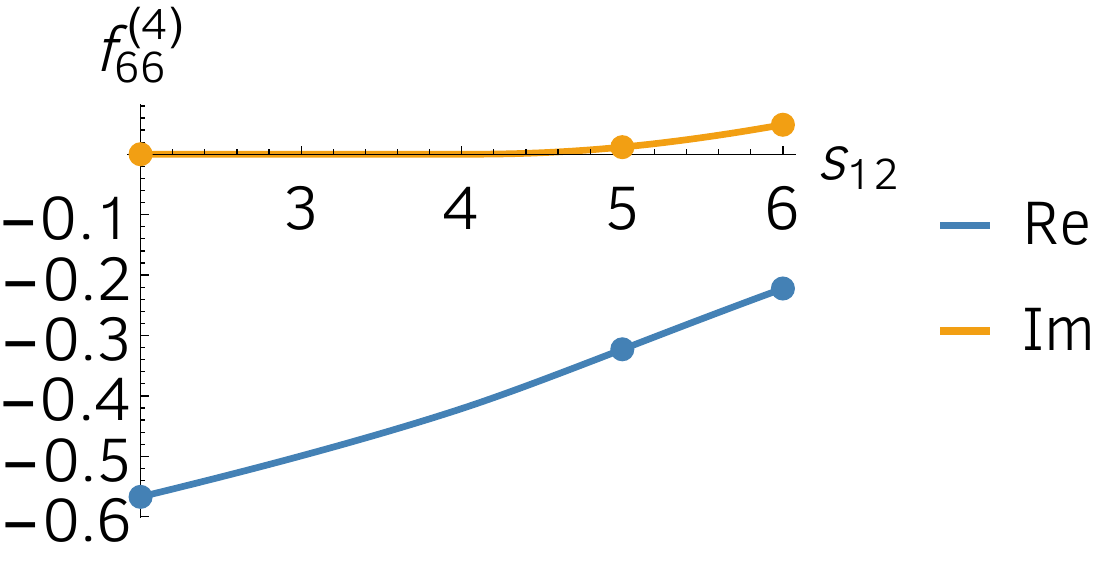}
\end{center}
\endminipage
\minipage{0.5\textwidth}
\begin{center}
  \includegraphics[trim={0 0cm 0 0cm},width=5.5cm]{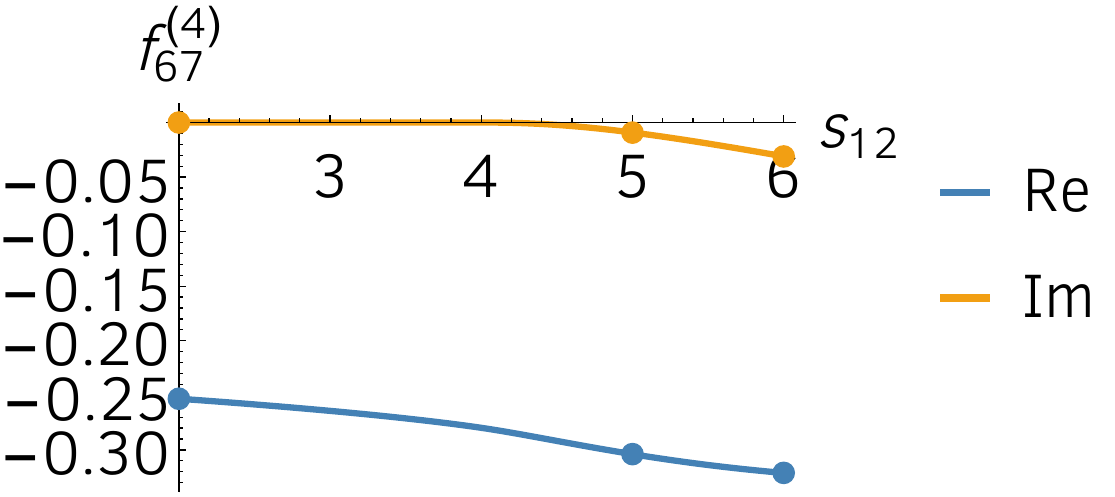}
\end{center}
\endminipage\\
\minipage{0.5\textwidth}
\begin{center}
  \includegraphics[trim={0cm 0cm 0cm 0cm},width=5.5cm]{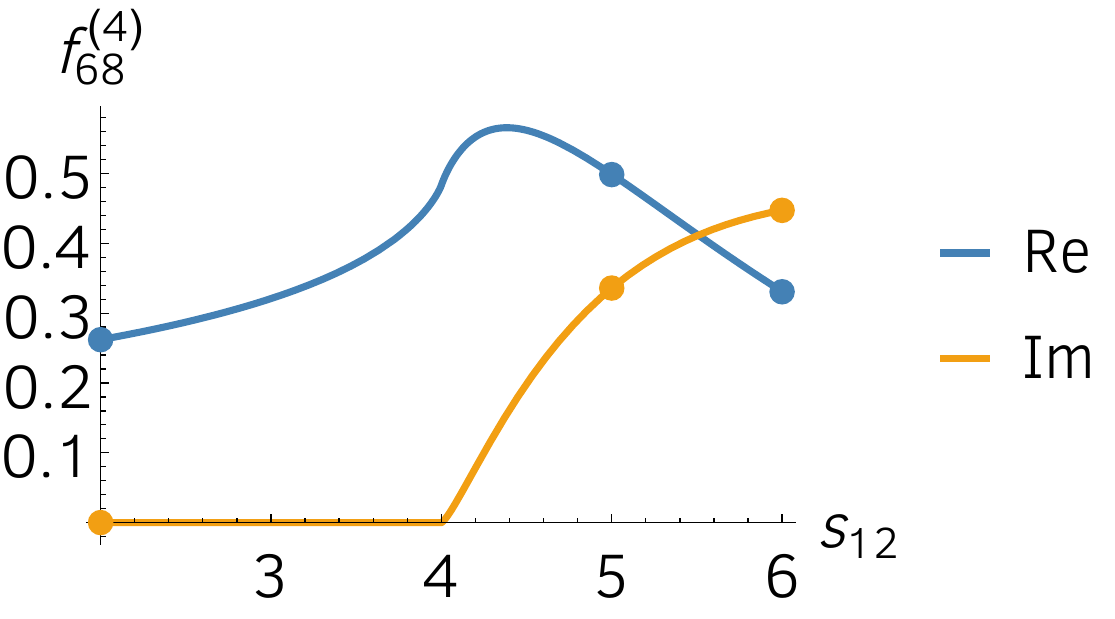}
\end{center}
\endminipage
\minipage{0.5\textwidth}
\begin{center}
  \includegraphics[trim={0 0cm 0 0cm},width=5.5cm]{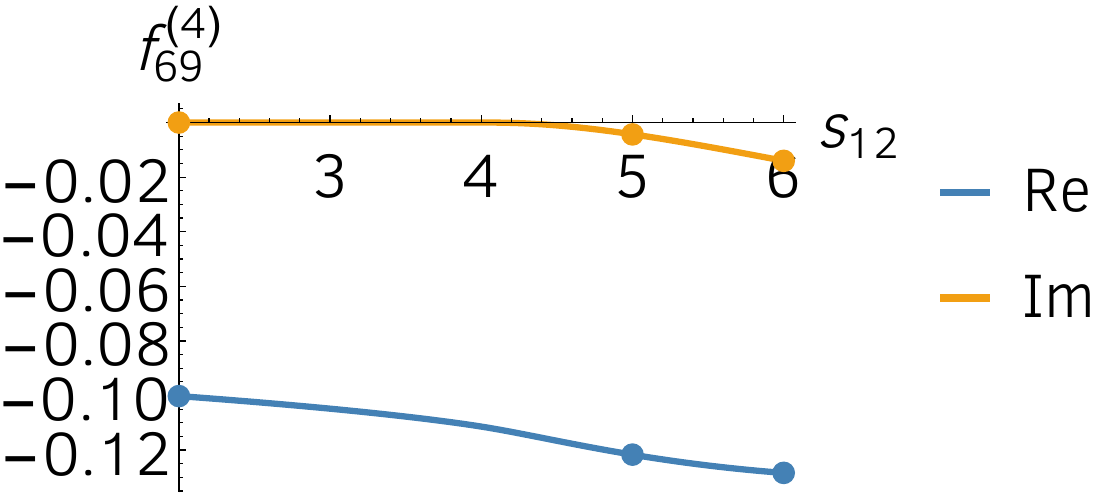}
\end{center}
\endminipage\\
\minipage{0.5\textwidth}
\begin{center}
  \includegraphics[trim={0cm 0cm 0cm 0cm},width=5.5cm]{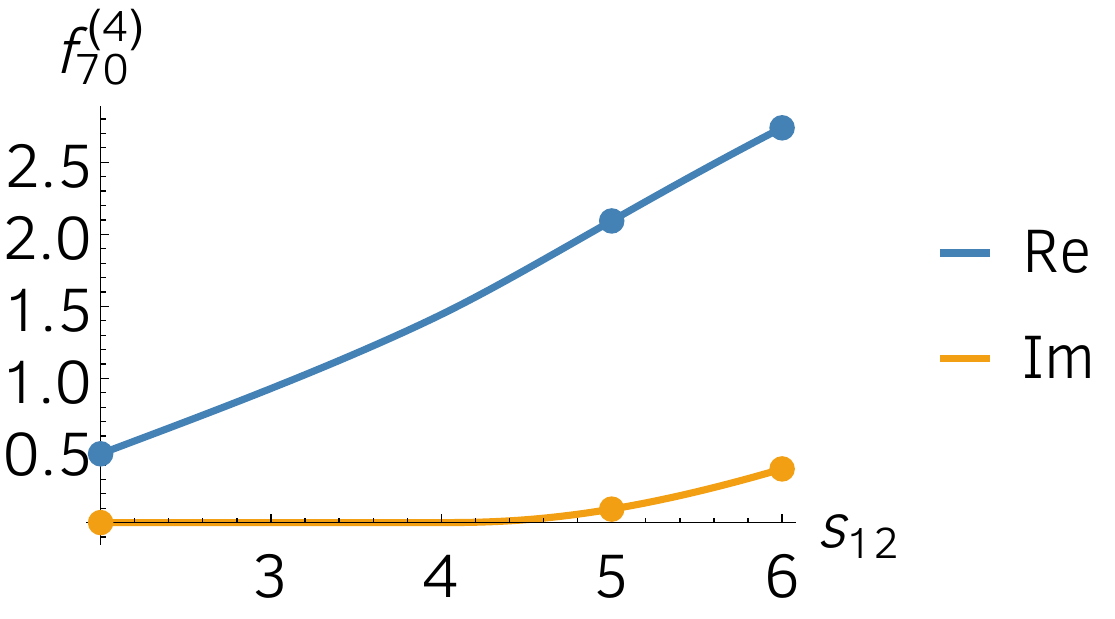}
\end{center}
\endminipage
\minipage{0.5\textwidth}
\begin{center}
  \includegraphics[trim={0 0cm 0 0cm},width=5.5cm]{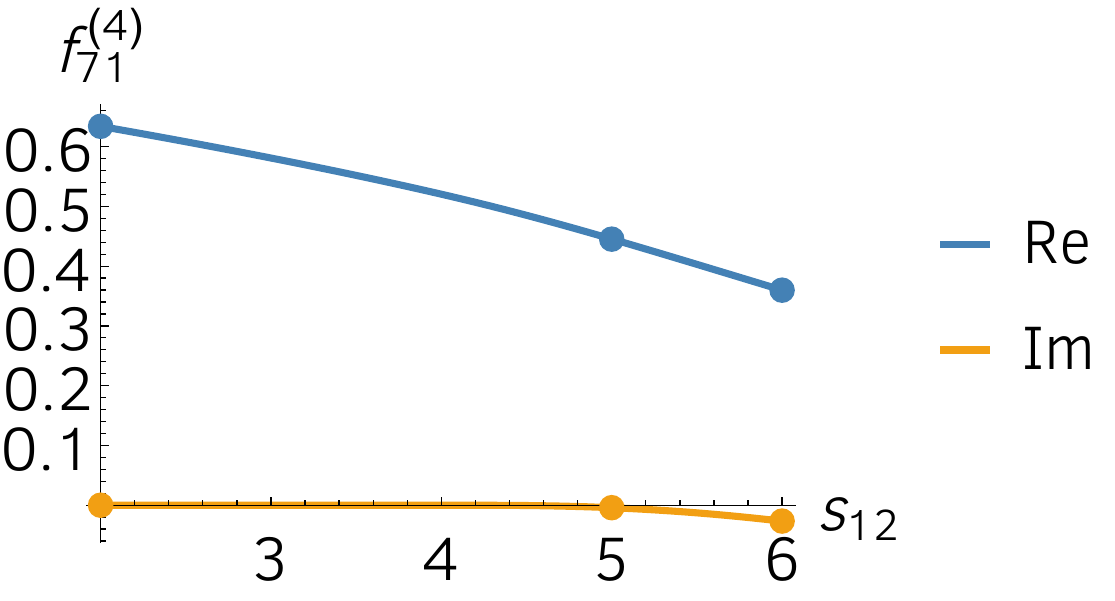}
\end{center}
\endminipage\\
\minipage{0.5\textwidth}
\begin{center}
  \includegraphics[trim={0cm 0cm 0cm 0cm},width=5.5cm]{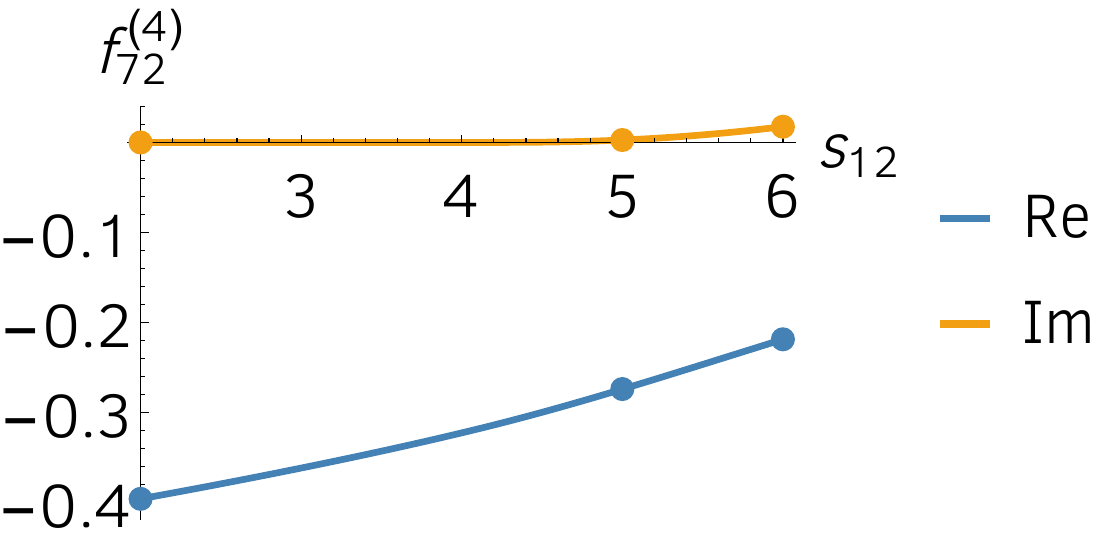}
\end{center}
\endminipage
\minipage{0.5\textwidth}
\begin{center}
  \includegraphics[trim={0 0cm 0 0cm},width=5.5cm]{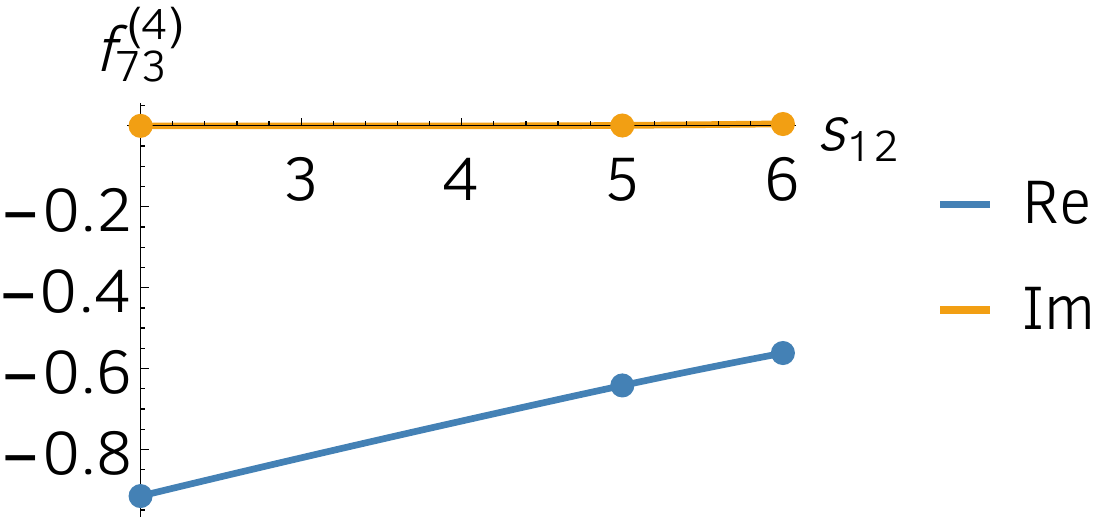}
\end{center}
\endminipage
\caption{Plots of the elliptic integral sectors, for $s_{13}=-1,\;p^2_4=\frac{13}{25},\;m^2=1$ as a function of $s_{12}$, at order $\epsilon^4$, obtained by series expanding along the contour $\gamma_{\text{thr}}$. The solid points represent values computed numerically with the software FIESTA \cite{Smirnov:2015mct}.}
\label{fig:1}
\end{figure}

The first class of singularities are the so-called physical singularities, which correspond to the singularities predicted by unitarity cuts.  In correspondence of physical singularities the solution develops branch cuts, and in order to analytically continue the solution across them we use Feynman prescription, i.e. we assign a small imaginary part to the contour parameter in such a way that the invariant crossing the singularity acquires a small positive imaginary part. In order to show this, let us write down explicitly the first few terms of the expansion around the heavy-quark threshold $t=\frac{1}{2}$, discussed in the previous section, for, e.g., the (elliptic) integral 68 at order $\epsilon^4$ which reads\footnote{In what follows we show truncated numerical coefficients, while the full coefficients are computed with hundreds of digits. Keeping the coefficients numeric allows to drastically speed up the series expansion, especially at high truncation order.},
\begin{align}
f^{(4)}_{68\;[\gamma_{\text{thr}}]}&(t)_{[0.19,\;0.5,\;0.81]}=\; 0.4832111029292333-2.941324925794234 \left(\frac{1}{2}-t\right)\nonumber\\
&-\;6.056389570960138 \left(\frac{1}{2}-t\right)^2-9.562648662124036 \left(\frac{1}{2}-t\right)^3\nonumber\\
&+ \left[1.698166315737334-2.075302321039630 \log \left(\frac{1}{2}-t\right)\right]\left(\frac{1}{2}-t\right)^{\frac{3}{2}} \nonumber\\
&+  \left[3.363752386195187-5.622399506540701 \log \left(\frac{1}{2}-t\right)\right]\left(\frac{1}{2}-t\right)^{\frac{5}{2}}\nonumber\\
&+\mathcal{O}\left(\left(\frac{1}{2}-t\right)^{\frac{7}{2}}\right).
\end{align}
When expanding along a path crossing a physical singularity, in this case $x_{1}=4$, the branch cuts are expressed by rational powers of the expansion parameter and logarithms. The analytic continuation is then performed by assigning a small imaginary part to the parameter $t$, consistently with Feynman prescription. In this case, since $x_1(t)=2+4t$, we have $t\rightarrow t+i\delta$, and,
\begin{align}
\log\left(\frac{1}{2}-t\right)=\log\left(t-\frac{1}{2}\right)- i \pi,\quad \left(\frac{1}{2}-t\right)^{\frac{1}{2}}=-i \left(t-\frac{1}{2}\right)^{\frac{1}{2}},\quad t>\frac{1}{2}.
\end{align} 
On the other hand, the expansion $\vec{f}_{[\gamma_{\text{thr}}]}(t,\epsilon)_{[0.81,\;1,\;1]}$ is a regular power series, as $t=1$ is a regular point, and the branch cut ambiguities of the solution in this region are fixed by imposing the boundary conditions from the analytically continued $\vec{f}_{[\gamma_{\text{thr}}]}(t)_{[0.19,\;0.5,\;0.81]}$, as in Eq.~(\ref{eq:BC Chain}). In Fig.~\ref{fig:1} we present the plots of integrals $\vec{f}_{66-73}$ along the contour $\gamma_{\text{thr}}$.

The second class of singularities are the so-called non-physical singularities, which are not physical (they are not predicted by unitarity) but they appear in the solution of a given integral basis. In order to explain the origin of these singularities let us consider a generic basis element of the integral basis\footnote{In this work we consider only integral basis with algebraic coefficients.},
\begin{equation}
\label{eq:genbas}
B_i(\vec{x})=\sum_{j} c_{ij}(\vec{x}) I_{a_{1ij},\dots,a_{nij}},
\end{equation}
where $I_{a_{1ij},\dots,a_{nij}}$ is a scalar Feynman integral of the form,
\begin{equation}
I_{a_1, \dots , a_n} = \int \!\!  \prod_{i=1}^l\frac{ d^d k_i}{(i \pi^{d/2})^l}\prod_{j=1}^n \frac{1}{d_j^{a_j} },
\end{equation}
If one of the coefficients of Eq.~(\ref{eq:genbas}), say $c_{ik}(\vec{x})$, is singular for $\vec{x}=\vec{x}_s$, where $\vec{x}_s$ is not a physical singularity and $I_{a_{1ik},\dots,a_{nik}}$ is non-zero in $\vec{x}_s$, $\vec{B}(\vec{x})$ is singular for $\vec{x}=\vec{x}_s$. This is what we call a non-physical singularity, since it originates only from the prefactors of the integral basis\footnote{In general, multiple coefficients $c_{ij}(\vec{x})$ might be singular in $\vec{x}_s$, or the corresponding $I_{a_{1ij},\dots,a_{nij}}$ might vanish in the same point. In this case it is not obvious whether or not $\vec{x}_s$ is a singular point for $\vec{B}(\vec{x})$, since the different singular terms might cancel out, and only after the series solution is obtained one can verify whether or not $\vec{x}_s$ is a singular point for $\vec{B}(\vec{x})$.}. For integral bases with algebraic coefficients (as the ones considered in this paper) there are two kind of non-physical singularities: poles and branching points of square roots. Poles do not give rise to cuts, and no analytic continuation is needed across them. On the other hand, for square roots giving rise to non-physical branch cuts, we can choose an arbitrary analytic continuation across their branching points, since these cuts cancel at the level of the integrals $I_{a_{1ij},\dots,a_{nij}}$ and at the level of the scattering amplitude. More specifically, for the integrals of family A the only physical singularities are for $x_{1}=4$ and $x_{2}=4$. According to Feynman prescription we define $\sqrt{4 -x_{1}}=-i\sqrt{x_{1}-4}$ and $\sqrt{4 -x_{2}}=-i\sqrt{x_{2}-4}$ when $x_{1},x_{2}>4 $. On the other hand, all the other square roots have no $x_{1}-4$,  $x_{2}-4$ factors and we choose, for simplicity, to define them as $\sqrt{a}=i\sqrt{-a}$ when $a<0$, for any $a$ and in any region. In Appendix \ref{app:1loopexample} we discuss a simple one-loop example and we show how non-physical singularities appear in the series solution. 


Finally we have the so-called spurious singularities. These are singularities of the differential equations that do not correspond to singularities of the solution. These singularities are not physical  and the coefficients of the integral basis are regular in these points. Therefore, since we consider contours entirely contained in the physical sheet, the solution is regular in these points, and the singular terms of the series solution corresponding to spurious singularities cancel within the truncation error, and no analytic continuation is needed. 

\subsection{Mapping the physical region to a finite region}
\label{sec:Compact phys}
\begin{figure}[!h]
\begin{center}
  \includegraphics[trim={0cm 0cm 0cm 0cm},width=4.5cm]{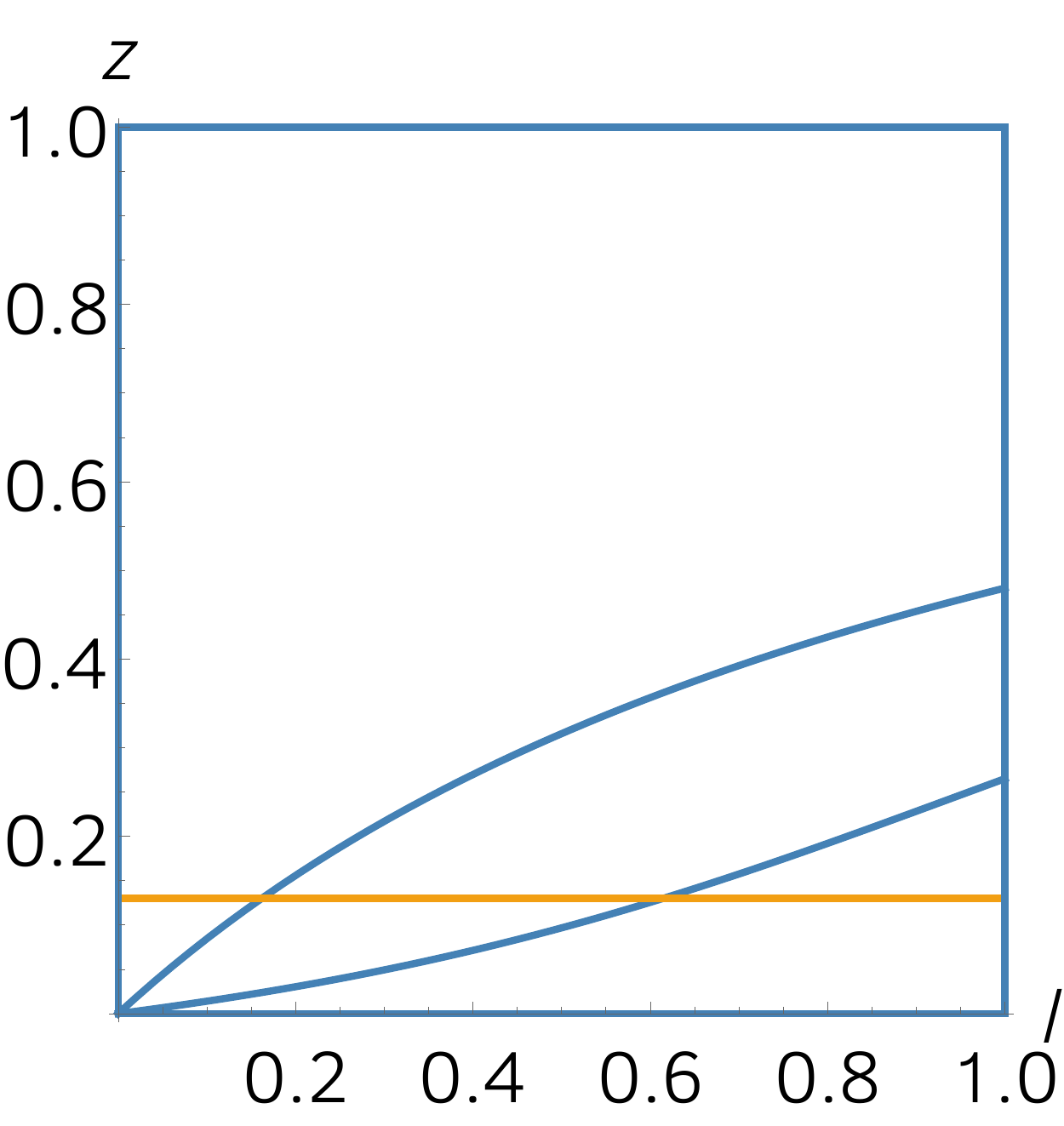}
\end{center}
\caption{The physical region  in the $l,z$ variables for fixed Higgs mass ($x_2=13/25$) and unit propagators mass (see text). The lines represent the singular points of the differential equations. The orange line corresponds to the physical threshold $x_{1}=4$.}
\label{fig:phys}
\end{figure}
\begin{figure}[!h]
\minipage{0.5\textwidth}
\begin{center}
  \includegraphics[trim={0cm 0cm 0cm 0cm},width=5.5cm]{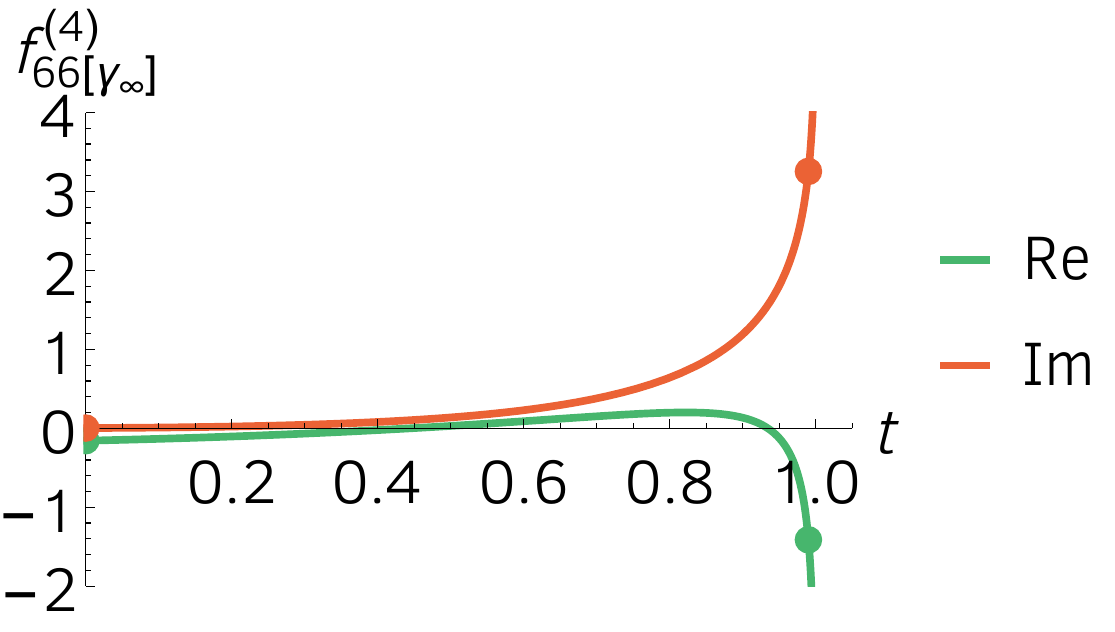}
\end{center}
\endminipage
\minipage{0.5\textwidth}
\begin{center}
  \includegraphics[trim={0 0cm 0 0cm},width=5.5cm]{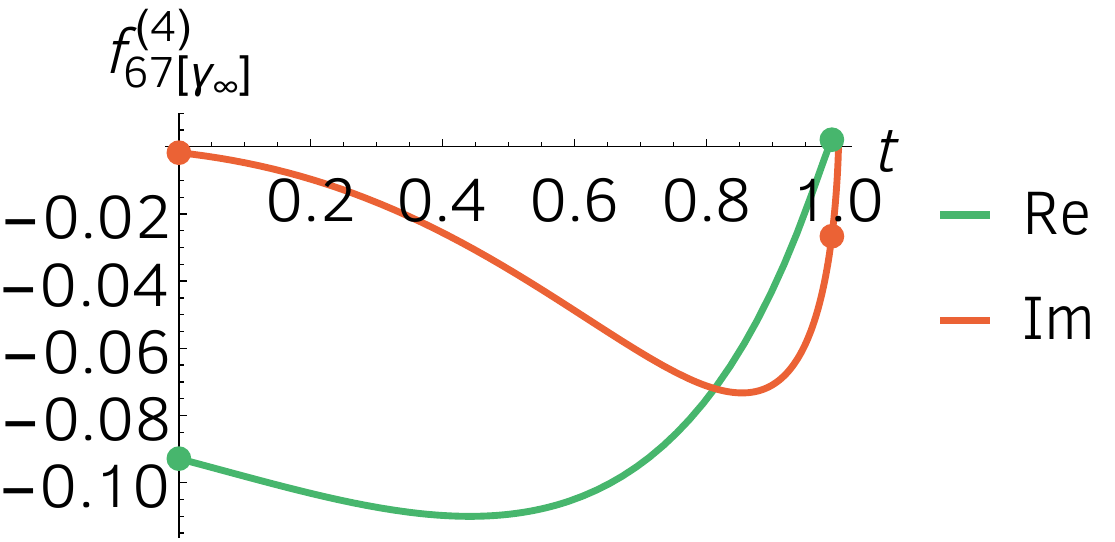}
\end{center}
\endminipage\\
\minipage{0.5\textwidth}
\begin{center}
  \includegraphics[trim={0cm 0cm 0cm 0cm},width=5.5cm]{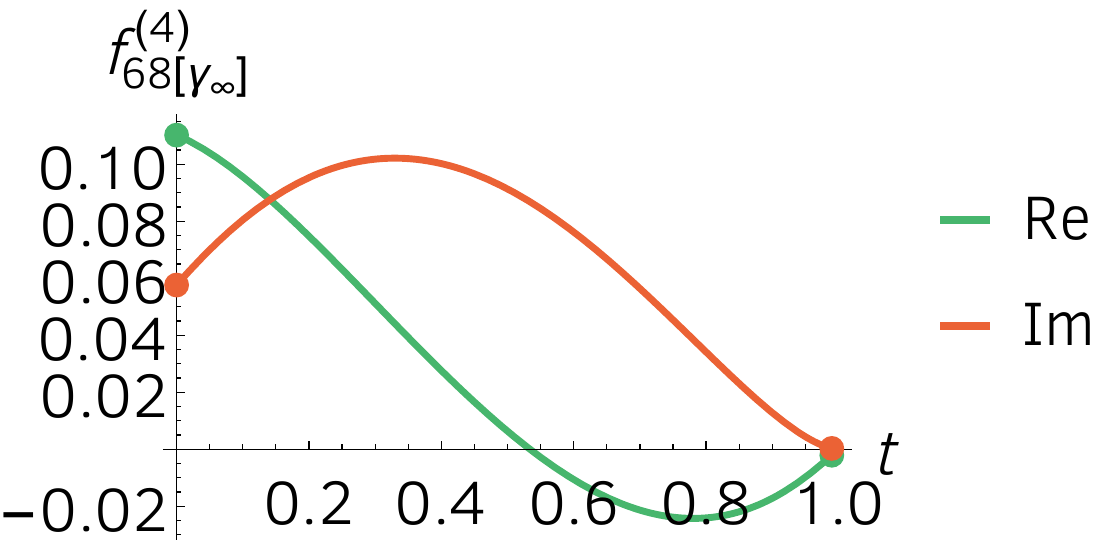}
\end{center}
\endminipage
\minipage{0.5\textwidth}
\begin{center}
  \includegraphics[trim={0 0cm 0 0cm},width=5.5cm]{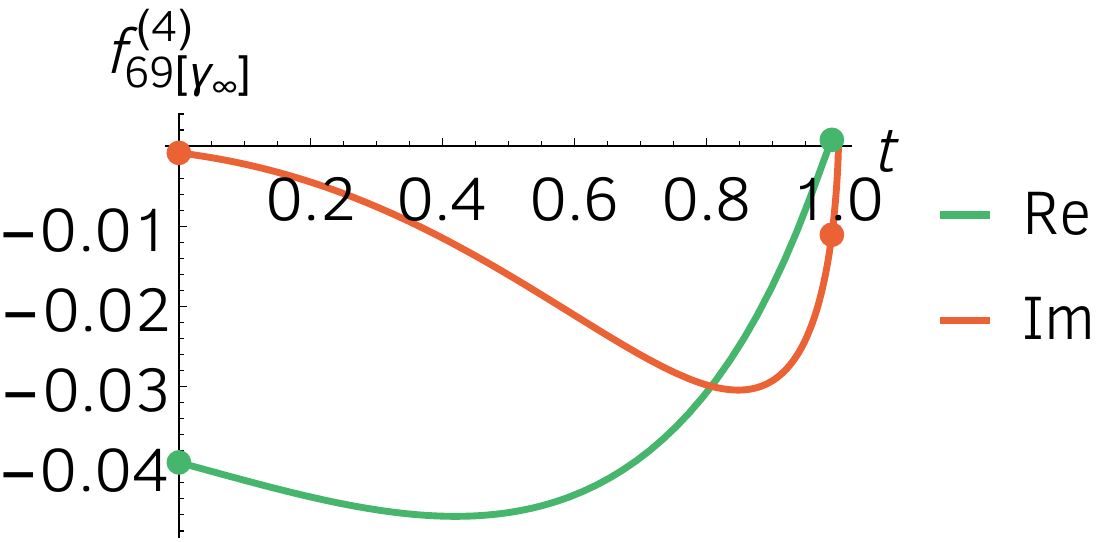}
\end{center}
\endminipage\\
\minipage{0.5\textwidth}
\begin{center}
  \includegraphics[trim={0cm 0cm 0cm 0cm},width=5.5cm]{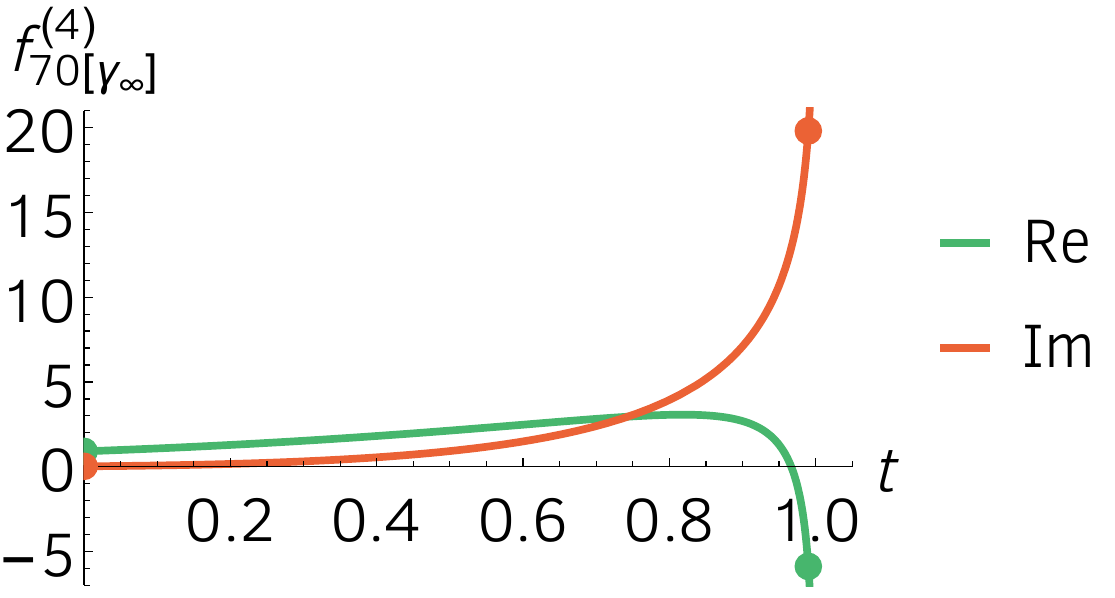}
\end{center}
\endminipage
\minipage{0.5\textwidth}
\begin{center}
  \includegraphics[trim={0 0cm 0 0cm},width=5.5cm]{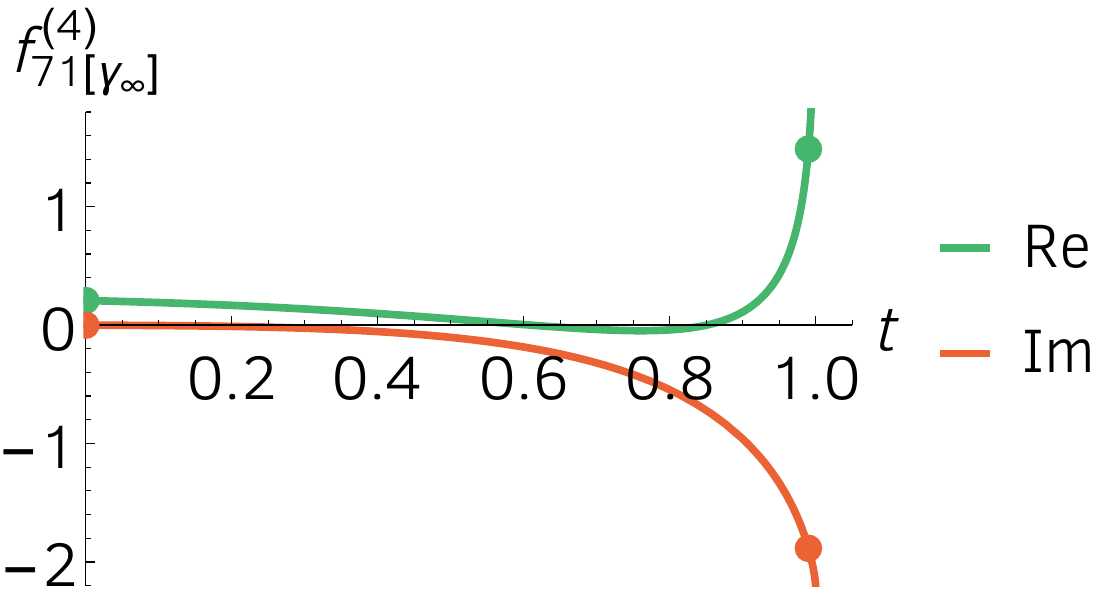}
\end{center}
\endminipage\\
\minipage{0.5\textwidth}
\begin{center}
  \includegraphics[trim={0cm 0cm 0cm 0cm},width=6cm]{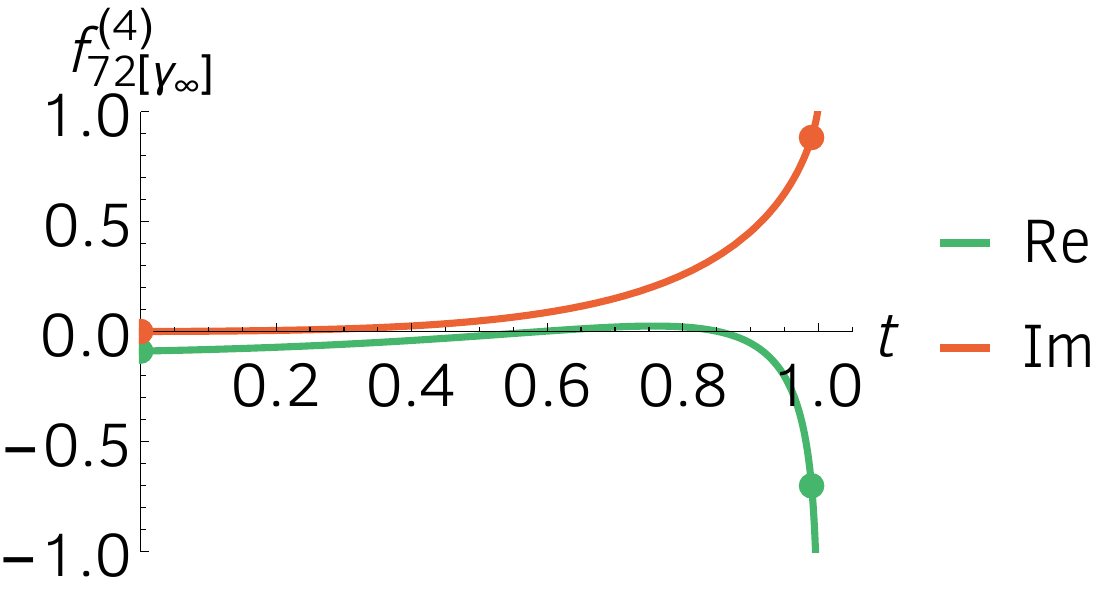}
\end{center}
\endminipage
\minipage{0.5\textwidth}
\begin{center}
  \includegraphics[trim={0 0cm 0 0cm},width=6cm]{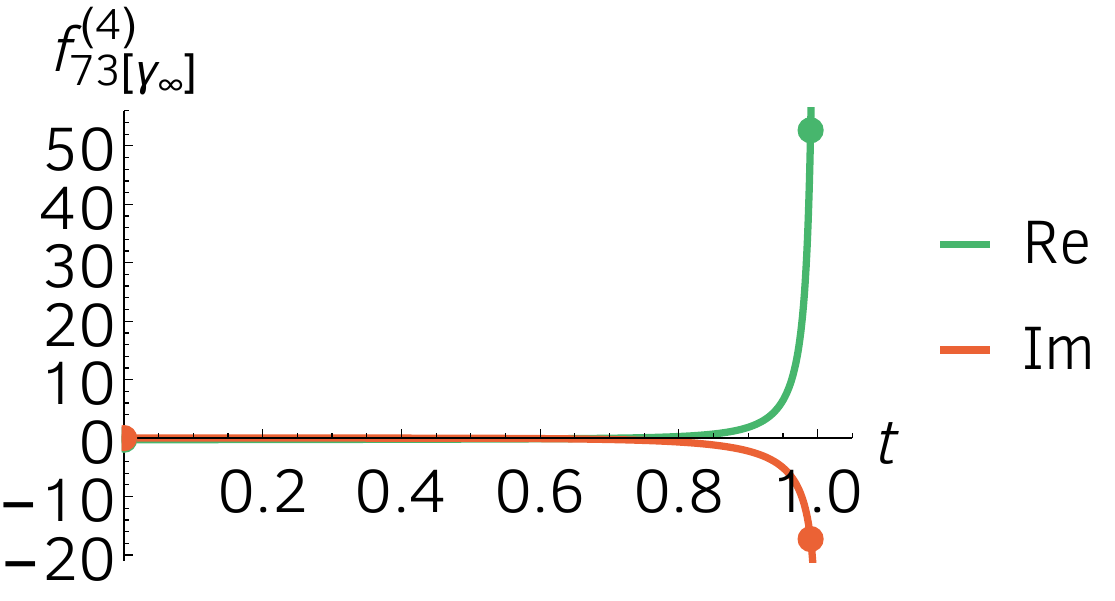}
\end{center}
\endminipage
\caption{Plots along the contour $\gamma_{\infty}$, for values of the contour parameter $t\in[0,1]$. The solid points represent values computed numerically with the software FIESTA \cite{Smirnov:2015mct}.}
\label{fig:Plots Inf}
\end{figure}
In physical applications it is often necessary to consider integrals for very large or small values of the external invariants. When working with series expansions it is then convenient to map the relevant physical region to a finite region, so that it is possible to expand around limiting points in a straightforward manner. For two-to-two processes a convenient set of variables is the following,
\begin{equation}
x_1(l,z)= \frac{x_2}{z},\quad x_3(l,z)= \frac{x_2 l (z-1)}{z},\qquad 0\leq l \leq 1,\; 0\leq z \leq 1,
\end{equation}
which, by keeping the Higgs mass fixed, maps the physical region~(\ref{eq:physical region}) to the unit square in the $l,z$ space. The $l,z$ variables are also convenient when studying the singularities of the differential equations. The singularities are the zeros of the denominators of the differential equations and the zeros of the square roots. For our differential equations, the singular points are for $l=0,\;l=1,\;z=0,\;z=1$ and along the orbits defined by the following equations,\footnote{The singular orbits can be found, e.g., by using the built-in Mathematica function Solve.}
\begin{equation}
\label{eq:SingOrbits}
 z=\frac{13}{100},\;\; z=\frac{12 l}{13 + 12 l},\;\; z=\frac{-50 + 37 l + 
 10 \sqrt{25 - 37 l + 25 l^2}}{87 l}.
\end{equation}
The orbit $z=0$ corresponds to points where $x_{1},x_{3}$ are infinity and these are singular points for our integral basis (see below). $l=0$ and $z=1$ correspond to $x_{3}=0$ and are not physical singularities but the series expansions are not analytic there because some of the square roots vanish.  $l=1$ is a spurious singularity. The first singularity  of Eq.~(\ref{eq:SingOrbits}) is the physical threshold. The second and last singular orbits of Eq.~(\ref{eq:SingOrbits}) are spurious. Indeed, the singular orbits do not correspond to physical singularities, and none of the square roots or the rational prefactors of the integral basis are singular along these orbits. The physical region, including the singular orbits, is represented in Fig.~\ref{fig:phys}.

The $l,z$ variables are also the natural variables to perform expansions along contours reaching very large or small values of the invariants. Let us consider for example the contour,
\begin{equation}
\gamma_{\infty}(t)=\left\{l(t), \; z(t) \right\}=\left\{\frac{845 (1-t)}{8208}, \;  \frac{13 (1-t)}{120} \right\}, \qquad t \in[0,1],
\end{equation}
which corresponds to a contour from $\{x_1= \frac{24}{5}, \; x_2= \frac{13}{25},\; x_3 = -\frac{18083}{41040}\}$ to $\{x_1 = \infty,\; x_2= \frac{13}{25}, \;x_3 = -\frac{169}{342}\}$. In this case the only singular point is for $t=1$, and it is sufficient to perform only one expansion around it to cover the full contour. $t=1$ corresponds to a (singular) point at infinity. Nonetheless, we never cross the points at infinity and possible branch cut ambiguities are fixed imposing boundary conditions at a finite point and by treating the square roots as explained in Section \ref{sec:Analytic continuation}.  The plots of the elliptic integral sectors along $\gamma_{\infty}$ are presented in Fig.~\ref{fig:Plots Inf}.

\subsection{Numerical results and timings}
\label{sec:Numerics}

\begin{table}[ht]
\centering
\begin{tabular}{lcccc}
          & Truncation & Relative error & Time (73 MIs) & $\frac{\text{Time}}{\text{integral}}$  \\
\hline
Expansion ($\gamma_{\text{thr}}$) & $\mathcal{O}(t^{85})$& $\leq 10^{-24}$ & 79 sec & 1.11 sec \\
Expansion ($\gamma_{\text{thr}}$) &$\mathcal{O}(t^{125})$ & $\leq 10^{-32}$ & 162 sec & 2.21 sec   \\
Expansion ($\gamma_{\text{thr}}$, 1 segment) & $\mathcal{O}(t^{85})$& $\leq 10^{-24}$ & 20 sec & 0.27 sec \\
Expansion ($\gamma_{\text{thr}}$, 1 segment) &$\mathcal{O}(t^{125})$ & $\leq 10^{-32}$ & 40 sec & 0.55 sec   \\
FIESTA 4.1 (Vegas) & &  $\leq 10^{-2}$    & $60000$ sec & $821$ sec    \\
pySecDec 1.4.3 (QMC)& & $\leq 10^{-2}$    & $25000$ sec & $342$ sec    \\
\end{tabular}
\caption{Timings for the computation of all the 73 master integrals in the point above threshold $s_{12}=5,s_{13}=-1,p_4^2=\frac{13}{15},m^2=1$ up to and including order $\epsilon^4$, on 1 CPU core. The point is reached by expanding along the contour $\gamma_{\text{thr}}$  defined in Section \ref{sec:series sol DE A}. We use FIESTA 4.1 with the Vegas \cite{Hahn:2004fe} integrator while pySecDec 1.4.3 with the quasi-Monte Carlo integrator. The '1 segment' lines correspond to the average time needed to perform the expansion along one of the four line segments of the contour $\gamma_{\text{thr}}$. This timing is more representative of the typical time needed to reach a given point when several points have been already computed. Indeed, once several points have been computed (for example when performing a phase space integration) one can use as a boundary point the closest point to the next target point, which will likely be 'close' to an already computed point.} 
\label{tab:compare FIESTA}
\end{table}

\begin{figure}[htb]
\begin{center}
  \includegraphics[trim={0cm 0cm 0cm 0cm},width=9cm]{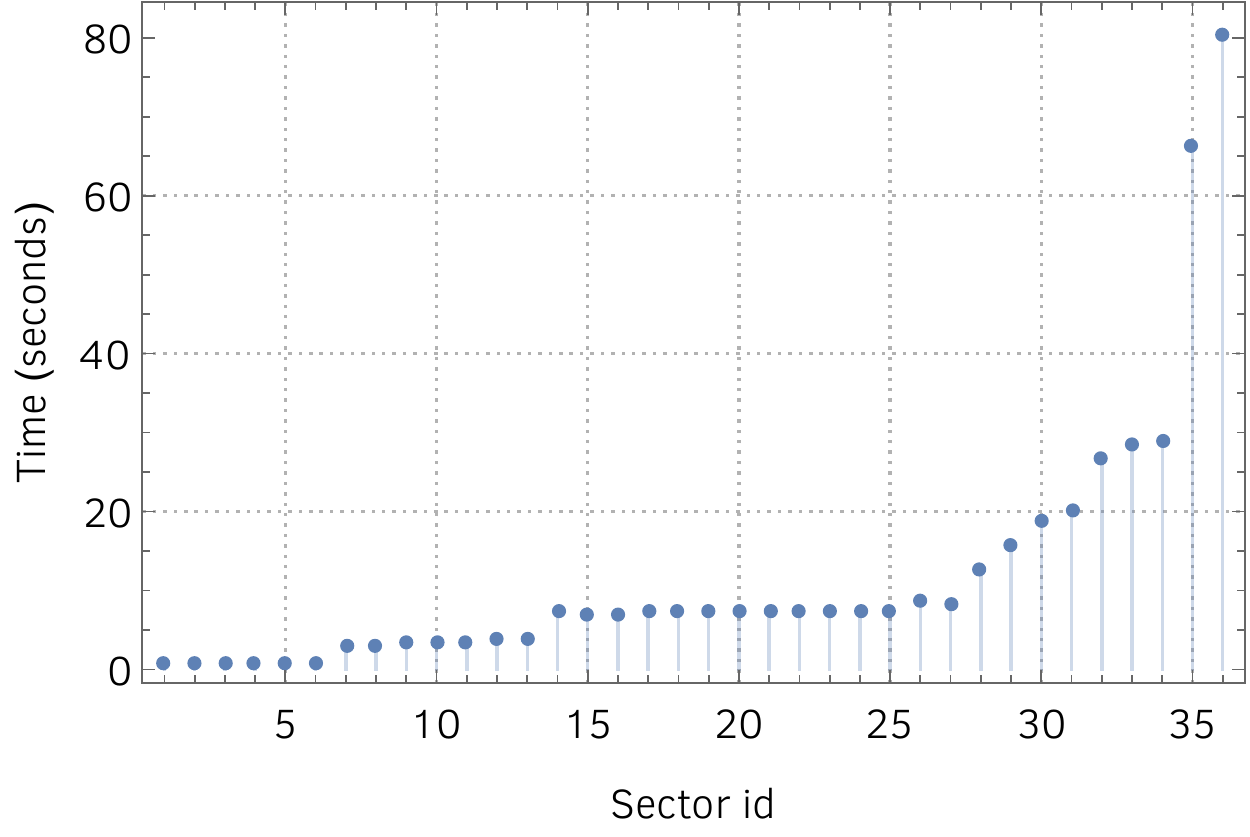}
\end{center}
\caption{Timings for the computation of the integrals up to and including a given sector at the point  $s_{12}=5,s_{13}=-1,p_4^2=\frac{13}{15},m^2=1$ up to and including order $\epsilon^4$, on 1 CPU core. The sector id is defined in Eq.~(\ref{eq:secid}) The point is reached by expanding along the contour $\gamma_{\text{thr}}$  defined in Section \ref{sec:series sol DE A}. The last two values on the right of the plot correspond to the elliptic sectors.}
\label{fig:sectime}
\end{figure}

\begin{table}[!]
\begin{tabular}{cllc}
\# & Real part & Imaginary part & Rel. err.\\
\hline
66 &  $-0.3229462567669706555224248824$ & $+0.0117911783550457146213317093 $ & $= 10^{-32}$ \\
67 & $-0.3039273344500928782645744918$ & $-0.0091549103610263866197445332$ & $= 10^{-32}$ \\
68 &  $+0.4989288717476986372753776549$ & $ +0.3357385827462031355699714492 $& $= 10^{-32}$ \\
69 & $-0.1216791054426520965987437711$ & $ -0.0043091895703585235915244984 $& $= 10^{-32}$\\
70 & $ +2.0936824603121477765883288075$ & $ +0.0934305088779206383805166362$ & $= 10^{-32}$ \\
71 & $+0.4453410820313442706096983608$ & $-0.0041620964011457311182442998 $& $= 10^{-32}$ \\
72 & $-0.2737950895536402954579381308$ &$+0.0027361587278654015689589630$ & $= 10^{-32}$\\
73 &  $-0.6420218523272138407244813424$ &$+ 0.0011438044096963059034512574$ & $= 10^{-32}$
\end{tabular}
\caption{Numerical results for the elliptic sectors in the point $s_{12}=5,s_{13}=-1,p_4^2=\frac{13}{15},m^2=1$ at order $\epsilon^4$.} 
\label{tab:HD res}
\end{table}
In this section we provide numerical results with accuracy and timings for the elliptic integrals $\vec{f}_{66-73}$, which are the most complicated ones.  We tested different ways to estimate the accuracy of the numerical results obtained by truncating the relevant series. One estimate of the error is obtained by expanding along two different paths reaching the same end point, and then taking the difference of the results. Another way is to keep the path fixed but truncating the series at different orders. In all the cases we analysed these methods give the same estimate of the error. In Table \ref{tab:compare FIESTA} we provide a comparison of our expansion method against the c++ version of FIESTA 4.1 \cite{Smirnov:2015mct} and pySecDec 1.4.3 \cite{Borowka:2017idc,Borowka:2018goh}. We remark that the timings include the time needed, starting from the set of differential equations~(\ref{eq:DE A poly}) and~(\ref{eq:DE A elliptic}), to define the differential equations along the contour, series expand the matrix elements and recursively integrate them up to the desired $\epsilon$ order, for all the expansion points along the contour. This is an accurate estimate of the timings for physical applications where, in principle, one has to repeat the procedure for each target point of interest. In Fig.~\ref{fig:sectime} we report the time required to compute all the integrals up to and including the sectors defined in Eq.~(\ref{eq:secid}). The sectors are ordered according to their position in the integral basis under consideration and, at the level of the differential equations, they are coupled to one or more sectors with lower indices, and none of the sectors with higher indices. We note that most of the time is spent expanding the elliptic sectors, due to the lack of a canonical (hence simpler) basis for those integrals. In Table \ref{tab:HD res} we provide high-precision numerical results for integrals $\vec{f}_{66-73}$ at order $\epsilon^4$. 

Finally, we remark that our method is easily parallelizable since, given one or more boundary points, the series expansion along different contours are completely independent operations.


\begin{align}
\label{eq:secid}
& 1\;:\; S_{0,0,0,0,1,0,1,0,0},\quad 2\;:\; S_{0,1,0,0,0,1,1,0,0},\quad 3\;:\; S_{0,1,1,0,0,1,0,0,0},\quad 4\;:\; S_{0,0,1,1,1,0,0,0,0},\nonumber\\
& 5\;:\; S_{0,0,0,1,1,0,1,0,0},\quad 6\;:\; S_{1,1,0,0,0,1,1,0,0},\quad 7\;:\; S_{0,0,0,0,1,1,1,0,0},\quad 8\;:\; S_{1,0,0,1,0,1,1,0,0},\nonumber\\
& 9\;:\; S_{1,0,0,0,1,1,1,0,0},\quad 10\;:\; S_{0,1,1,0,0,1,1,0,0},\quad 11\;:\; S_{0,1,1,0,1,1,0,0,0},\quad 12\;:\; S_{0,1,0,0,1,1,1,0,0},\nonumber\\
& 13\;:\; S_{0,0,1,1,1,1,0,0,0},\quad 14\;:\; S_{0,0,0,1,1,1,1,0,0},\quad 15\;:\; S_{0,0,1,1,1,0,1,0,0},\quad 16\;:\; S_{1,1,1,1,0,0,0,0,0},\nonumber\\
& 17\;:\; S_{1,1,0,1,0,0,1,0,0},\quad 18\;:\; S_{1,1,1,1,1,0,0,0,0},\quad 19\;:\; S_{1,1,0,0,1,0,1,0,0},\quad 20\;:\; S_{1,1,1,1,0,0,1,0,0},\nonumber\\
& 21\;:\; S_{1,1,1,0,0,1,1,0,0},\quad 22\;:\; S_{1,1,0,1,1,0,1,0,0},\quad 23\;:\; S_{1,0,1,0,1,1,1,0,0},\quad 24\;:\; S_{1,0,1,1,1,1,0,0,0},\nonumber\\
& 25\;:\; S_{1,1,0,1,0,1,1,0,0},\quad 26\;:\; S_{0,1,0,1,1,1,1,0,0},\quad 27\;:\; S_{1,1,1,1,1,0,1,0,0},\quad 28\;:\; S_{0,0,1,1,1,1,1,0,0},\nonumber\\
&29\;:\; S_{0,1,1,0,1,1,1,0,0},\quad 30\;:\; S_{1,1,0,0,1,1,1,0,0},\quad 31\;:\; S_{0,1,1,1,1,1,1,0,0},\quad 32\;:\; S_{1,0,0,1,1,1,1,0,0},\nonumber\\
& 33\;:\; S_{1,0,1,1,1,1,1,0,0},\quad 34\;:\; S_{1,1,1,0,1,1,1,0,0},\quad 35\;:\; S_{1,1,0,1,1,1,1,0,0},\quad 36\;:\; S_{1,1,1,1,1,1,1,0,0}.
\end{align}

\section{Conclusion}
\label{sec:conclusion}
We showed that by defining the differential equations for a set of multi-scale Feynman integrals along contours connecting two generic points of the kinematic regions, it is possible to systematically obtain one-dimensional series expansions for the integrals along the contours. Specifically, we applied this method to obtain new results for a planar family of integrals relevant for the two-loop QCD corrections to Higgs + jet production.  When the expansion is performed along a contour such that only one invariant changes along it, the analytic continuation above the physical thresholds becomes straightforward. We demonstrated that performing an expansion along a contour is fast, and makes it possible to repeat the procedure to compute the integrals over the entire kinematic regions, with arbitrary numerical precision.  Our approach is algorithmic, and it seems plausible to implement it in a computer code that can be applied to solve complicated integrals of phenomenological integrals.

\section*{Acknowledgements}
We thank Roberto Bonciani, Vittorio Del Duca, Hjalte Frellesvig, Johannes Henn, Martijn Hidding,  Leila Maestri, Giulio Salvatori and Volodya Smirnov for useful discussions and collaboration on related work. We thank Roman Lee for insightful discussions about series expansion methods. We thank Armin Schweitzer for useful discussions about the analytic continuation of Feynman integrals of elliptic type, and for providing the (unpublished) canonical basis of the one-loop box example discussed in Appendix \ref{app:1loopexample}. This research received funding from the European Research Council (ERC) under grant agreement No. 694712 (pertQCD) and by the Swiss National Science Foundation project No. 177632 (ElliptHiggs).

\appendix

\clearpage

\section{General formulation of the Frobenius method}
\label{App:Frobenious}
The Frobenius method has general validity, and in this appendix we consider a generic order-$k$ linear differential equation for an unknown function $f(t)$ (the discussion closely follows \cite{Coddington}). The Frobenius method can be used to find a complete set of homogeneous series solutions to the equation in the vicinity of a singular (or regular) point that we assume being located at $t=0$. The homogeneous equation can be written in general as:
\begin{equation}
\label{eq:L[f(t)]=0}
L[f(t)] \equiv t^{k}\frac{\partial^{k}f(t)}{\partial t^{k}}+t^{k-1}b_{1}(t)\frac{\partial^{k-1}f(t)}{\partial t^{k-1}}+\cdots+b_k(t)f(t)=0.
\end{equation}
Since the $b_i(t)$ are analytic in $t=0$ they admit a Taylor series representation of the form,
\begin{equation}
b_i(t)=\sum_{j=0}^\infty b_i^{(j)} t^j,\quad \{j=1\dots k\}.
\end{equation}
We look for a solution of the form,
\begin{equation}
f(t)= t^\lambda \sum_{i=0}^\infty c^{(i)} t^i,
\end{equation}
where lambda is a (complex in general) parameter to be fixed. By substituting the formal series solution into the equation we obtain,
\begin{equation}
\label{eq:D[formal solution]}
L[f(t)]=\sigma(\lambda)c^{(0)} t^\lambda+[\sigma(\lambda+1)c^{(1)}-g^{(1)}]t^{\lambda+1}+\cdots+[\sigma(\lambda+j)c^{(j)}-g^{(j)}]t^{\lambda+j}+\cdots,
\end{equation}
Where the $g(j)$ are linear in the $c^{(1)},c^{(2)},\cdots, c^{(j-1)}$  with polynomial coefficients in $\lambda$, while the polynomial $f(\lambda)$ is called the indicial polynomial and it reads,
\begin{equation}
\sigma(\lambda)=\lambda (\lambda-1)\cdots(\lambda-k+1)+b_{1}^{(0)} \lambda (\lambda-1)\cdots (\lambda-k+2)+b_{k}^{(0)}.
\end{equation}
In order for $f(t)$ to be a solution, the coefficients of the right-hand side of (\ref{eq:D[formal solution]}) need to satisfy
\begin{equation}
\label{eq:Frob recursion}
c^{(j)}=\frac{g^{(j)}}{\sigma(\lambda+j)},
\end{equation} 
which is a recursion relation that can be solved for $c^{(j)}$  with $j\geq 1$ if $\sigma(\lambda+j)\neq 0$ for any $j\geq 1$. The $c^{(j)}$, $j\geq 1$ are then uniquely determined functions of $c^{(0)}$. In this case we get
\begin{equation}
\label{eq:L[f(t)]=indicEq}
L[f(t)]=c^{(0)} \sigma(\lambda)  t^{\lambda}.
\end{equation}
It is convenient to fix the value of $c^{(0)}$, and we conventionally choose $c^{(0)}=1$. If $\lambda=\lambda_1$ where $\lambda_1$ is a root of the indicial polynomial, $\sigma(\lambda_1)=0$, and $\sigma(\lambda+j)\neq 0$ for any $j\geq 1$ then, $f_{\lambda_1,1}(t)\equiv f(t)|_{\lambda=\lambda_1}$, is a solution of the equation $L[f(t)]=0$. If $\lambda_1$ has multiplicity 2, we need to find a second solution associated to it. If we differentiate Eq.~(\ref{eq:L[f(t)]=indicEq}) with respect to $\lambda$ we get,
\begin{equation}
\frac{\partial L[f(t)]}{\partial \lambda}=L\left[\frac{\partial f(t)}{\partial \lambda}\right]=\left(\frac{\partial\sigma(\lambda) }{\partial \lambda}+\log(t)\sigma(\lambda)\right) t^\lambda.
\end{equation}
If $\lambda_1$ is a root of multiplicity 2 then $\sigma(\lambda_1)=\left. \frac{\partial\sigma(t) }{\partial \lambda}\right|_{\lambda=\lambda_1}=0$, and $\frac{\partial f(t)}{\partial \lambda}$ is also a solution of Eq.~(\ref{eq:L[f(t)]=0}). We have then that the second solution associated to the root $\lambda_1$ is 
\begin{equation}
f_{\lambda_1,2}(t)=\log(t) f_{\lambda_1,1}(t)+\sum_{j=1}^\infty\left.\frac{\partial c^{(j)}}{\partial \lambda}\right|_{\lambda =\lambda_1} t^{j}.
\end{equation}
If $\lambda_1$ has multiplicity $m_1$, all the $m_1$ solutions associated to $\lambda_1$ are obtained by taking $m_1-1$ derivatives. 

Let us now suppose that $\lambda_1+k_2=\lambda_2$, for a positive integer $k_2$, i.e. the root $\lambda_1$ differs by an integer $k_2$ from another root $\lambda_2$. Moreover we assume that $\lambda_2$ is the only root that differs from $\lambda_1$ by an integer. In this case the solution associated to $\lambda_1$ cannot be determined as explained above, since the recursion (\ref{eq:Frob recursion}) becomes ill defined when $j=k_2$. In order to find a solution we set  $c^{(0)}=(\lambda-\lambda_1)^{m_2}$, where $m_2$ is the multiplicity of the root $\lambda_2$,
\begin{equation}
\label{eq:f(t) diff integer}
f(t)=(\lambda-\lambda_1)^{m_2} t^{\lambda}+t^{\lambda} \sum_{j=1}^\infty c^{(j)} t^j.
\end{equation}
An explicit computation shows that Eq.~(\ref{eq:L[f(t)]=indicEq}) becomes
\begin{equation}
L[f(t)]=(\lambda-\lambda_1)^{m_2} \sigma(\lambda)  t^{\lambda},
\end{equation}
and the recursion (\ref{eq:Frob recursion}) can be written as,
\begin{equation}
c^{(i)}=(\lambda-\lambda_1)^{m_2} G^{(i)}, \quad \quad 1\leq i \leq k_2-1,
\end{equation}
while
\begin{equation}
\label{eq:c^(k2)}
c^{(k_2)}=\frac{(\lambda-\lambda_1)^{m_2}}{\sigma(\lambda+k_2)}  G^{(k_2)},
\end{equation}
and 
\begin{equation}
c^{(i)}= G^{(i)}, \quad \quad i > k_2,
\end{equation}
where the functions $G^{(i)}$ are non-zero for $\lambda=\lambda_1$. We note that the right hand side of (\ref{eq:c^(k2)}) is now well defined when $\lambda$ approaches $\lambda_1$ since $\sigma(\lambda_1+k_2)=\sigma(\lambda_2)=0$ and the denominator has a factor $(\lambda-\lambda_1)^{m_2}$ that cancels against the numerator. This shows that $f(t)|_{\lambda=\lambda_1}$ with $f(t)$ defined in Eq.~(\ref{eq:f(t) diff integer}) is a solution of Eq.~(\ref{eq:L[f(t)]=0}). However we notice that $c^{(i)}=0$ for all $i\in\{1,\dots,k_2-1\}$ so that the solution has the form
\begin{equation}
f(t)=t^{\lambda_1+k_2} \sum_{j=0}^\infty c^{(j+k_2)} t^j=t^{\lambda_2} \sum_{j=0}^\infty c^{(j+k_2)} t^j.
\end{equation}
The leading term of the series is $t^{\lambda_2}$, therefore this solution is linearly dependent from the solution one would obtain by considering the root $\lambda_2$. In order to find a solution truly associated to $\lambda_1$ we take the $m_2$-th derivative of $f(t)$, which satisfies
\begin{equation}
L\left[\frac{\partial^{m_2} f(t)}{\partial \lambda^{m_2}}\right]=m_2! \sigma(\lambda)  t^{\lambda}+(\lambda-\lambda_1)\Psi,
\end{equation}
and $\frac{\partial^{m_2} f(t)}{\partial \lambda^{m_2}}|_{\lambda=\lambda_1}$ is indeed a solution. Moreover its leading term is $m! t^{\lambda_1}$ so that $\frac{\partial^{m_2} f(t)}{\partial \lambda^{m_2}}|_{\lambda=\lambda_1}$ is a genuine solution associated to $\lambda_1$. If $\lambda_1$ has multiplicity $m_1>1$ the other solutions are obtained by taking $m_1-1$ derivatives of $\frac{\partial^{m_2} f(t)}{\partial \lambda^{m_2}}$ with respect to $\lambda$ evaluated at $\lambda_1$.

The last case we need to consider is when there are more than one root differing from $\lambda_1$ by an integer. More precisely, we consider $r-1$ roots $\lambda_2,\dots,\lambda_r$ with $Re(\lambda_2)<\dots<Re(\lambda_r)$ such that $\lambda_1+k_i=\lambda_i$ with $k_i$ a positive integer. Each root has multiplicity $m_i$ and we define $M=\sum_{i=2}^r m_i$. We set $c^{(0)}=(\lambda-\lambda_1)^M$, so that 
\begin{equation}
\label{eq:f(t) diff integer multiple}
f(t)=(\lambda-\lambda_1)^M t^{\lambda}+t^{\lambda} \sum_{j=1}^\infty c^{(j)} t^j.
\end{equation}
By proceeding as in the case of only two roots differing by an integer, one obtains that a solution associated to $\lambda_1$ is given by
\begin{equation}
\left.\frac{\partial^{M} f(t)}{\partial \lambda^{M}}\right|_{\lambda=\lambda_1},
\end{equation}
while if $m_1>1$ the other solutions are obtained by taking $m_1-1$ derivatives.
\clearpage

\section{One-loop example}
\label{app:1loopexample}
\begin{figure}[!h]
\centering
\vspace{3mm}
\includegraphics[width=3.2cm]{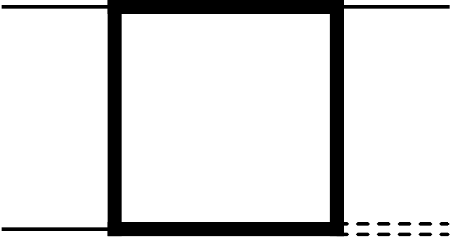} \;
\caption{The integral family contributing to the one-loop $H{+}j$-production in QCD.}
\label{fig:family1l}
\end{figure}
In this appendix we apply the methods of Sec.~\ref{sec:series along countour} to a one-loop integral family, and we discuss in detail the derivation of the series expansion along a given contour. Specifically, we consider the integral family relevant for the leading order corrections to Higgs + one jet production in QCD depicted in Fig.~\ref{fig:family1l},
\begin{equation}
I_{a_1,a_2,a_3,a_4}=\int \frac{d^d k_1}{i \pi^{d/2}} \frac{1}{d_1^{a_1}d_2^{a_2}d_3^{a_3}d_4^{a_4}},
\end{equation}
with
\begin{equation}
d_1=m^2-k_1^2,\;d_2=m^2-(k_1-p_1)^2,\; d_3=m^2-(k_1+p_2)^2,\; d_4=m^2-(k_1-p_1-p_3),
\end{equation}
while the kinematics is $p_1^2 = p_2^2 = p_3^2 = 0$ with,
\begin{align}
s_{12}&=(p_1{+}p_2)^2, & s_{13} &= (p_1{+}p_3)^2, & s_{23} &= (p_2{+}p_3)^2, & p_4^2 &= (p_1{+}p_2{+}p_3)^2 = s_{12}+s_{13}+s_{23},
\end{align}
where $p_4^2$ is the squared Higgs-mass, and $m^2$ is the squared mass of the propagators.
We solve the kinematic constraints for $s_{12},s_{13},p_4^2$. A canonical basis for the integral family is given by,
\begin{align}
\label{eq:basis1loop}
&B_1=+\epsilon \; I_{2,0,0,0},\nonumber\\
&B_2=-\epsilon\; r_1 r_2 \; I_{0,2,1,0} ,\nonumber\\
&B_3=- \epsilon \; r_5 r_6\; I_{2,0,0,1} ,\nonumber\\
&B_4=-\epsilon \;r_3 r_4 \; I_{0,0,2,1},\nonumber\\
&B_5= +\epsilon ^2 s_{13} \; I_{1,1,0,1},\nonumber\\
&B_6= +\epsilon ^2 s_{12} \; I_{1,1,1,0},\nonumber\\
&B_7=+\epsilon ^2(p_4^2-s_{13}) \;I_{1,0,1,1} ,\nonumber\\
&B_8=+\epsilon ^2 (p_4^2-s_{12})\; I_{0,1,1,1} ,\nonumber\\
&B_9=-\epsilon ^2 \;r_1 r_5 r_7\; I_{1,1,1,1},
\end{align}
where we have used the following definitions for the square roots,
\begin{align}
&r_1=\sqrt{-s_{12}},\;r_2=\sqrt{4 m^2-s_{12}},\;r_3=\sqrt{-p_4^2},\;r_4=\sqrt{4 m^2-p_4^2},\nonumber\\
&r_5=\sqrt{-s_{13}},\;r_6=\sqrt{4 m^2-s_{13}},\;r_7=\sqrt{s_{12} s_{13}-4 m^2
   \left(-p_4^2+s_{12}+s_{13}\right)}.
\end{align}
The corresponding differential equations are,
\begin{equation}
d \vec{B}=\epsilon d\tilde{A}\vec{B},
\end{equation}
where the matrix $\tilde{A}$ is,

\begin{equation}
\tilde{A}=\left(
\begin{array}{ccccccccc}
 a_{1,1} & 0 & 0 & 0 & 0 & 0 & 0 & 0 & 0 \\
 a_{2,1} & a_{2,2} & 0 & 0 & 0 & 0 & 0 & 0 & 0 \\
 a_{3,1} & 0 & a_{3,3} & 0 & 0 & 0 & 0 & 0 & 0 \\
 a_{4,1} & 0 & 0 & a_{4,4} & 0 & 0 & 0 & 0 & 0 \\
 0 & 0 & a_{5,3} & 0 & a_{5,5} & 0 & 0 & 0 & 0 \\
 0 & a_{6,2} & 0 & 0 & 0 & a_{6,6} & 0 & 0 & 0 \\
 0 & 0 & a_{7,3} & a_{7,4} & 0 & 0 & a_{7,7} & 0 & 0 \\
 0 & a_{8,2} & 0 & a_{8,4} & 0 & 0 & 0 & a_{8,8} & 0 \\
 0 & a_{9,2} & a_{9,3} & a_{9,4} & a_{9,5} & a_{9,6} & a_{9,7} & a_{9,8} & a_{9,9} \\
\end{array}
\right)
\end{equation}
and its matrix elements,
\begin{align}
&a_{1,1}=-l_4,\;a_{2,1}=l_1+l_4-2 l_{14},\;a_{2,2}=-l_5,\;a_{3,1}=l_2+l_4-2 l_{13},\;a_{3,3}=-l_6,\nonumber\\
&a_{4,1}=l_3+l_4-2 l_{12},\;a_{4,4}=-l_7,\;a_{5,3}=-l_2-l_4+2 l_{13},\;a_{5,5}=-l_4,\nonumber\\
&a_{6,2}=-l_1-l_4+2
   l_{14},\;a_{6,6}=-l_4,\;a_{7,3}=l_2+l_4-2 l_{13},\;a_{7,4}=-l_3-l_4+2 l_{12},\nonumber\\
   &a_{7,7}=-l_4,\;a_{8,2}=l_1+l_4-2 l_{14},\;a_{8,4}=-l_3-l_4+2 l_{12},\;a_{8,8}=-l_4,\nonumber\\
   &a_{9,2}=2 l_4+2 l_9-2 l_{17},\;a_{9,3}=2 l_4+2
   l_8-2 l_{16},a_{9,4}=-2 l_4-2 l_8-2 l_9+2 l_{18},\nonumber\\
   &a_{9,5}=-l_1-l_2-l_4-l_{10}+2 l_{15},\;a_{9,6}=-l_1-l_2-l_4-l_{10}+2 l_{15},\nonumber\\
   &a_{9,7}=l_1+l_2+l_4+l_{10}-2 l_{15},\;a_{9,8}=l_1+l_2+l_4+l_{10}-2
   l_{15},\;a_{9,9}=l_{10}-l_{11}.
\end{align}
The letters are defined as 
\begin{align}
l_1&=\log \left(s_{12}\right),\;l_2=\log \left(s_{13}\right),\;l_3=\log \left(p_4^2\right),\;l_4=\log \left(m^2\right),\;l_5=\log \left(4 m^2-s_{12}\right),\nonumber\\
l_6&=\log \left(4 m^2-s_{13}\right),\;l_7=\log \left(4 m^2-p_4^2\right),\;l_8=\log \left(s_{13}-p_4^2\right),\;l_9=\log \left(s_{12}-p_4^2\right),\nonumber\\
l_{10}&=\log \left(-p_4^2+s_{12}+s_{13}\right),\;l_{11}=\log \left(s_{12} s_{13}-4 m^2
   \left(-p_4^2+s_{12}+s_{13}\right)\right),\nonumber\\
   l_{12}&=\log \left(r_3 r_4-p_4^2\right),\;l_{13}=\log \left(r_5 r_6-s_{13}\right),\;l_{14}=\log \left(r_1 r_2-s_{12}\right),\nonumber\\
   l_{15}&=\log \left(r_1 r_5 r_7+s_{12}
   s_{13}\right),\;l_{16}=\log \left(2 m^2 \left(-p_4^2+2 s_{12}+s_{13}\right)+r_1 r_6 r_7-s_{12} s_{13}\right),\nonumber\\
   l_{17}&=\log \left(2 m^2 \left(-p_4^2+s_{12}+2 s_{13}\right)+r_2 r_5 r_7-s_{12}
   s_{13}\right),\nonumber\\
   l_{18}&=\log \left(-2 m^2 \left(p_4^2 \left(s_{13}-p_4^2\right)+s_{12} \left(p_4^2+s_{13}\right)\right)+p_4^2 s_{12} s_{13}+r_1 r_3 r_4 r_5 r_7\right).
\end{align}
In order to discuss the solution we define, as in the main text, the following scaleless variables,
\begin{equation}
x_1=\frac{s_{12}}{m^2},\quad x_2=\frac{p_4^2}{m^2},\quad x_3=\frac{s_{13}}{m^2},
\end{equation}
and we set $m^2=1$. A convenient boundary point is $\vec{x}=\vec{0}$, where we have

\begin{equation}
\label{eq:BC 1 loop}
B_i(\vec{0},\epsilon)=\left\{\begin{array}{cl}
1+\frac{\pi ^2 \epsilon ^2}{12}-\frac{\zeta (3) \epsilon ^3}{3}+\frac{\pi ^4 \epsilon ^4}{160}+\mathcal{O}(\epsilon^5) & \qquad  \text{if $i=1$},\\
0 & \qquad \text{otherwise}.
\end{array}\right.
\end{equation}
We consider the contour,
\begin{equation}
\label{gamma1}
\gamma_1(t)=\{x_1(t),x_2(t),x_3(t)\}=\left\{6 t,-t,\frac{13}{25}t\right\},\quad t\in[0,1].
\end{equation}
The differential equations along the contour are,
\begin{equation}
\frac{\partial \vec{B}_{[\gamma_1]}(t,\epsilon)}{\partial t}=\epsilon \frac{\partial \tilde{A}(\vec{x}(t))}{\partial t} \vec{B}_{[\gamma_1]}(t,\epsilon).
\end{equation}
It is easy to see that the differential equations along the contour are singular for $t=0$ and $t=2/3$, the latter being the physical threshold $s_{12}=4$ while the former is a non-physical singularity (see discussion below). Following the prescription of Sec.~\ref{sec:matching} we have ,
\begin{equation}
\vec{B}_{[\gamma_1]}(t,\epsilon)=\vec{B}_{[\gamma_1]}(t,\epsilon)_{[0,0,1/3]}+\vec{B}_{[\gamma_1]}(t,\epsilon)_{[1/3,2/3,1]}.
\end{equation}
We start with the computation of $\vec{B}_{[\gamma_1]}(t,\epsilon)_{[0,0,1/3]}$. We series expand $\frac{\partial \tilde{A}(\vec{x}(t))}{\partial t}$ around $t=0$. By using the shorthand $a'_{i,j}\equiv \frac{\partial a_{ij}}{\partial t} $ we have\footnote{As in the main text, we show truncated numerical coefficients, while the full coefficients are computed with hundreds of digits.},
\begin{align}
\label{eq:Aexp0}
&a'_{2,1}=-1.22474487139158 i \; t^{-1/2}-0.918558653543691 i \; t^{1/2}+\mathcal{O}(t^{3/2}),\nonumber\\
&a'_{2,2}=+1.500000000000000+2.25000000000000 \; t+\mathcal{O}(t^{2}),\nonumber\\
&a'_{3,1}=-0.500000000000000 \; t^{-1/2}+0.0625000000000000
   \; t^{1/2}+\mathcal{O}(t^{3/2}),\nonumber\\
&a'_{3,3}=-0.250000000000000+0.0625000000000000 \; t+\mathcal{O}(t^{2}),\nonumber\\
&a'_{4,1}=-0.36055512754639 i \; t^{-1/2}-0.0234360832905159 i \; t^{1/2}+\mathcal{O}(t^{3/2}),\nonumber\\
&a'_{4,4}=+0.130000000000000+0.0169000000000000 \; t+\mathcal{O}(t^{2}),\nonumber\\
&a'_{5,3}=+0.500000000000000 \; t^{-1/2}-0.0625000000000000 \; t^{1/2}+\mathcal{O}(t^{3/2}),\nonumber\\
&a'_{6,2}=+1.22474487139158 i \; t^{-1/2}+0.918558653543691 i
   \; t^{1/2}+\mathcal{O}(t^{3/2}),\nonumber\\
&a'_{7,3}=-0.500000000000000 \; t^{-1/2}+0.0625000000000000 \; t^{1/2}+\mathcal{O}(t^{3/2}),\nonumber\\
&a'_{7,4}=+0.36055512754639 i
   \; t^{-1/2}+0.0234360832905159 i \; t^{1/2}+\mathcal{O}(t^{3/2}),\nonumber\\
&a'_{8,2}=-1.22474487139158 i \; t^{-1/2}-0.918558653543691 i \; t^{1/2}+\mathcal{O}(t^{3/2}),\nonumber\\
&a'_{8,4}=+0.36055512754639 i\; t^{-1/2}+0.0234360832905159 i \; t^{1/2}+\mathcal{O}(t^{3/2}),\nonumber\\
&a'_{9,2}=+1.41736677378460 i+0.825742696334332 i \; t+\mathcal{O}(t^{2}),\nonumber\\
&a'_{9,3}=+0.578637562357844-0.169199822921601 \; t+\mathcal{O}(t^{2}),\nonumber\\
&a'_{9,4}=-0.417261480198140 i+0.0427320462310059 i
   \; t+\mathcal{O}(t^{2}),\nonumber\\
&a'_{9,5}=+0.578637562357844 \; t^{-1/2}-0.0968701276268713 \; t^{1/2}+\mathcal{O}(t^{3/2}),\nonumber\\
&a'_{9,6}=+0.578637562357844 \; t^{-1/2}-0.0968701276268713
   \; t^{1/2}+\mathcal{O}(t^{3/2}),\nonumber\\
&a'_{9,7}=-0.578637562357844 \; t^{-1/2}+0.0968701276268713 \; t^{1/2}+\mathcal{O}(t^{3/2}),\nonumber\\
&a'_{9,8}=-0.578637562357844
   \; t^{-1/2}+0.0968701276268713 \; t^{1/2}+\mathcal{O}(t^{3/2}),\nonumber\\
&a'_{9,9}=-0.334821428571428+0.112105389030612 \; t+\mathcal{O}(t^{2}),
\end{align}
while the other matrix elements are identically zero.
We are now able to find the series solution $\vec{B}_{[\gamma_1]}(t,\epsilon)_{[0,0,1/3]}$ by using the recursion,
\begin{equation}
\label{eq:solrecexp}
\vec{B}^{(i)}_{[\gamma_1]}(t)_{[0,0,1/3]}=\int dt \frac{\partial \tilde{A}(\vec{x}(t))}{\partial t}\vec{B}^{(i-1)}_{[\gamma_1]}(t)_{[0,0,1/3]} +\vec{C}^{(i)}_{[0,0,1/3]},
\end{equation}
where $C^{(i)}$ is the boundary vector and the integral on the right-hand-side is an indefinite integral. We remark that by series expanding $\frac{\partial \tilde{A}(\vec{x}(t))}{\partial t}$ all the integrals in (\ref{eq:solrecexp}) are elementary and can be performed analytically. At order $\epsilon^1$ we have
\begin{equation}
\label{eq:solrecexp1}
\vec{B}^{(1)}_{[\gamma_1]}(t)_{[0,0,1/3]}=\int dt \frac{\partial \tilde{A}(\vec{x}(t))}{\partial t}\vec{B}^{(0)}_{[\gamma_1]}(t)_{[0,0,1/3]} +\vec{C}^{(1)}_{[0,0,1/3]}.
\end{equation}
The order $\epsilon^0$ of the solution is given by the boundary conditions~(\ref{eq:BC 1 loop}). By performing the integrals on the right-hand-side and by imposing the boundary condition $\vec{B}^{(1)}_{[\gamma_1]}(0)_{[0,0,1/3]}\equiv \vec{0}$ we get,
\begin{align}
&B^{(1)}_{1,[\gamma_1]}(t)_{[0,0,1/3]},=0,\nonumber\\
&B^{(1)}_{2,[\gamma_1]}(t)_{[0,0,1/3]}=-2.44948974278318 i \; t^{1/2}-0.612372435695795 i \;t^{3/2}+\mathcal{O}(t^{5/2}),\nonumber\\
&B^{(1)}_{3,[\gamma_1]}(t)_{[0,0,1/3]}=-1.00000000000000 \; t^{1/2}+0.0416666666666667 \;t^{3/2}+\mathcal{O}(t^{5/2}),\nonumber\\
&B^{(1)}_{4,[\gamma_1]}(t)_{[0,0,1/3]}=-0.72111025509279 i \; t^{1/2}-0.015624055527010 i\; t^{3/2}+\mathcal{O}(t^{5/2}),\nonumber\\
   &B^{(1)}_{5,[\gamma_1]}(t)_{[0,0,1/3]}=0,\;B^{(1)}_{6,[\gamma_1]}(t)_{[0,0,1/3]}=0,\;B^{(1)}_{7,[\gamma_1]}(t)_{[0,0,1/3]}=0,\nonumber\\
   &B^{(1)}_{8,[\gamma_1]}(t)_{[0,0,1/3]}=0,\;B^{(1)}_{9,[\gamma_1]}(t)_{[0,0,1/3]}=0.
\end{align}
Proceeding in the same way, at order $\epsilon^2$ we get,
\begin{align}
&B_{1,\left[\gamma _1\right]}^{\text{(2)}}(t){}_{\text{[0,0,1/3]}}=+0.82246703342411,\nonumber\\
&B_{2,\left[\gamma _1\right]}^{\text{(2)}}(t){}_{\text{[0,0,1/3]}}=-2.44948974278318 i\;t^{3/2}+\mathcal{O}(t^{5/2}),\nonumber\\
&B_{3,\left[\gamma _1\right]}^{\text{(2)}}(t){}_{\text{[0,0,1/3]}}=+0.166666666666667 \;t^{3/2}+\mathcal{O}(t^{5/2}),\nonumber\\
&B_{4,\left[\gamma _1\right]}^{\text{(2)}}(t){}_{\text{[0,0,1/3]}}=-0.062496222108042 i\;t^{3/2}+\mathcal{O}(t^{5/2}),\nonumber\\
&B_{5,\left[\gamma _1\right]}^{\text{(2)}}(t){}_{\text{[0,0,1/3]}}=-0.5000000000000000\; t+0.04166666666666\; t^2+\mathcal{O}(t^{3}),\nonumber\\
&B_{6,\left[\gamma
   _1\right]}^{\text{(2)}}(t){}_{\text{[0,0,1/3]}}=+3.0000000000000000\; t+1.50000000000000\; t^2+\mathcal{O}(t^{3}),\nonumber\\
&B_{7,\left[\gamma _1\right]}^{\text{(2)}}(t){}_{\text{[0,0,1/3]}}=+0.7600000000000000\; t-0.03040000000000\;t^2+\mathcal{O}(t^{3}),\nonumber\\
&B_{8,\left[\gamma _1\right]}^{\text{(2)}}(t){}_{\text{[0,0,1/3]}}=-2.7400000000000000\; t-1.48873333333333\; t^2+\mathcal{O}(t^{3}),\nonumber\\
&B_{9,\left[\gamma _1\right]}^{\text{(2)}}(t){}_{\text{[0,0,1/3]}}=+1.72819751957543\;
   t^{3/2}+\mathcal{O}(t^{5/2}).
\end{align}
It is straightforward to iterate the procedure up to arbitrary order of $\epsilon$. We see that the solution is singular in $t=0$, since it develops a brunch cut for $t<0$. This is what we call, in Section~\ref{sec:Analytic continuation}, a non-physical singularity, since the only physical singularity along the contour is $s_{12}=4m^2$. The appearance of this singularity is due to the prefactors defining the canonical basis (\ref{eq:basis1loop}) and we see that, by inverting the basis, the integrals $I_{a_1,a_2,a_3,a_4}$ admit a regular Taylor series expansion around $t=0$. We remark that, according to (\ref{eq:basis1loop}), the differential equations depend originally on $\sqrt{-t}$ and, as discussed in Section~\ref{sec:Analytic continuation}, we consider the standard brunch for $\sqrt{-t}$, i.e. $\sqrt{-t}>0$ for $t<0$ and $\sqrt{-t}=i\sqrt{t}$ for $t>0$.  

The computation of $\vec{B}_{[\gamma_1]}(t,\epsilon)_{[1/3,2/3,1]}$ follows the same steps. One first expands $\frac{\partial \tilde{A}(\vec{x}(t))}{\partial t}$ around $t=2/3$. The series solution is then obtained by,
\begin{equation}
\label{eq:solrecexpphys}
\vec{B}^{(i)}_{[\gamma_1]}(t)_{[1/3,2/3,1]}=\int dt \frac{\partial \tilde{A}(\vec{x}(t))}{\partial t}\vec{B}^{(i-1)}_{[\gamma_1]}(t)_{[1/3,2/3,1]} +\vec{C}^{(i)}_{[1/3,2/3,1]},
\end{equation}
and by using as boundary condition the value of the integrals in the contact point of the two series, in this case $t=1/3$,
\begin{equation}
\vec{B}^{(i)}_{[\gamma_1]}(1/3)_{[1/3,2/3,1]}\equiv \vec{B}^{(i)}_{[\gamma_1]}(1/3)_{[0,0,1/3]},
\end{equation}
where the right-hand-side is known from the previous expansion. The first non-trivial $\epsilon$ orders of the solution are,
\begin{align}
B_{1,\left[\gamma _1\right]}^{\text{(1)}}(t){}_{\text{[1/3,2/3,1]}}=&0,\nonumber\\
B_{2,\left[\gamma _1\right]}^{\text{(1)}}(t){}_{\text{[1/3,2/3,1]}}=&-3.14159265358979 i+2.44948974278318 i\; t'{}^{1/2}\nonumber\\
&+0.612372435695795 i \; t'{}^{3/2}+\mathcal{O}(t'{}^{5/2}),\nonumber\\
B_{3,\left[\gamma _1\right]}^{\text{(1)}}(t){}_{\text{[1/3,2/3,1]}}=&-0.795365461223906+0.566946709513841
\;   t'\nonumber\\
   &+0.242977161220218 \; t'{}^{2}+\mathcal{O}(t'{}^{3}),\nonumber\\
B_{4,\left[\gamma _1\right]}^{\text{(1)}}(t){}_{\text{[1/3,2/3,1]}}=&-0.597638586883139 i+0.462064551372464 i
  \; t'\nonumber\\
   &+0.156832128750508 i \; t'{}^{2}+\mathcal{O}(t'{}^{3}),\nonumber\\
B_{5,\left[\gamma _1\right]}^{\text{(1)}}(t){}_{\text{[1/3,2/3,1]}}=&0,\;B_{6,\left[\gamma
   _1\right]}^{\text{(1)}}(t){}_{\text{[1/3,2/3,1]}}=0,\;B_{7,\left[\gamma _1\right]}^{\text{(1)}}(t){}_{\text{[1/3,2/3,1]}}=0,\nonumber\\
   B_{8,\left[\gamma
   _1\right]}^{\text{(1)}}(t){}_{\text{[1/3,2/3,1]}}=&0,\; B_{9,\left[\gamma _1\right]}^{\text{(1)}}(t){}_{\text{[1/3,2/3,1]}}=0
\end{align}
and 
\begin{align}
B_{1,\left[\gamma _1\right]}^{\text{(2)}}(t){}_{\text{[1/3,2/3,1]}}=&+0.822467033424113,\nonumber\\
B_{2,\left[\gamma _1\right]}^{\text{(2)}}(t){}_{\text{[1/3,2/3,1]}}=&+3.14159265358979 i \log
   \left(t'\right)+5.62897838552680 i\nonumber\\
   &-4.89897948556636 i \; t'{}^{1/2}-0.408248290463863 i \; t'{}^{3/2}\nonumber\\
   &+\mathcal{O}(t'{}^{5/2}),\nonumber\\
   B_{3,\left[\gamma_1\right]}^{\text{(2)}}(t){}_{\text{[1/3,2/3,1]}}=&+0.0813205590990830-0.170435455976551 \;t'\nonumber\\
   &+0.0424833485932810 \; t'{}^{2}+\mathcal{O}(t'{}^{3}),\nonumber\\
   B_{4,\left[\gamma _1\right]}^{\text{(2)}}(t){}_{\text{[1/3,2/3,1]}}=&-0.0362286434252506 i+0.0850653463081840 i\; t'\nonumber\\
   &-0.0389380766597542 i \; t'{}^{2}+\mathcal{O}(t'{}^{3}),\nonumber\\
   B_{5,\left[\gamma_1\right]}^{\text{(2)}}(t){}_{\text{[1/3,2/3,1]}}=&-0.316303108453958+0.450929831101852 \;t'\nonumber\\
   &+0.0325413561865079 \; t'{}^{2}+\mathcal{O}(t'{}^{3}),\nonumber\\
   B_{6,\left[\gamma_1\right]}^{\text{(2)}}(t){}_{\text{[1/3,2/3,1]}}=&+4.93480220054468-7.69529898097118 \; t'{}^{1/2}\nonumber\\
   &+3.00000000000000 \;t'-1.92382474524280 \; t'{}^{3/2}\nonumber\\
   &+1.50000000000000
   \; t'{}^{2}+\mathcal{O}(t'{}^{5/2}),\nonumber\\
   B_{7,\left[\gamma _1\right]}^{\text{(2)}}(t){}_{\text{[1/3,2/3,1]}}=&+0.494889048719796-0.727077436632883\; t'\nonumber\\
   &-0.0195184631733177
   \; t'{}^{2}+\mathcal{O}(t'{}^{3}),\nonumber\\
   B_{8,\left[\gamma _1\right]}^{\text{(2)}}(t){}_{\text{[1/3,2/3,1]}}=&-4.75621626027884+7.69529898097118 \; t'{}^{1/2}\nonumber\\
   &-3.27614760553103\; t'+1.92382474524280 \; t'{}^{3/2}\nonumber\\
   &-1.48697710698681 \; t'{}^{2}+\mathcal{O}(t'{}^{5/2}),\nonumber\\
   B_{9,\left[\gamma
   _1\right]}^{\text{(2)}}(t){}_{\text{[1/3,2/3,1]}}=&+3.19991525486188-6.57453761281272 \; t'{}^{1/2}\nonumber\\
   &+3.18425714889066 \;t'-0.299933285256055 \; t'{}^{3/2}\nonumber\\
   &+0.322016612766481
   \; t'{}^{2}+\mathcal{O}(t'{}^{5/2}),
\end{align}
where $t'=2/3-t$. We see that for $t>2/3$ the solution develops a brunch cut, and the cut ambiguity is resolved by Feynman prescription which, by (\ref{gamma1}), is $t\rightarrow t+i \delta$ with $\delta$ a small positive imaginary part.
\clearpage

\section{Plots}
In this appendix we show plots of all the 73 master integrals at order $\epsilon^4$. The plots are obtained by series expanding along the contour $\gamma_{\text{thr}}$ defined in section \ref{sec:series sol DE A} ($s_{13}=-1,\;p_4^2=\frac{13}{25},\;m^2=1$, $2\leq s_{12}\leq 6$). The solid dots represent numerical values computed with FIESTA 4.1. The real part of the integrals is in blue, the imaginary part is orange.
\label{app:plots}
\begin{table}[ht]
	\begin{tabular}{ p{4.75cm}  p{4.75cm}  p{4.75cm}}
\img{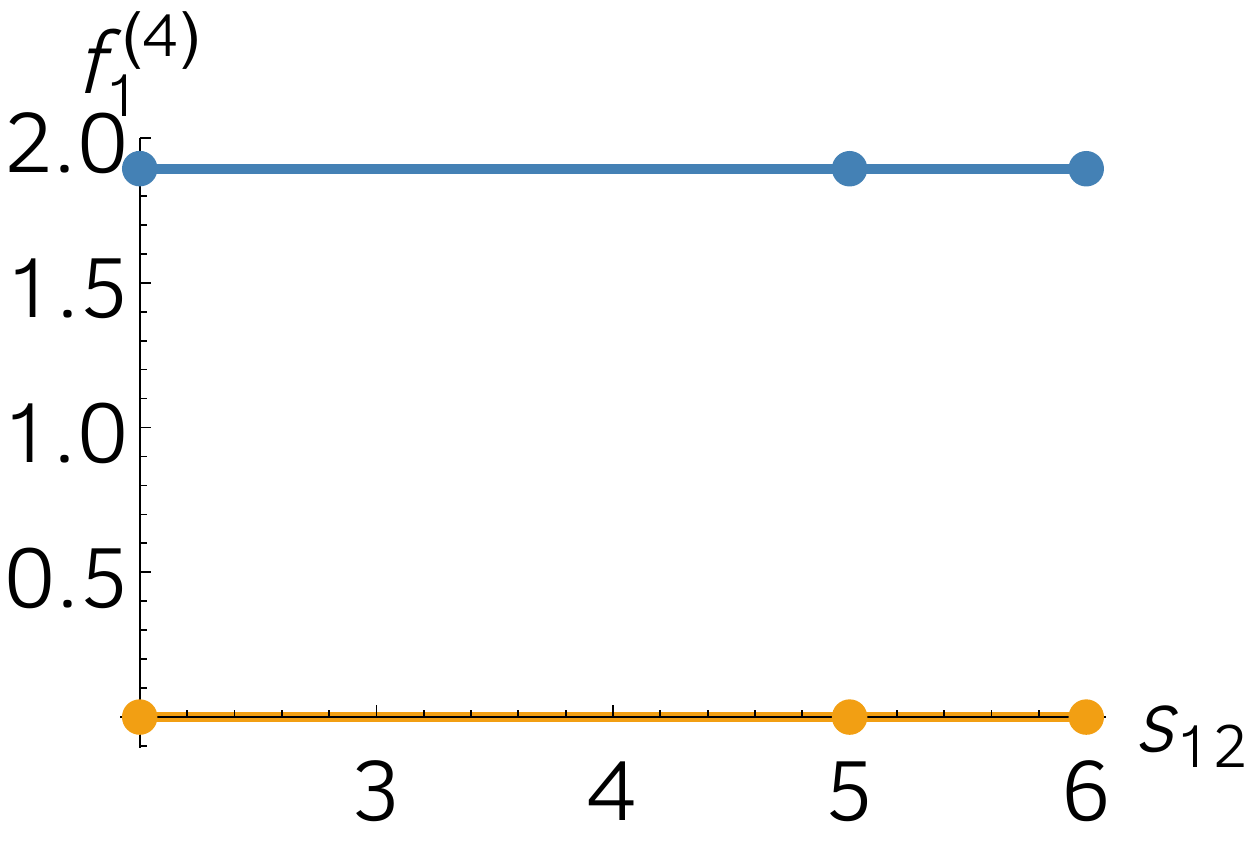} & \img{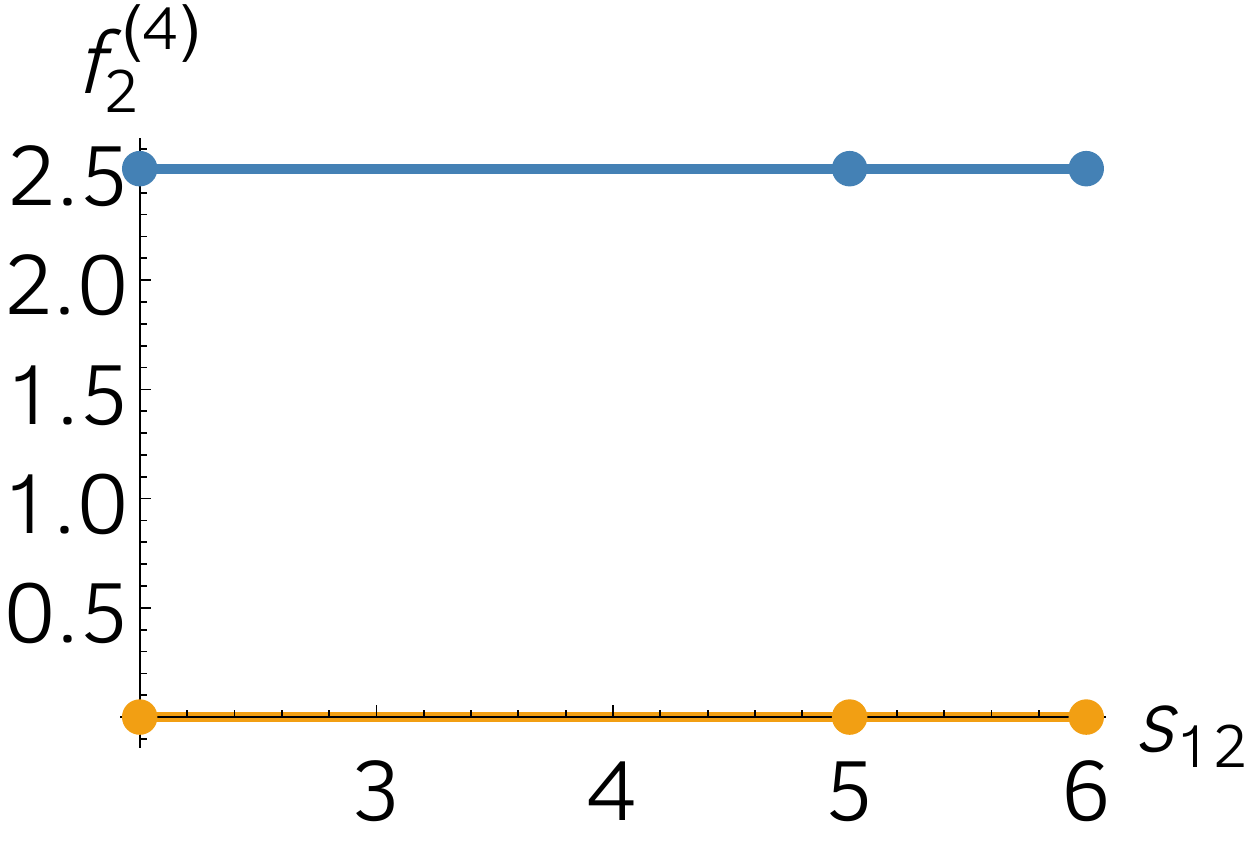} & \img{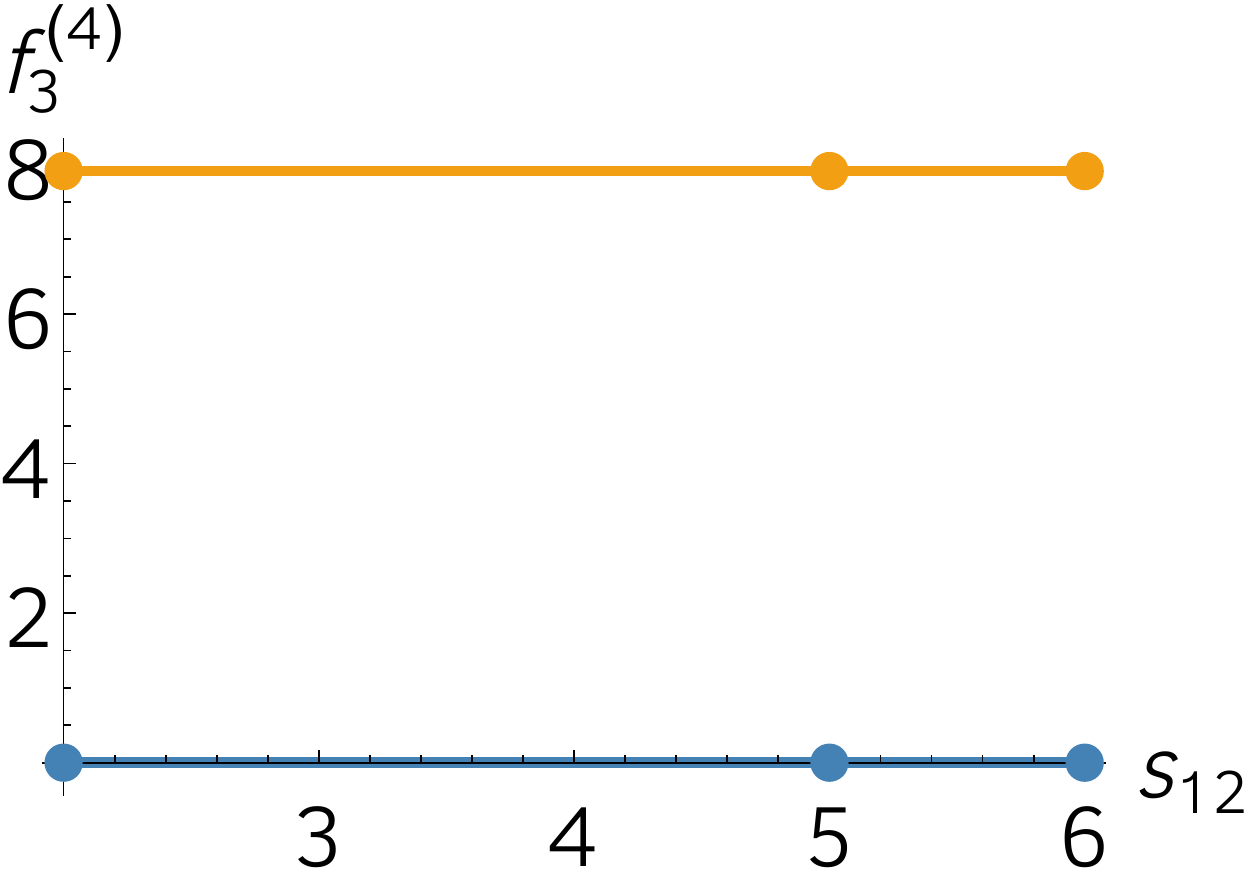} \\
\img{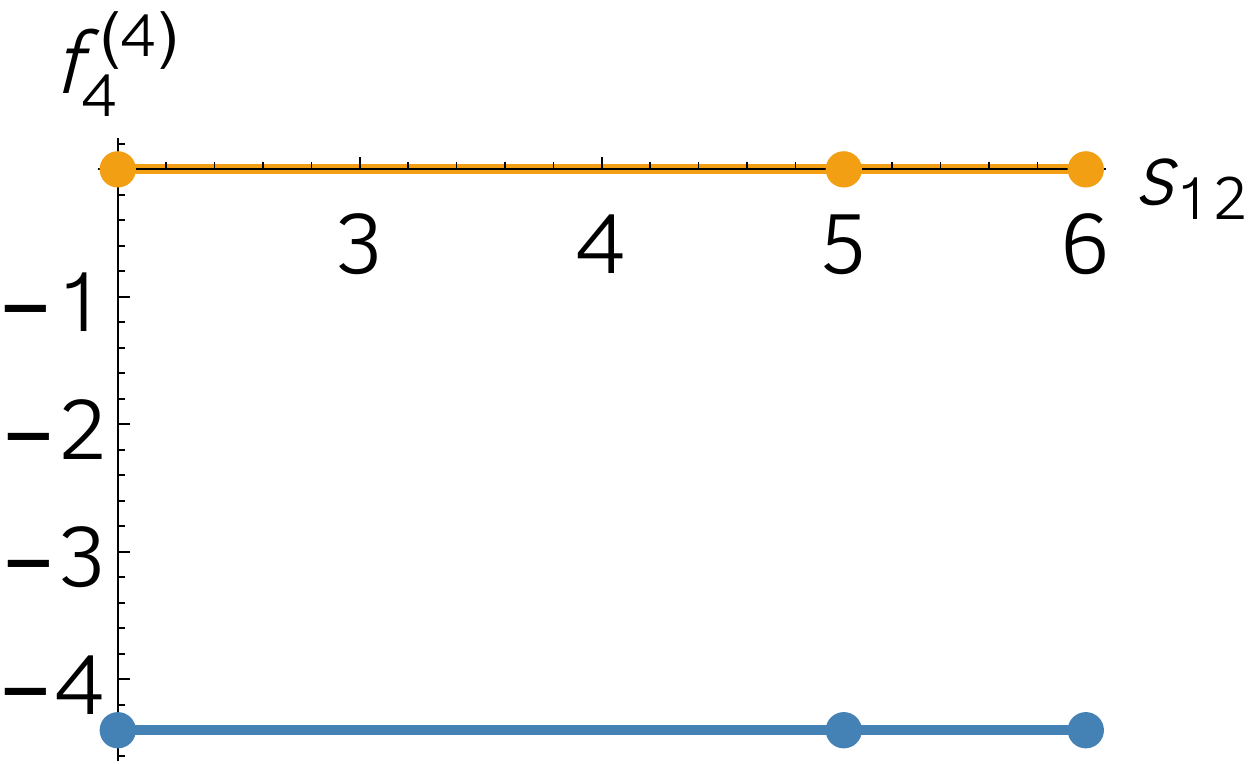} & \img{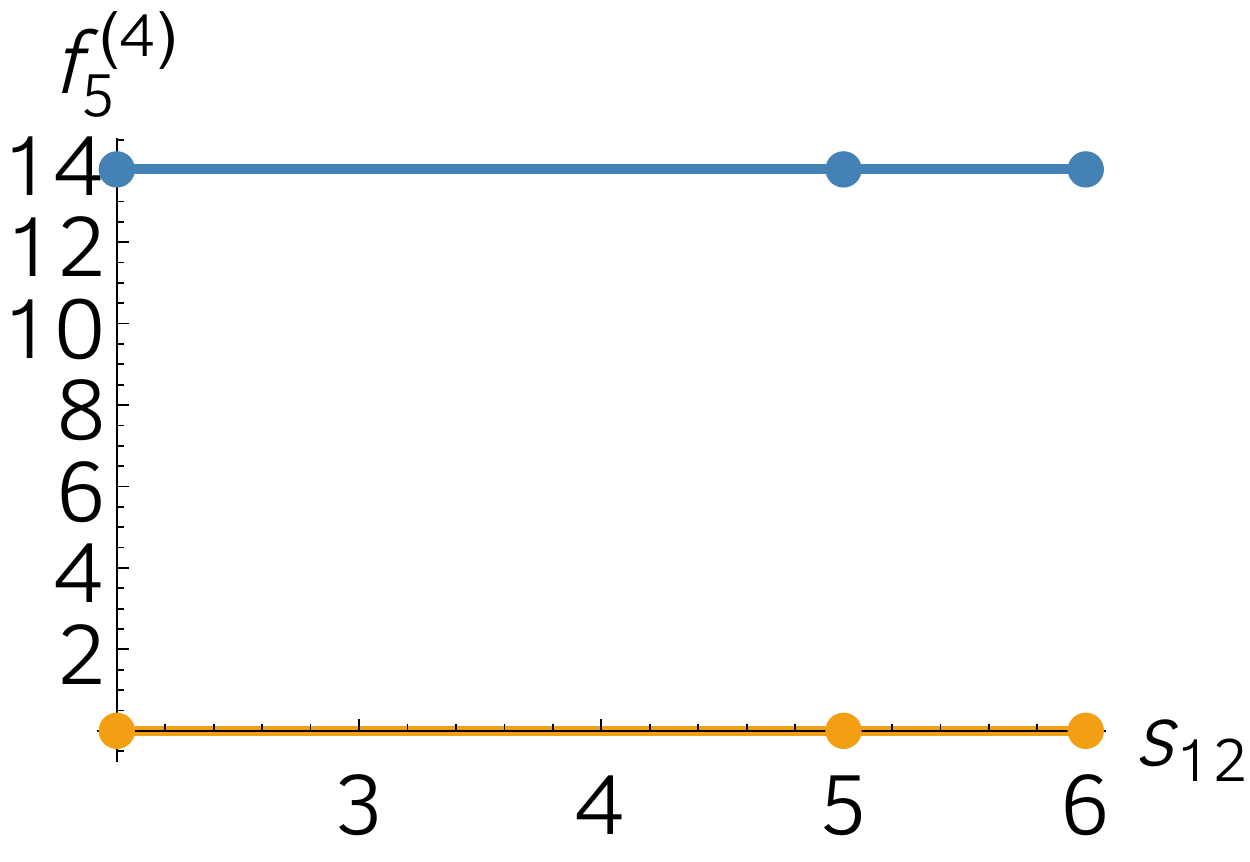} & \img{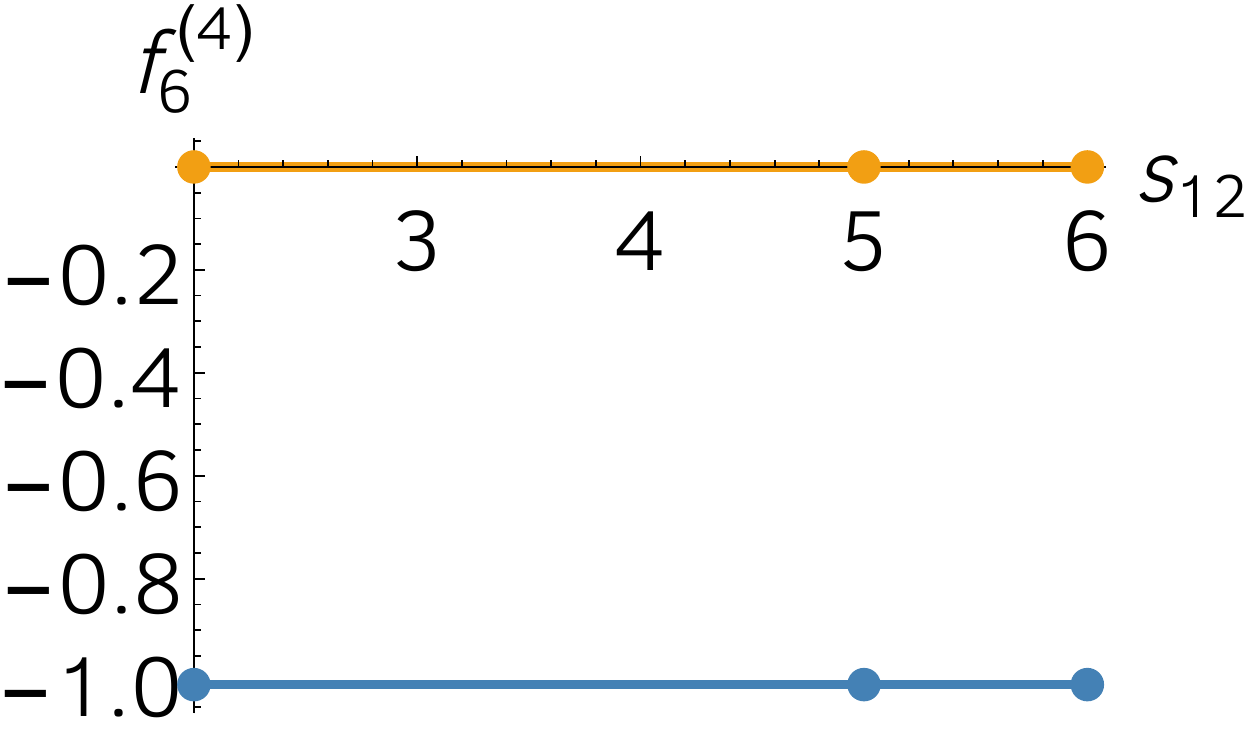} \\
\img{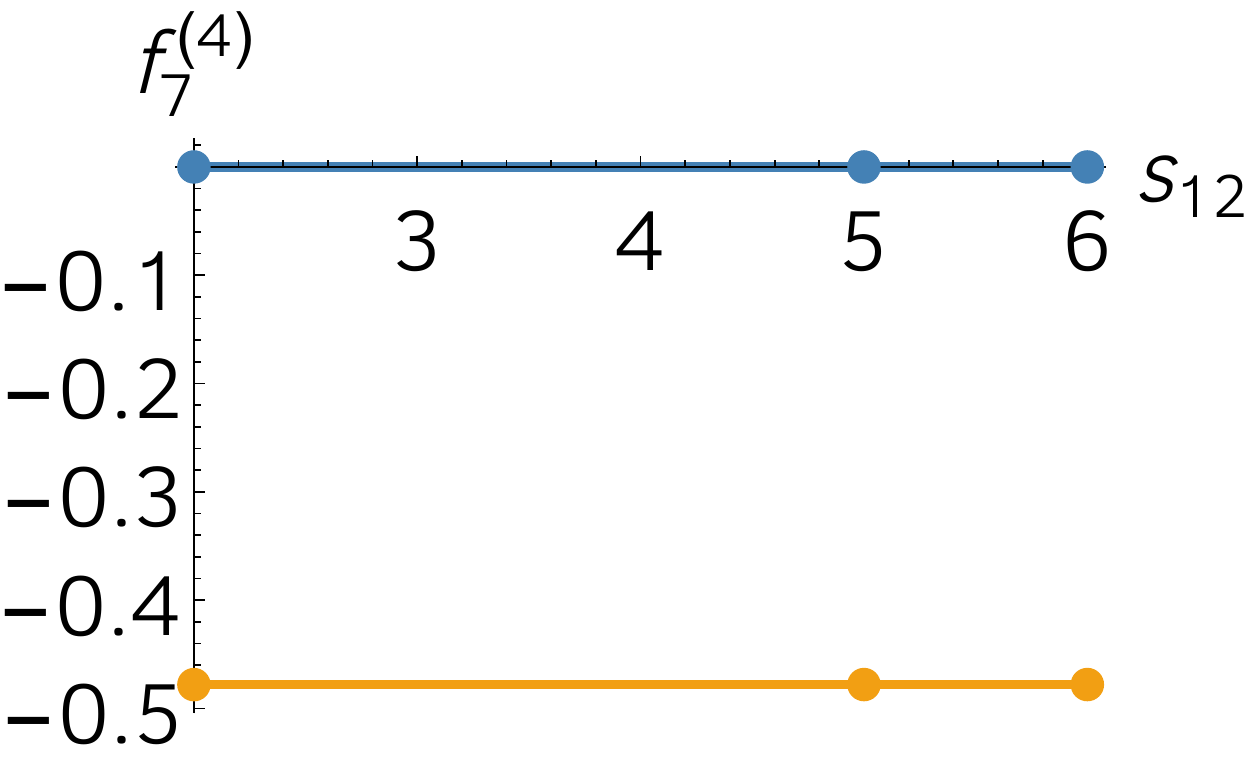} & \img{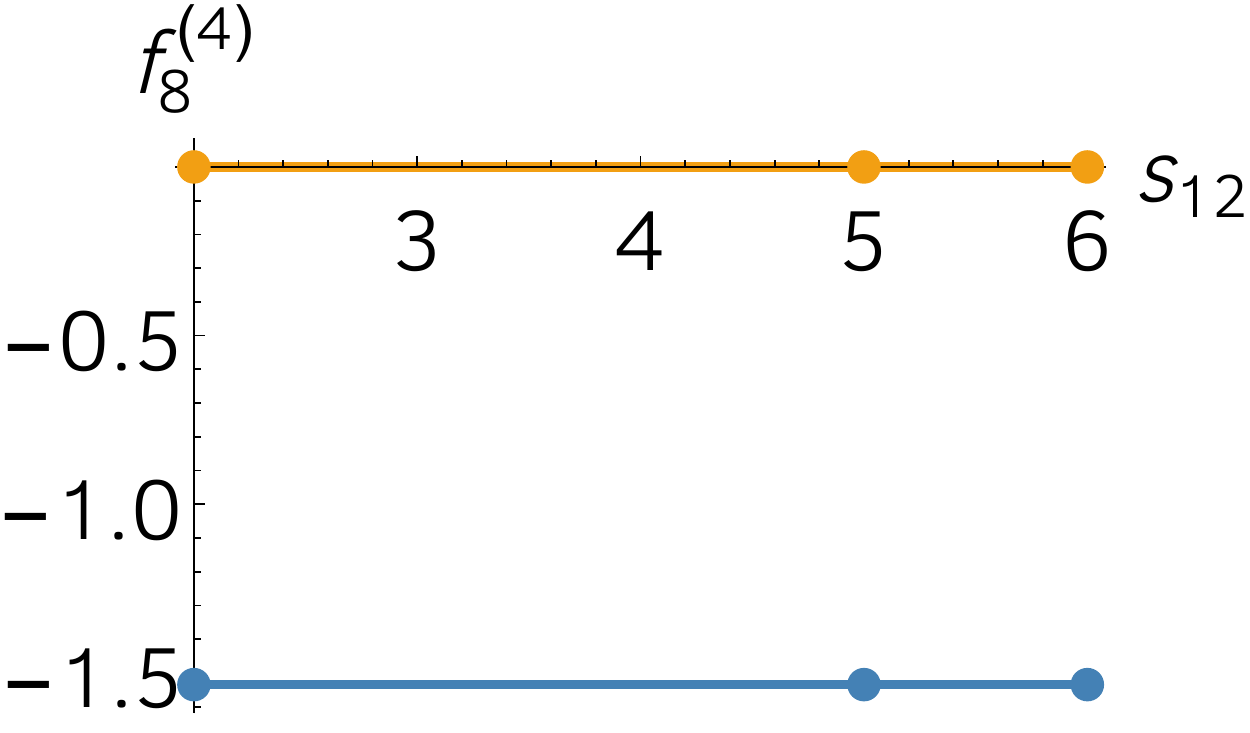} & \img{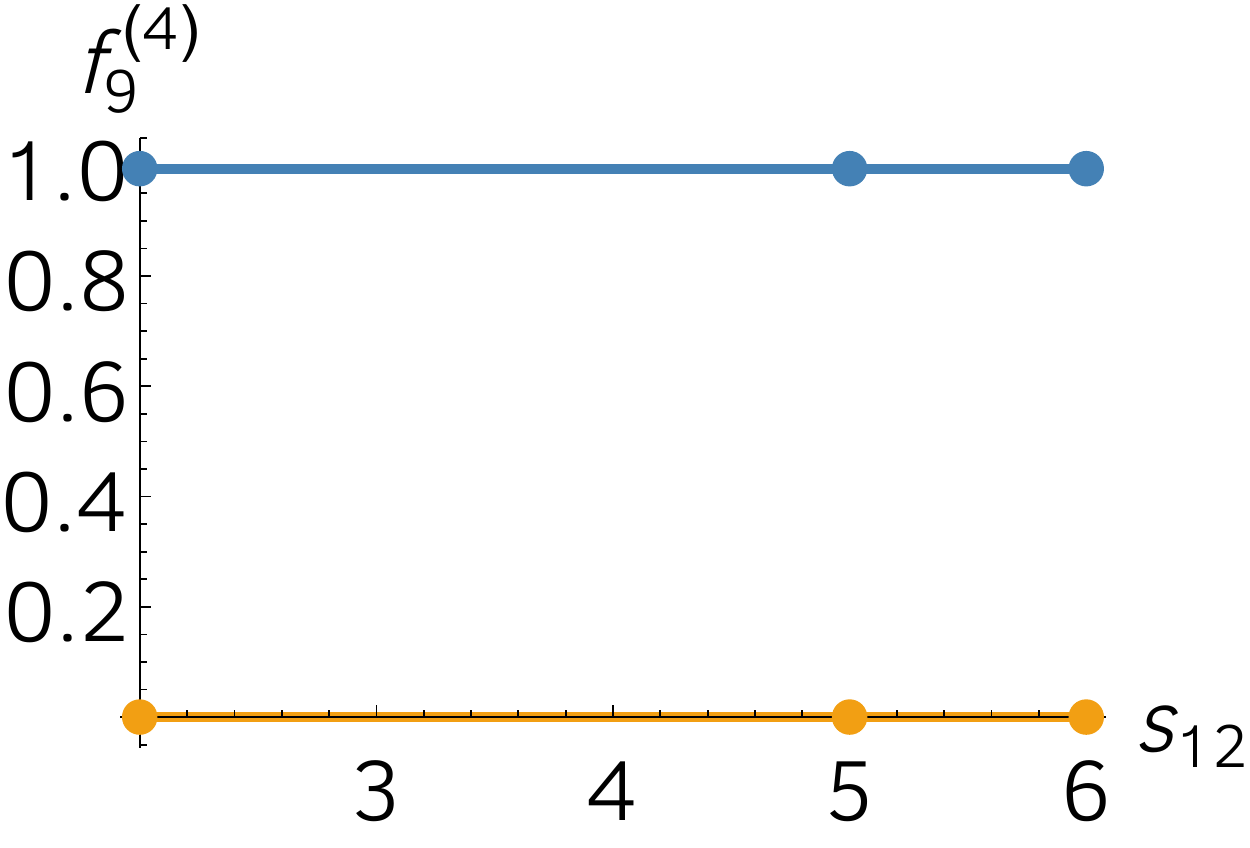} \\
\img{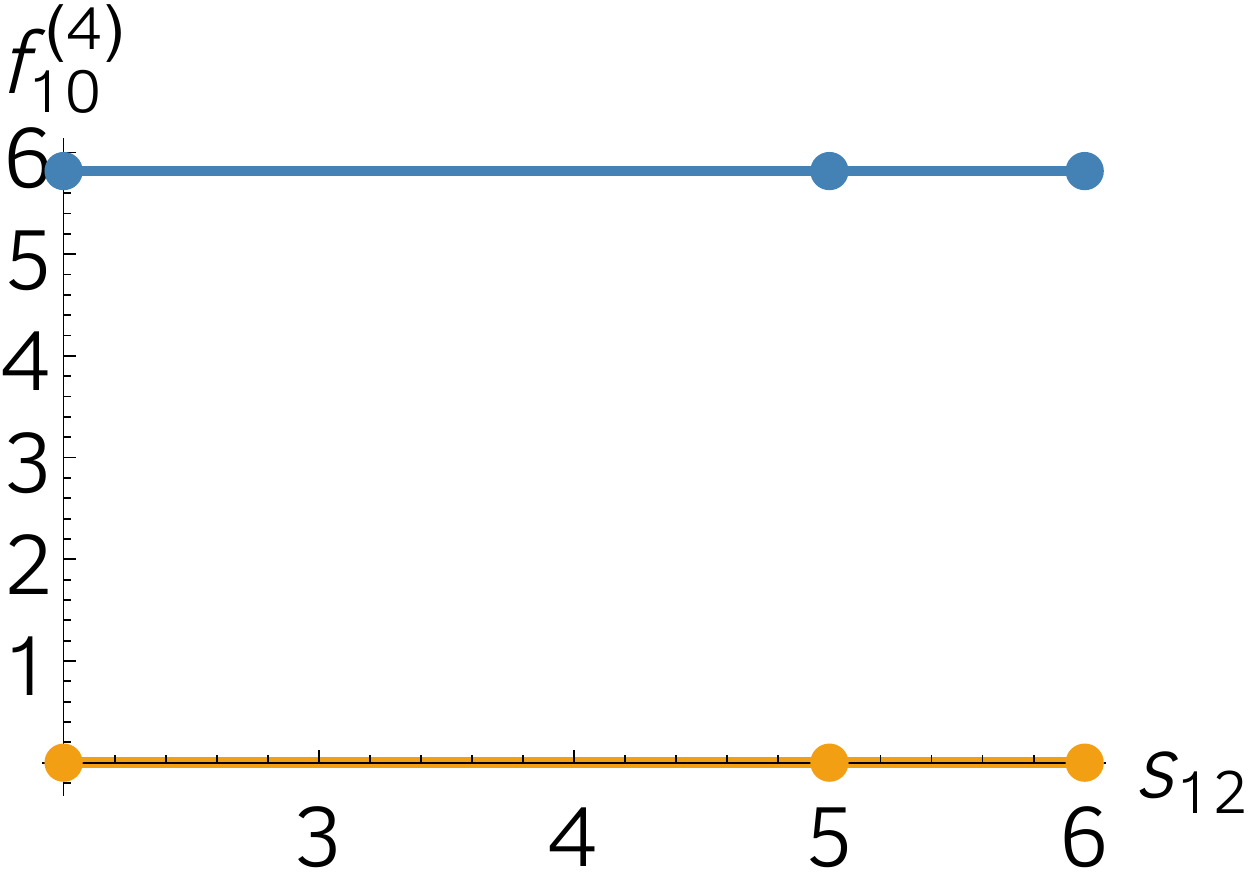} & \img{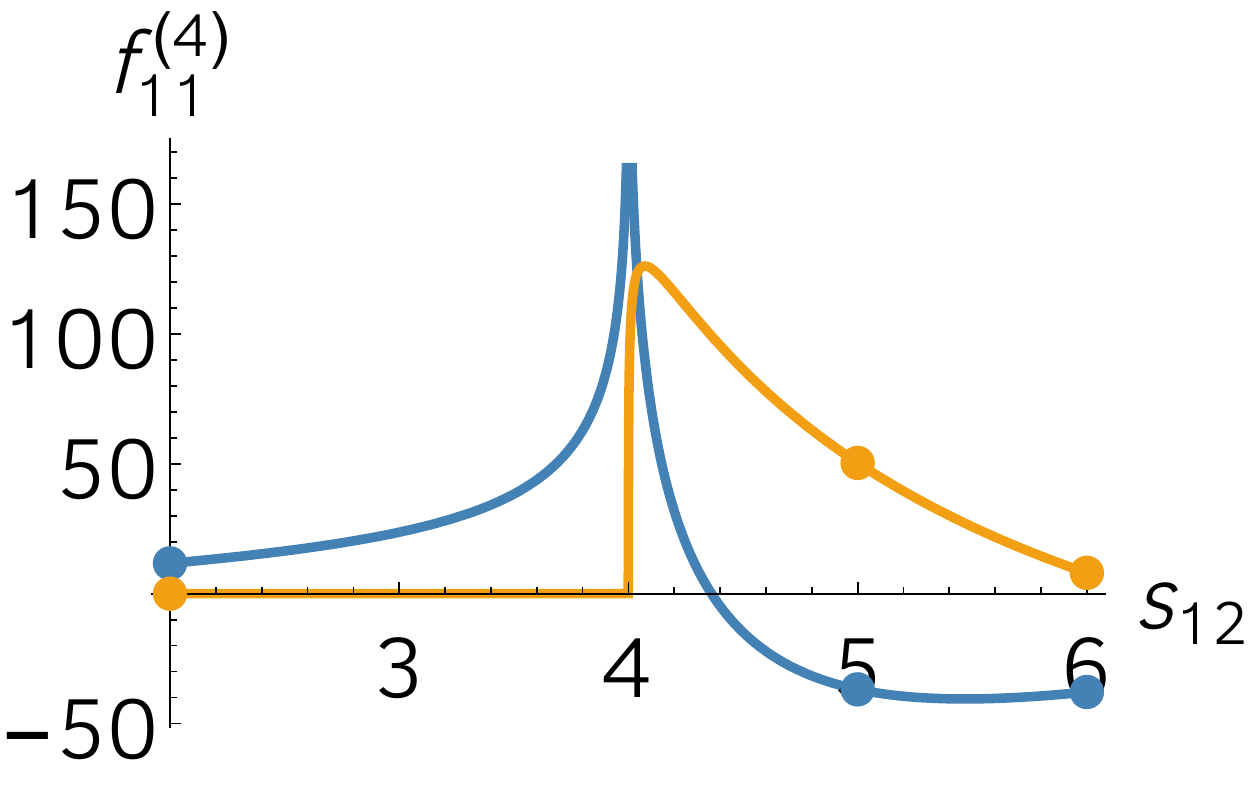} & \img{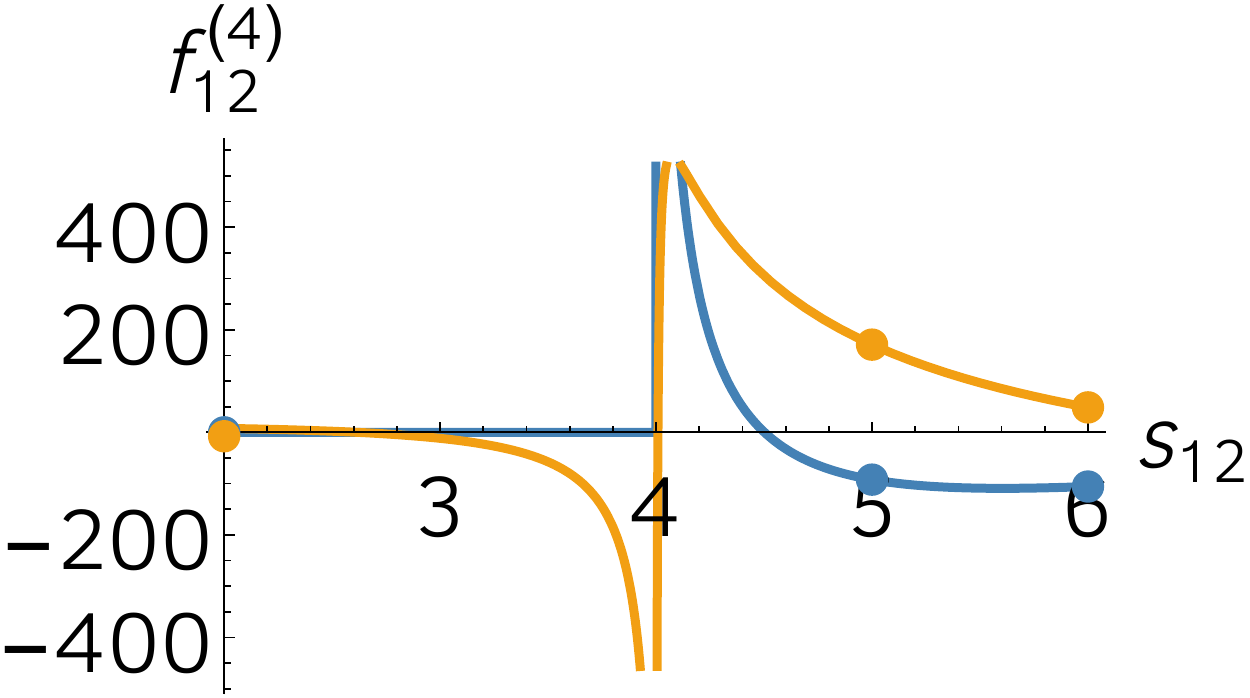} \\
\img{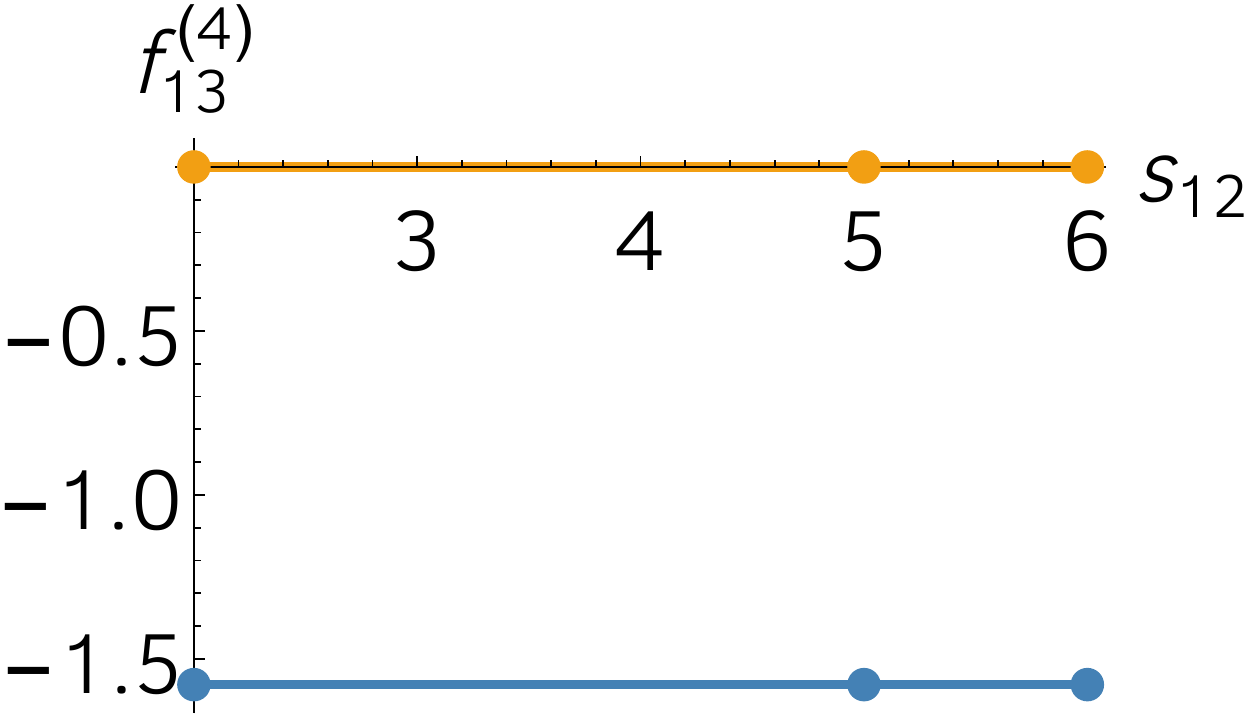} & \img{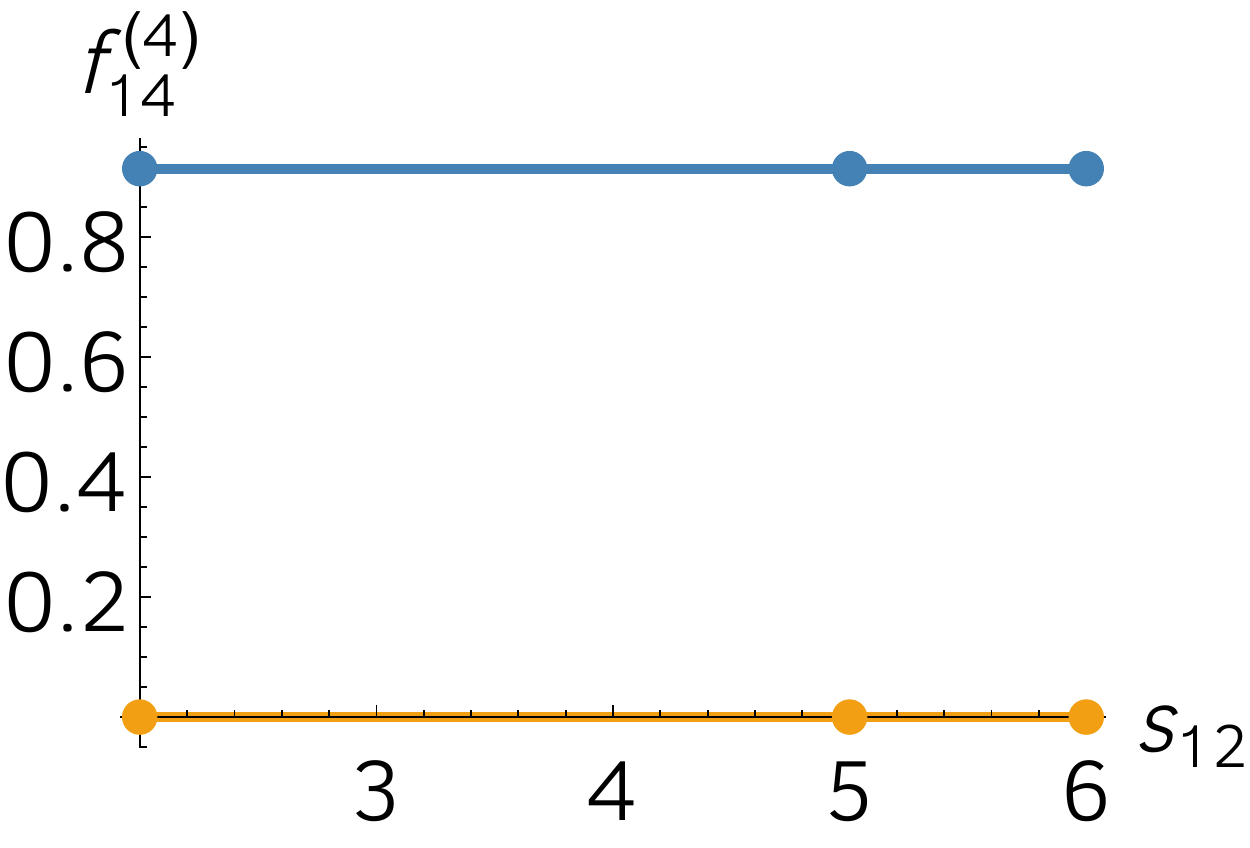} & \img{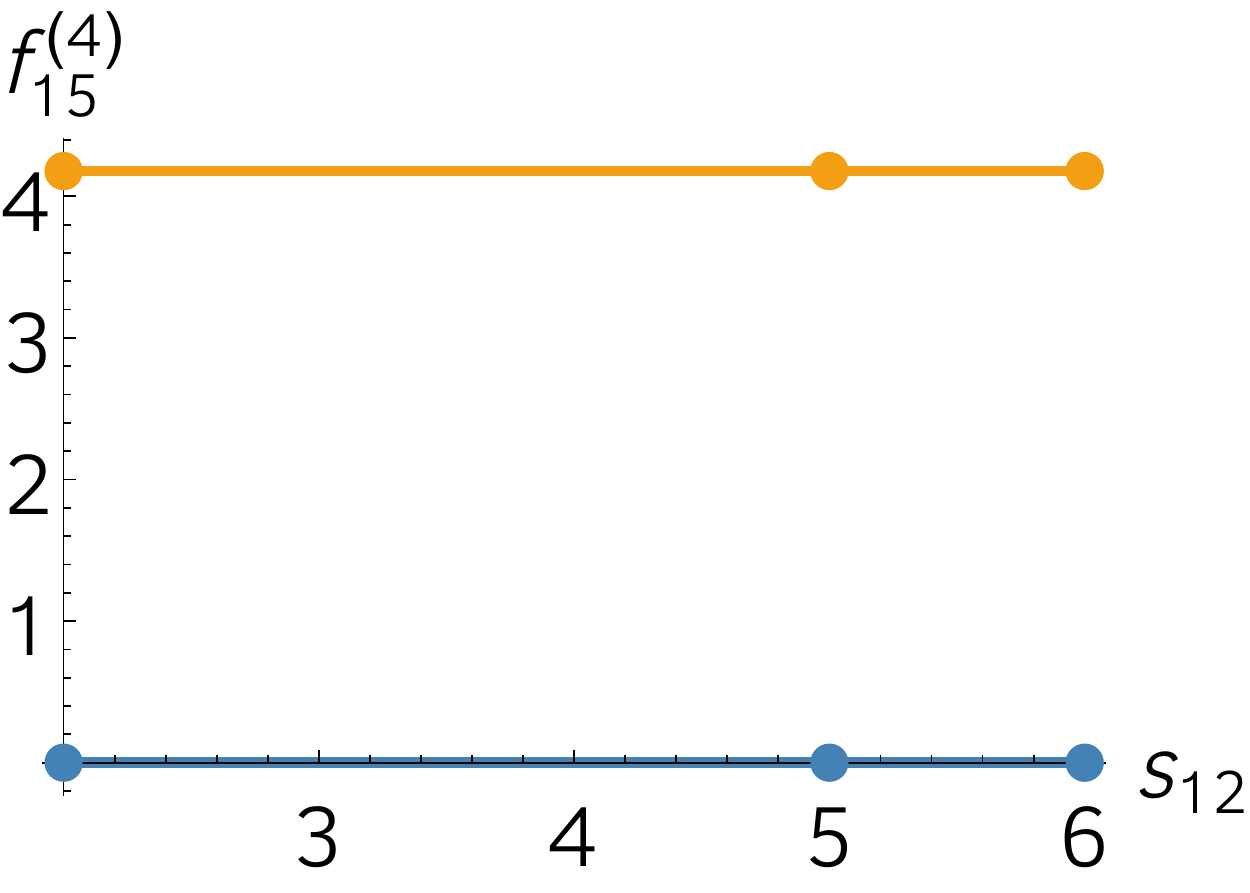} \\
\img{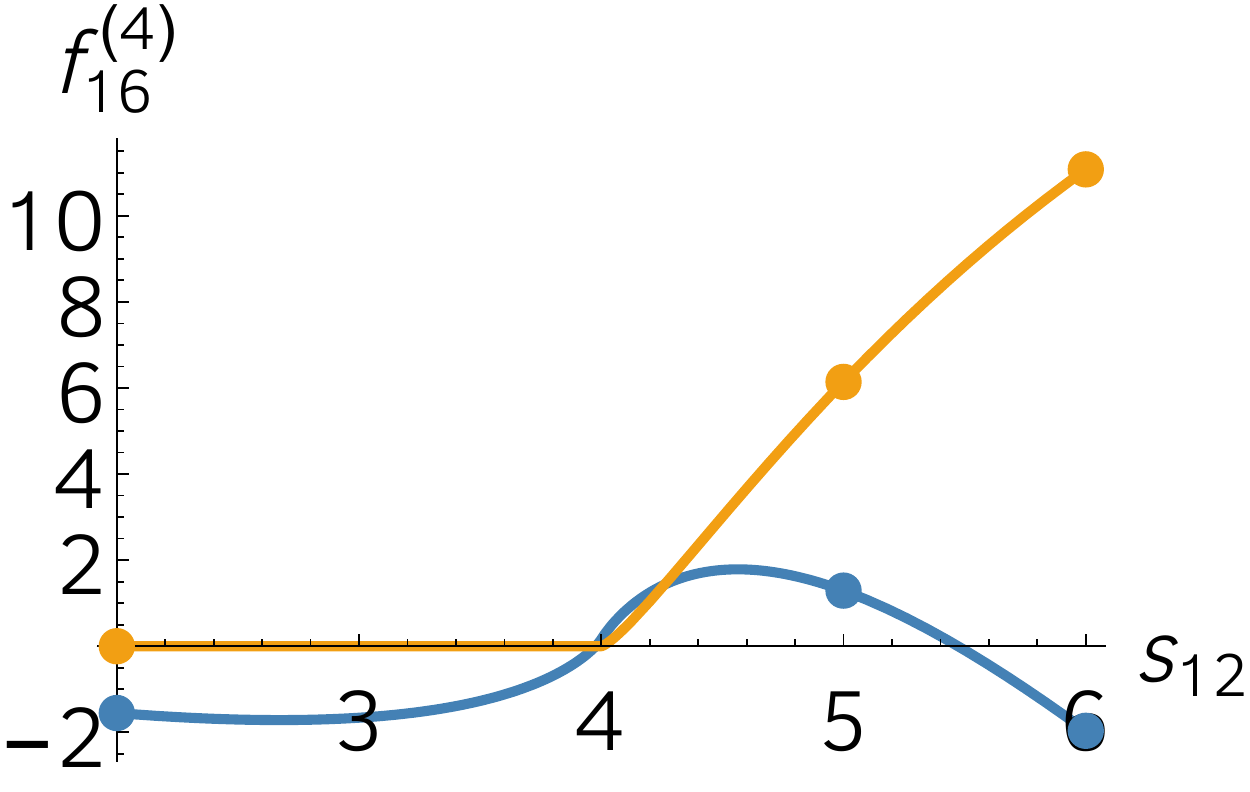} & \img{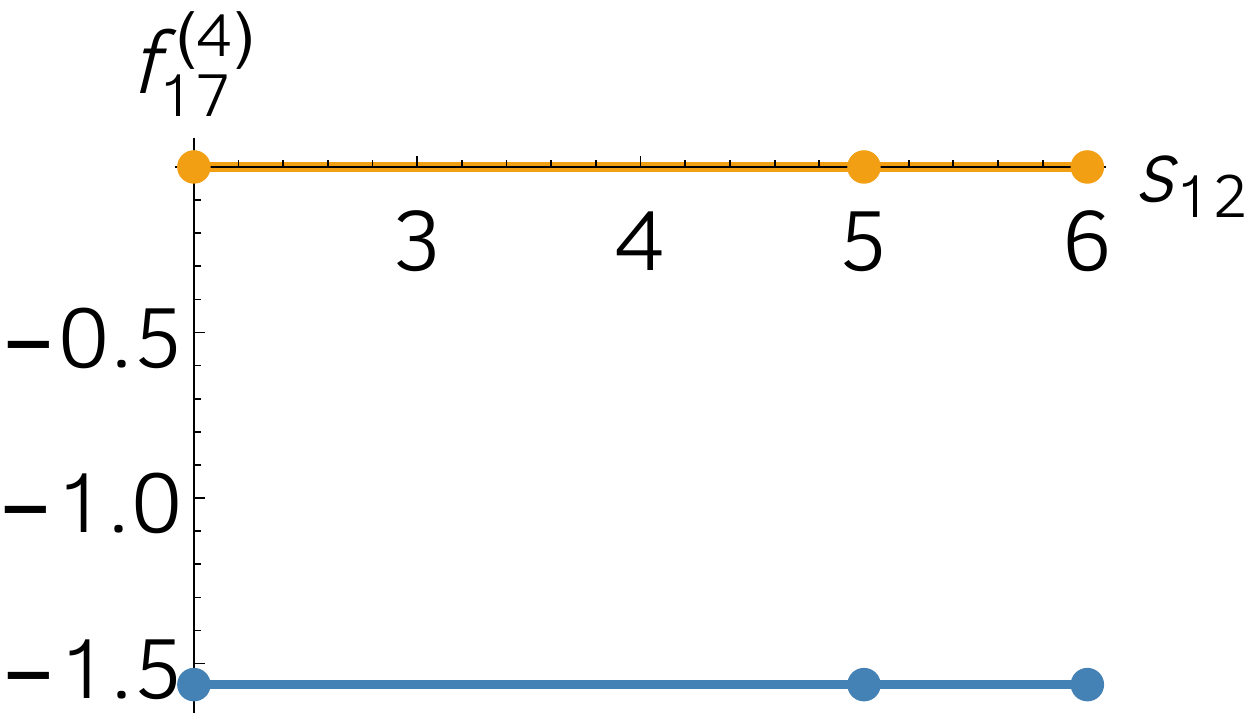} & \img{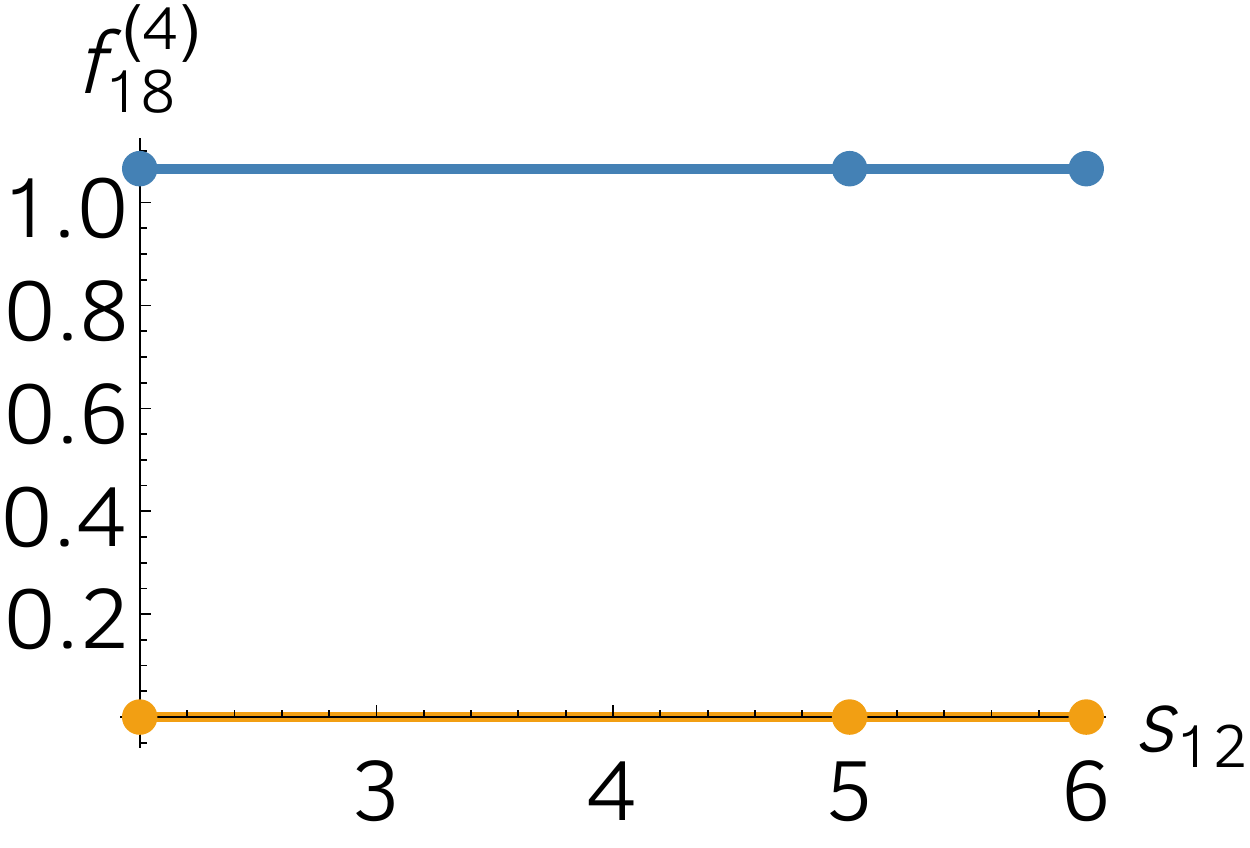} \\
\img{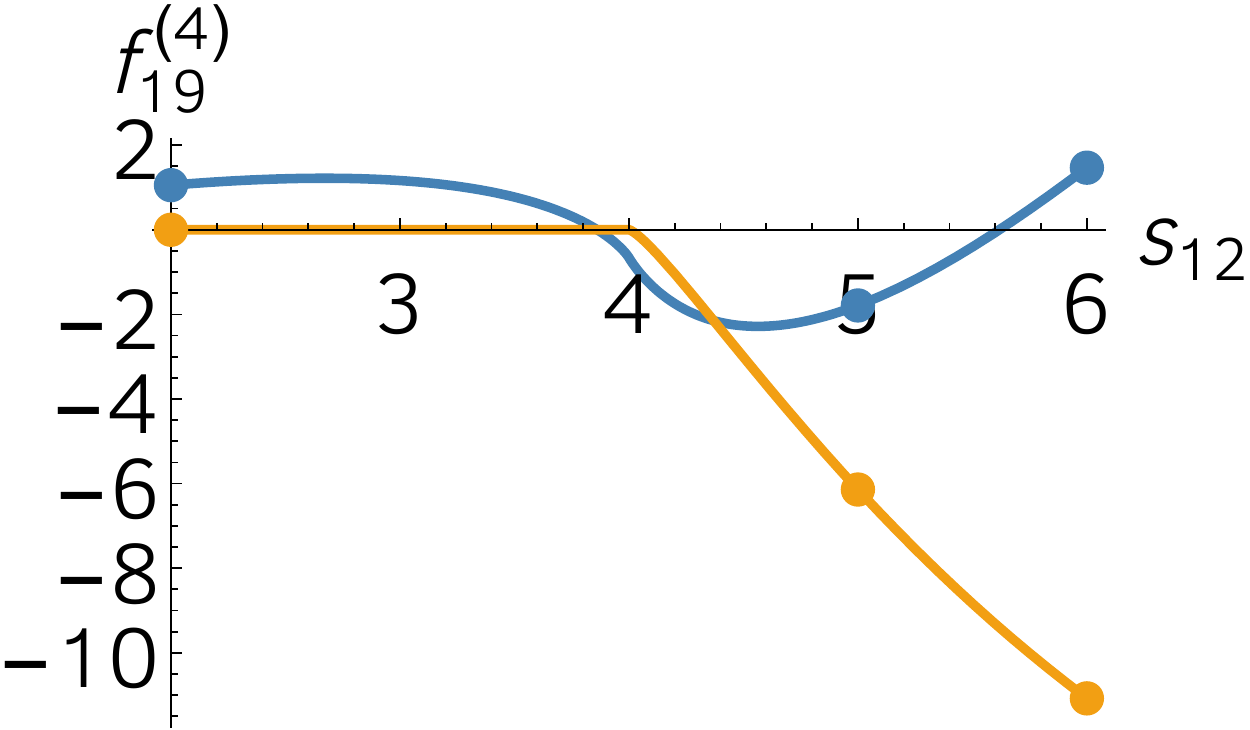} & \img{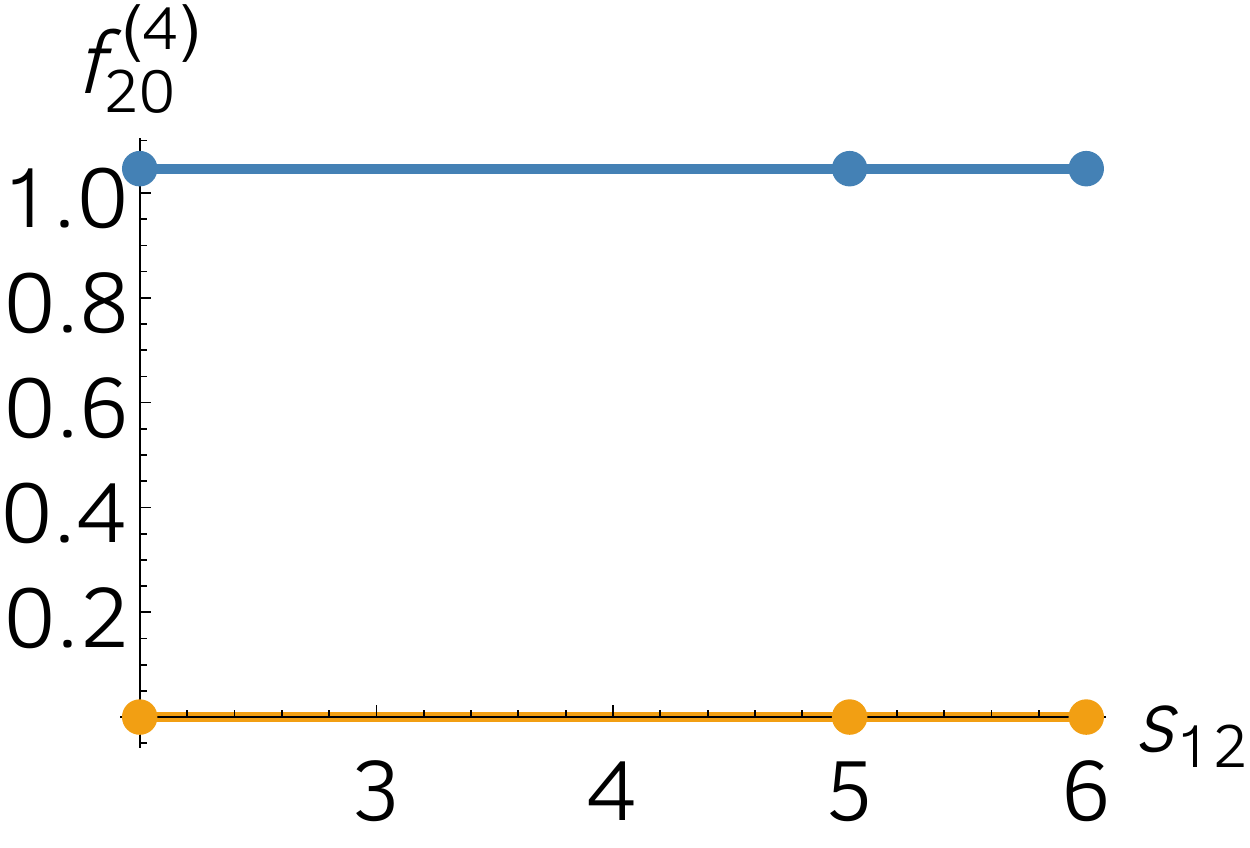} & \img{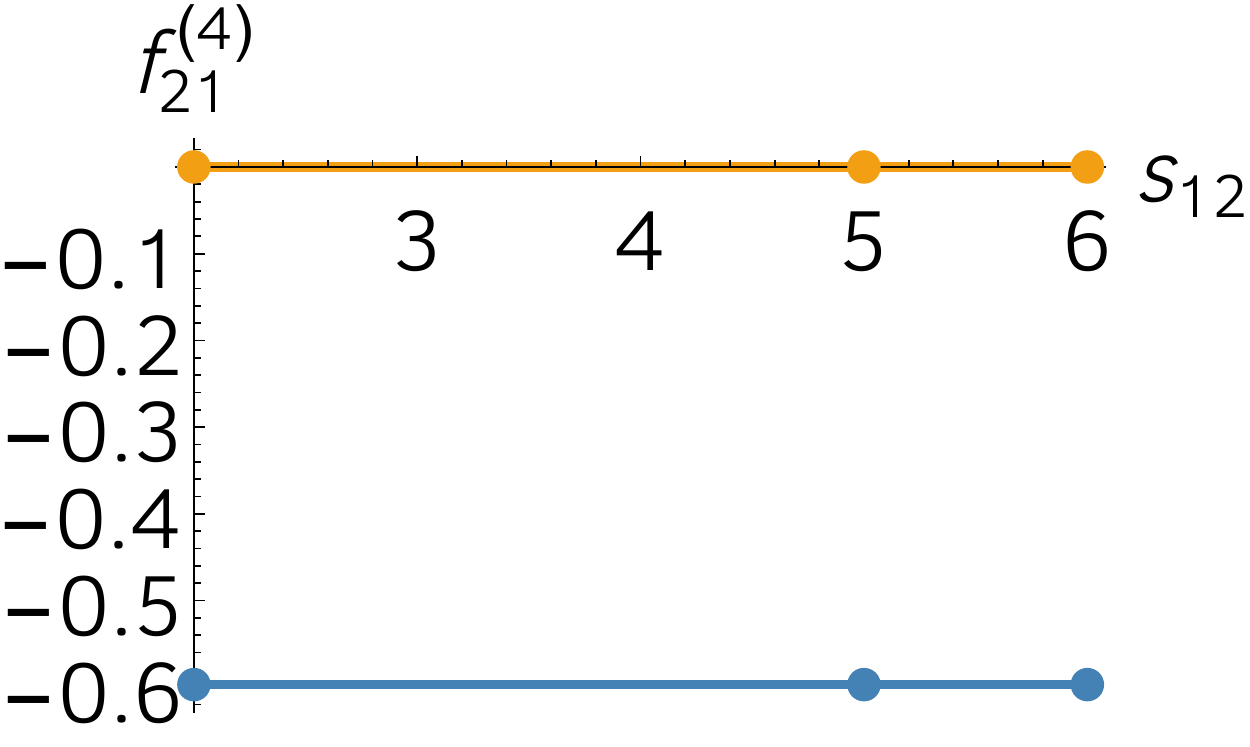} 
\end{tabular}
\end{table}
\begin{table}[ht]
	\begin{tabular}{ p{4.75cm}  p{4.75cm}  p{4.75cm}}
	\img{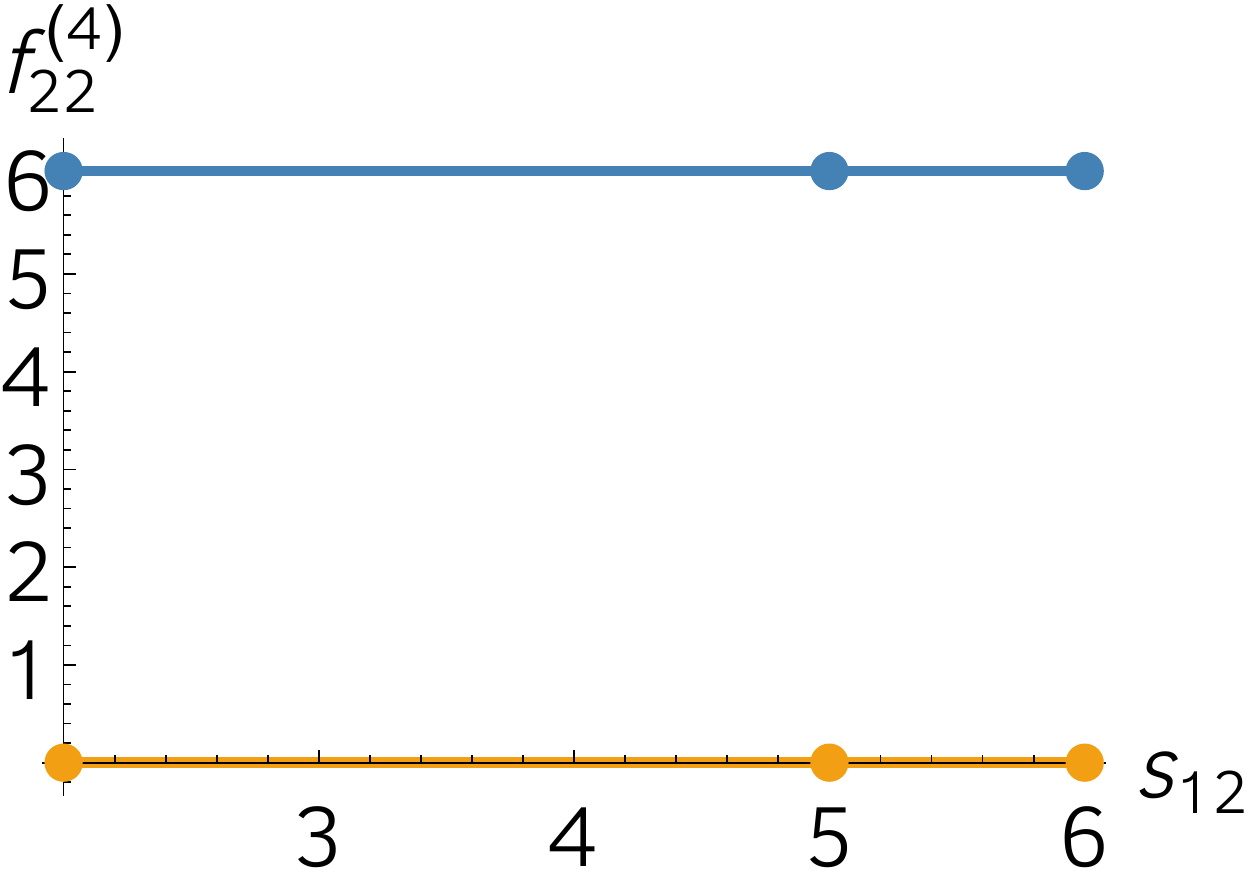} & \img{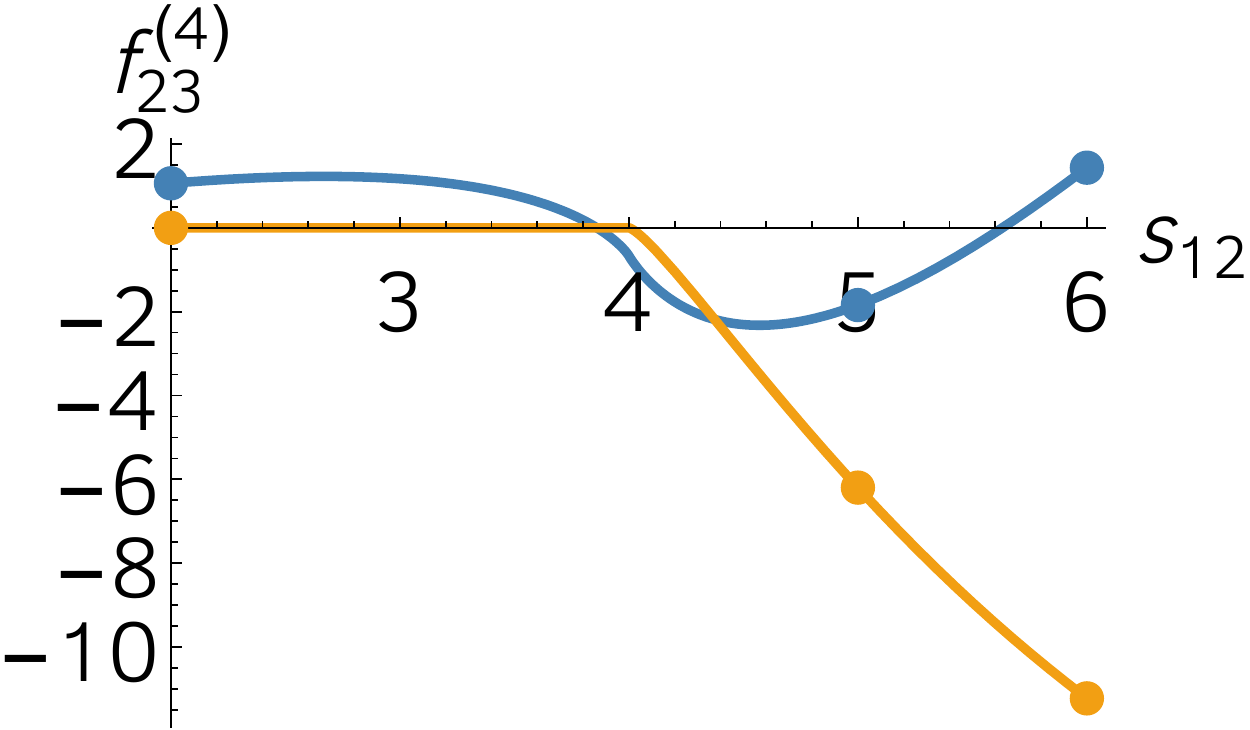} & \img{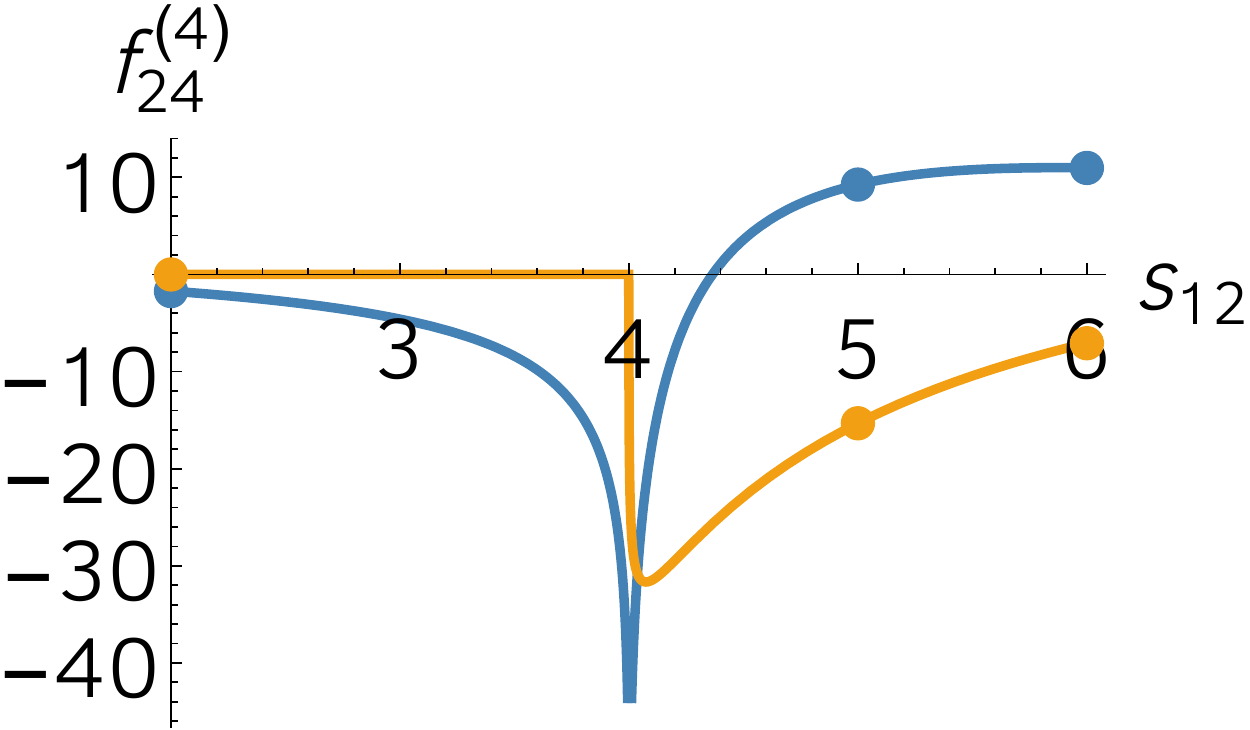} \\
\img{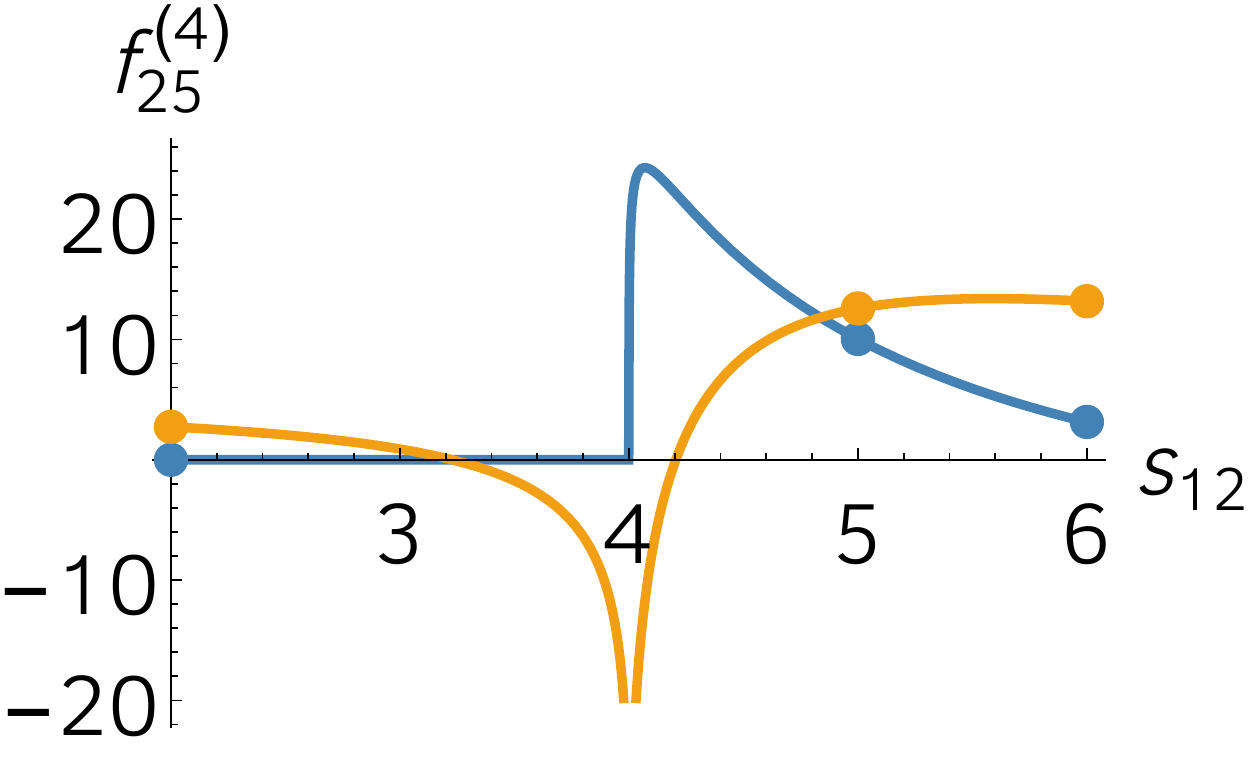} & \img{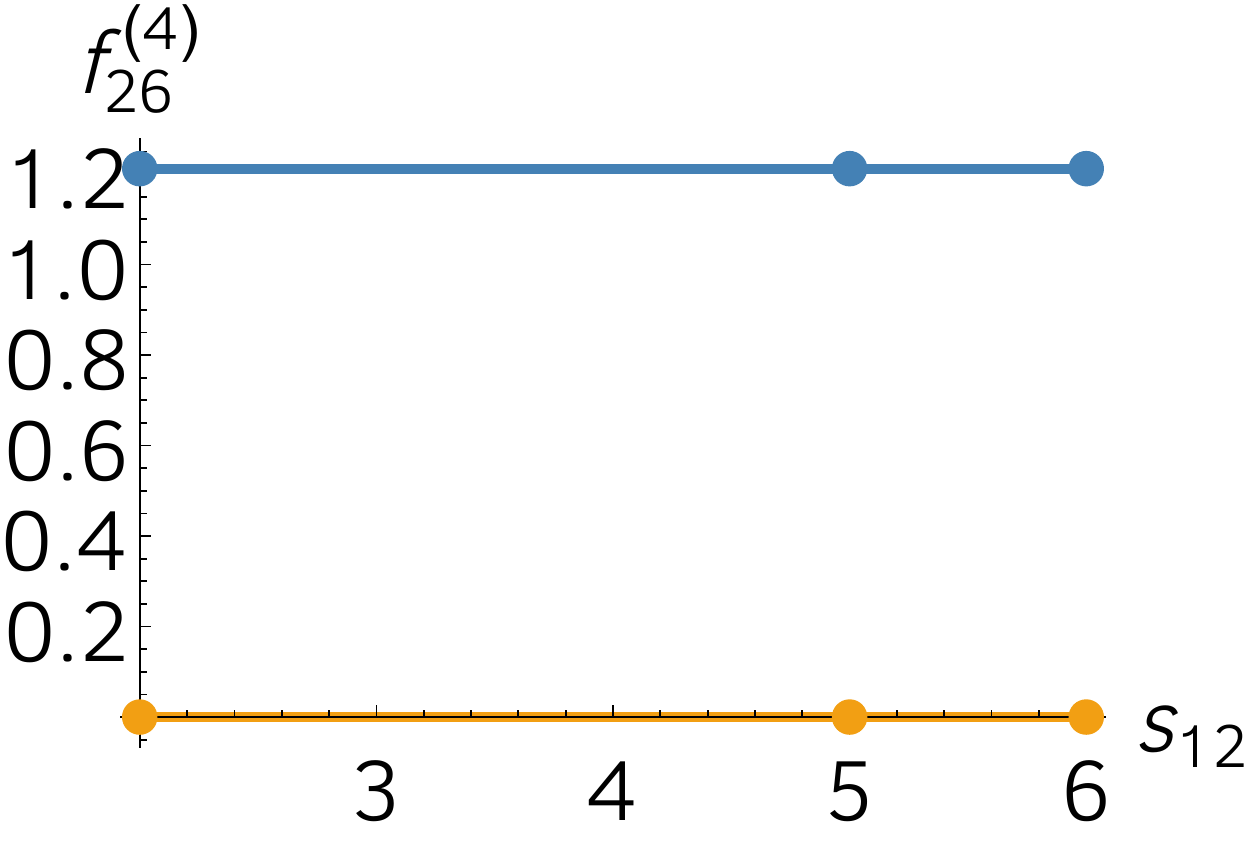} & \img{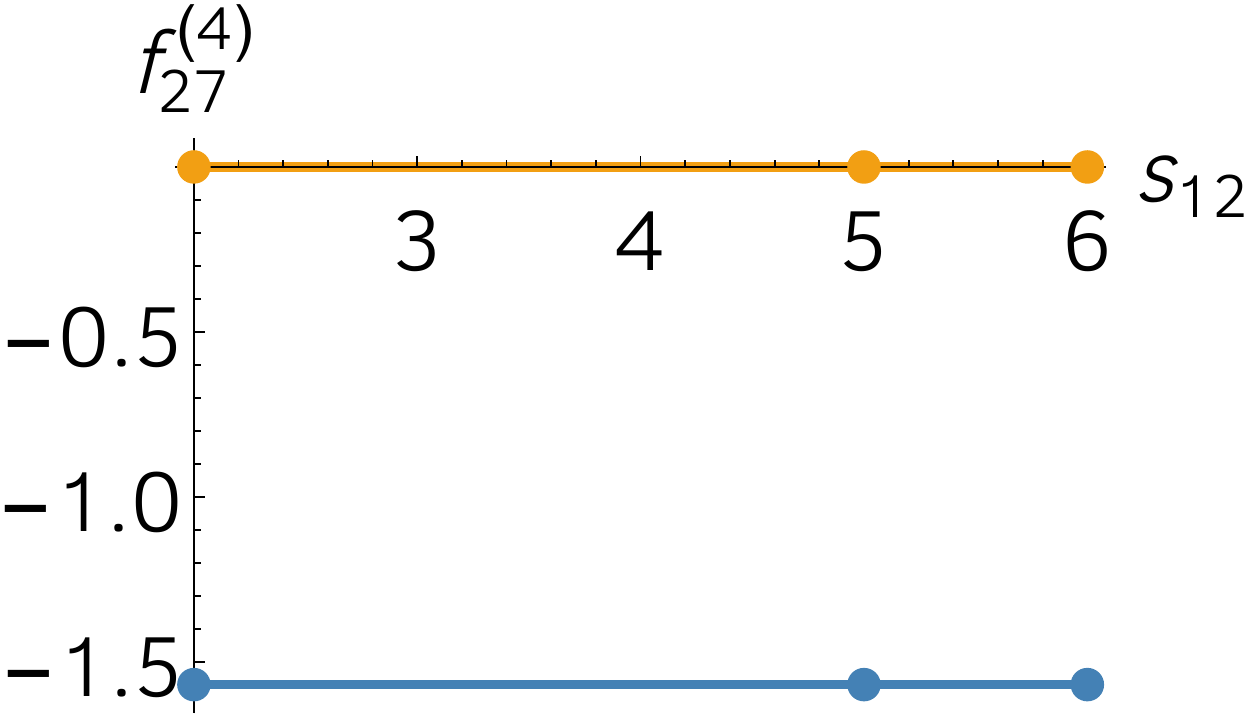} \\
\img{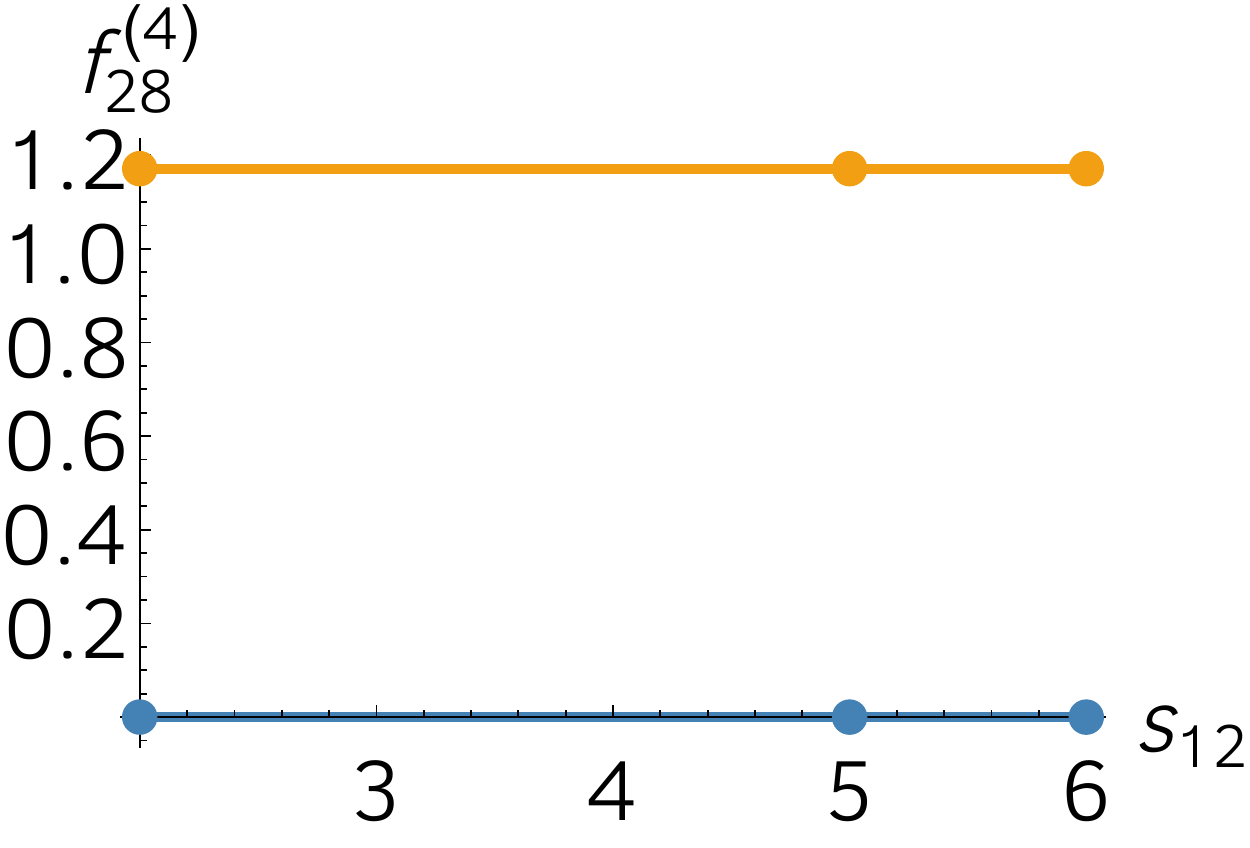} & \img{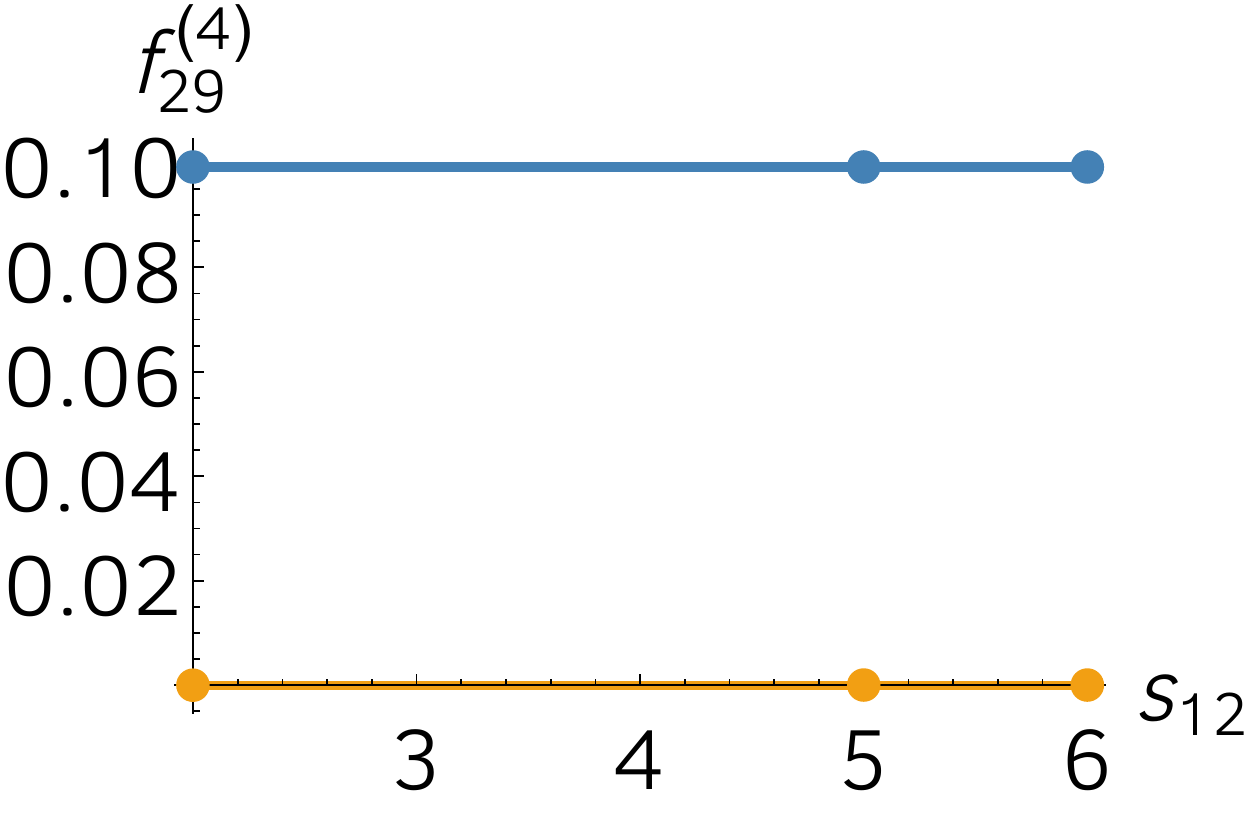} & \img{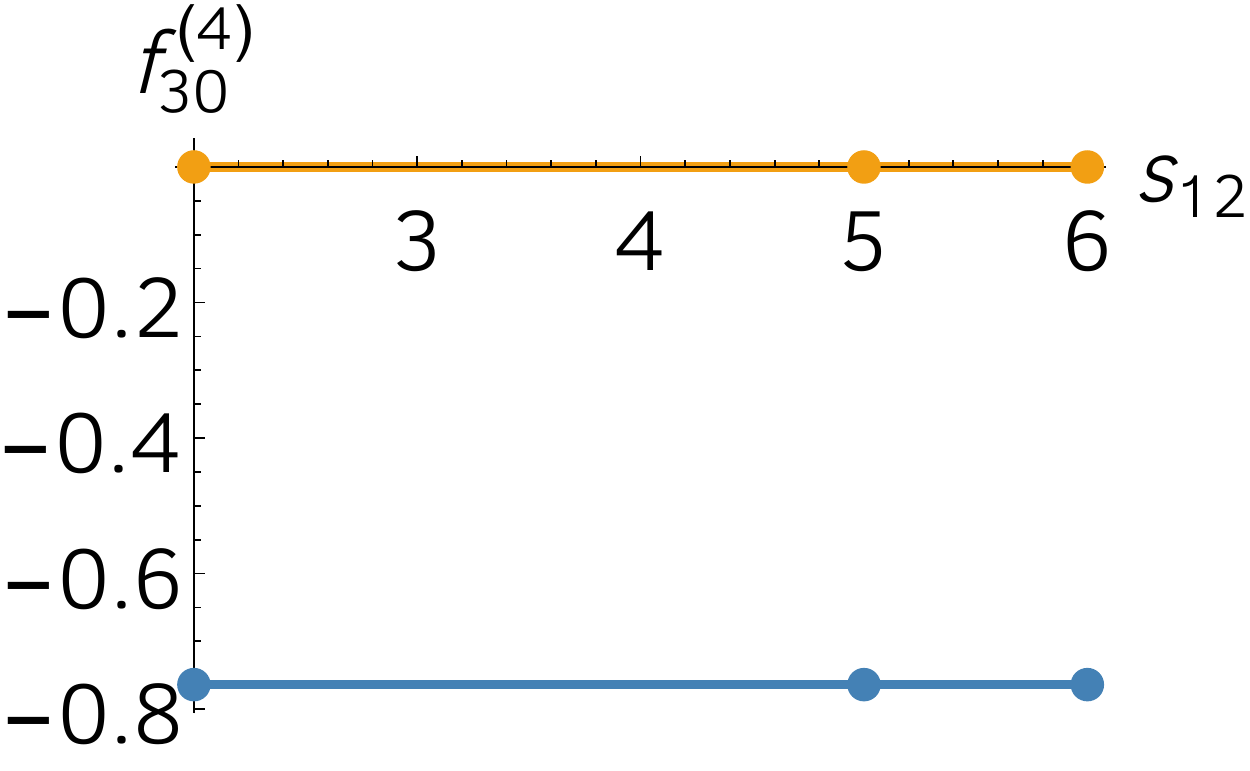} \\
\img{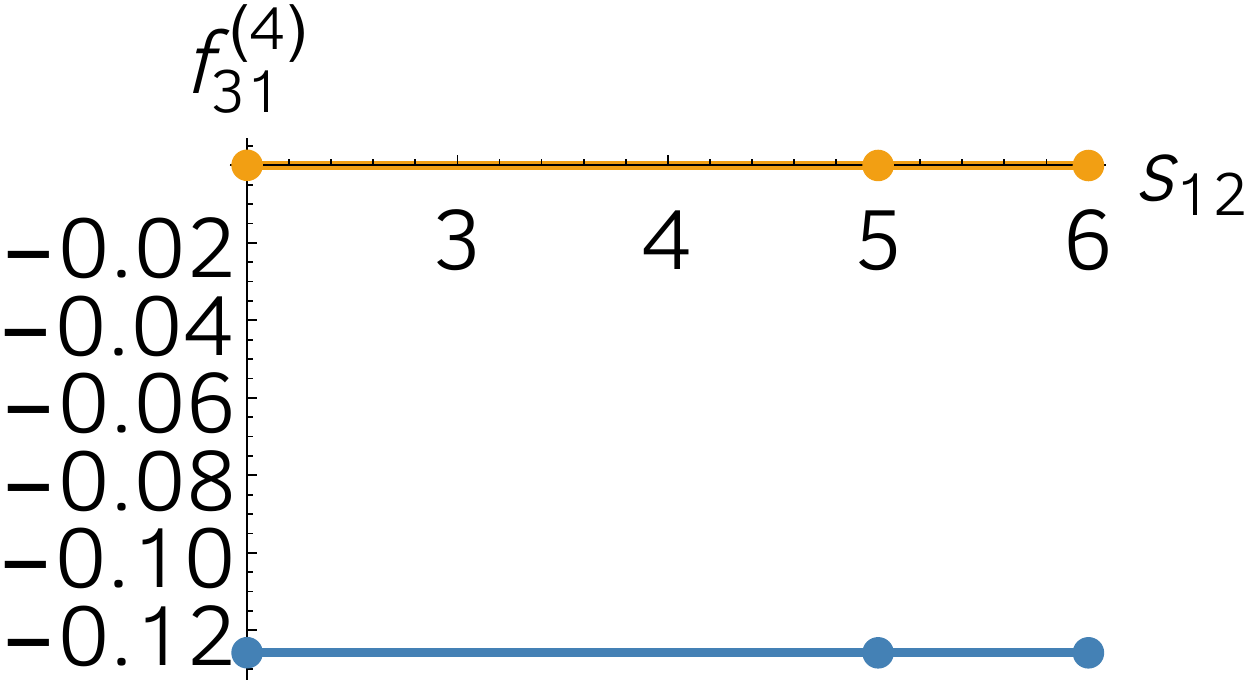} & \img{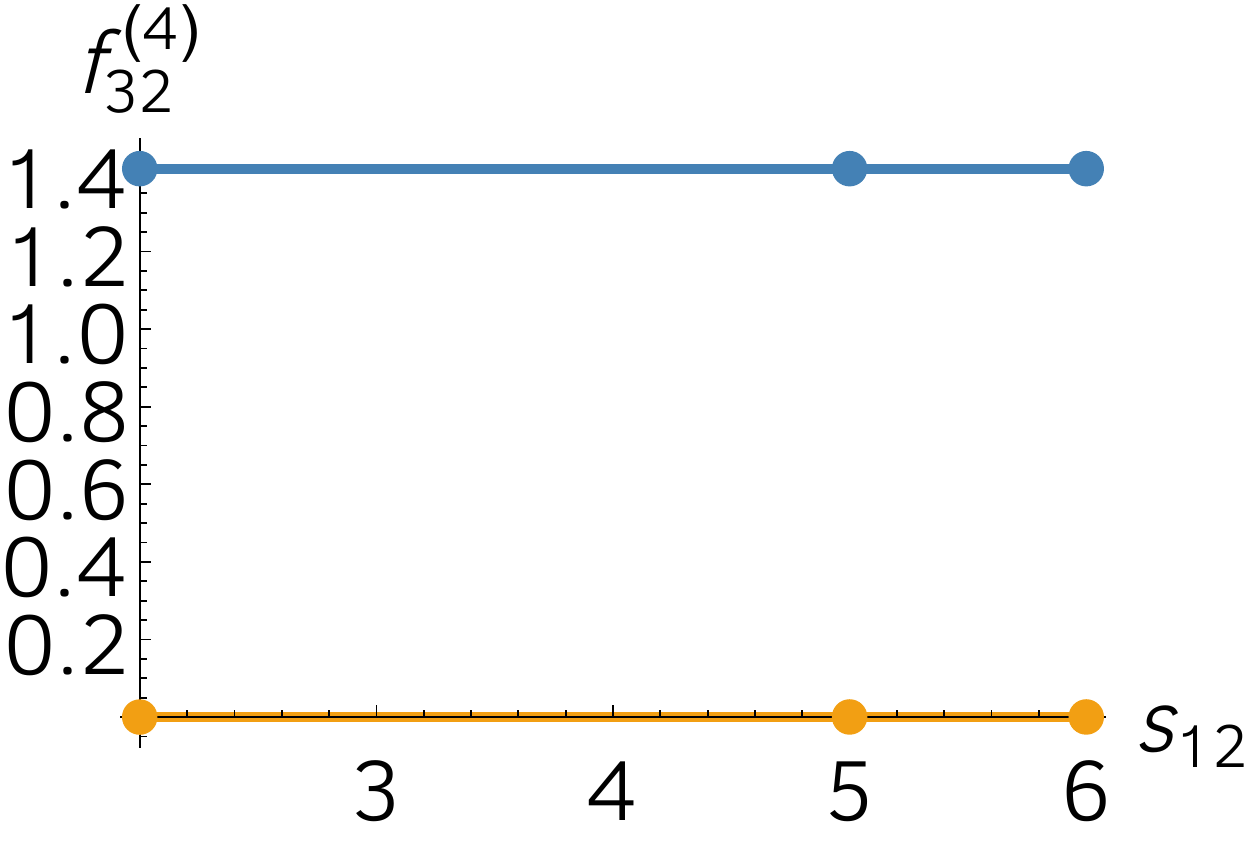} & \img{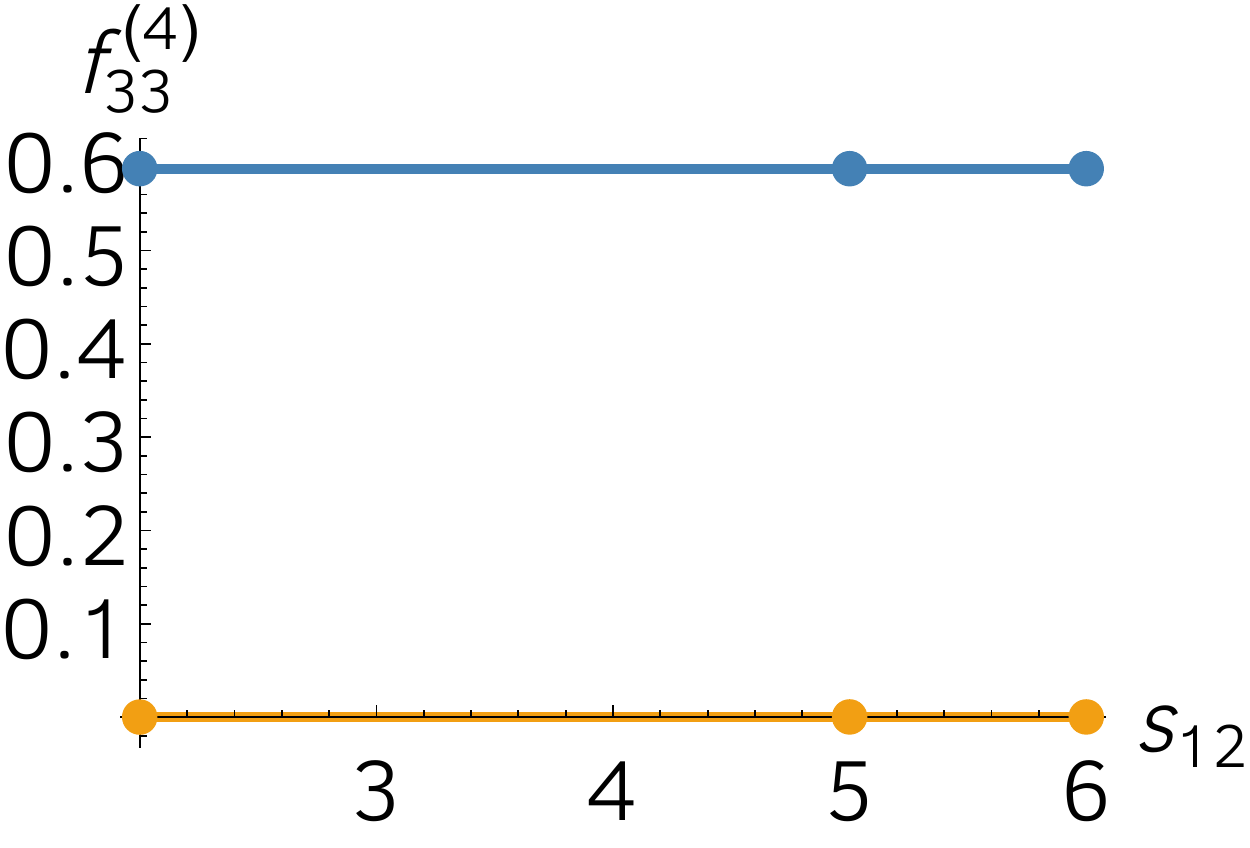} \\
\img{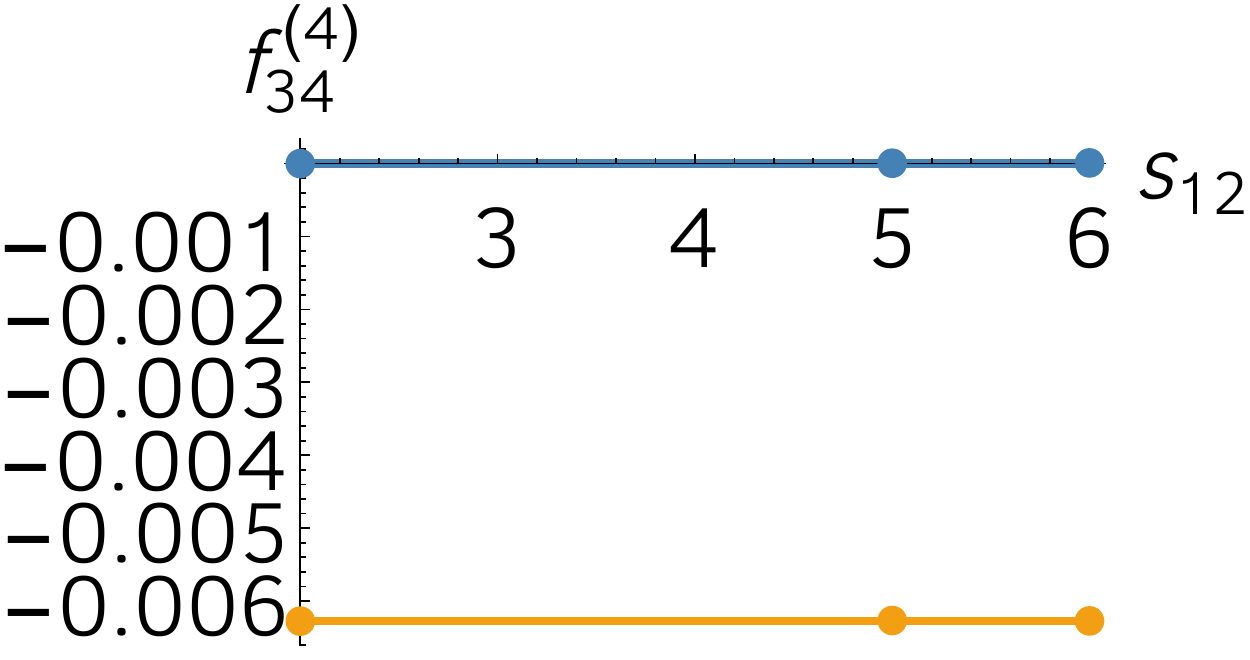} & \img{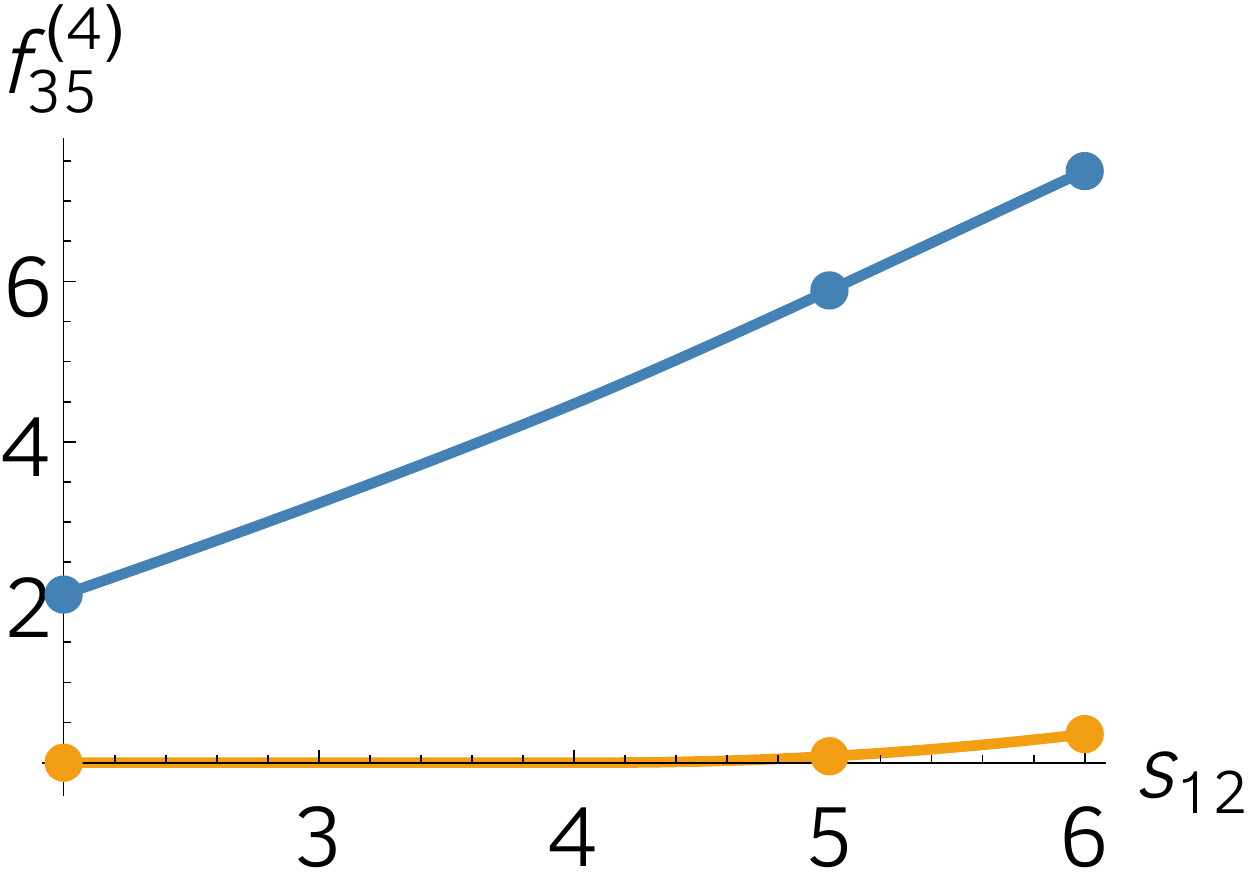} & \img{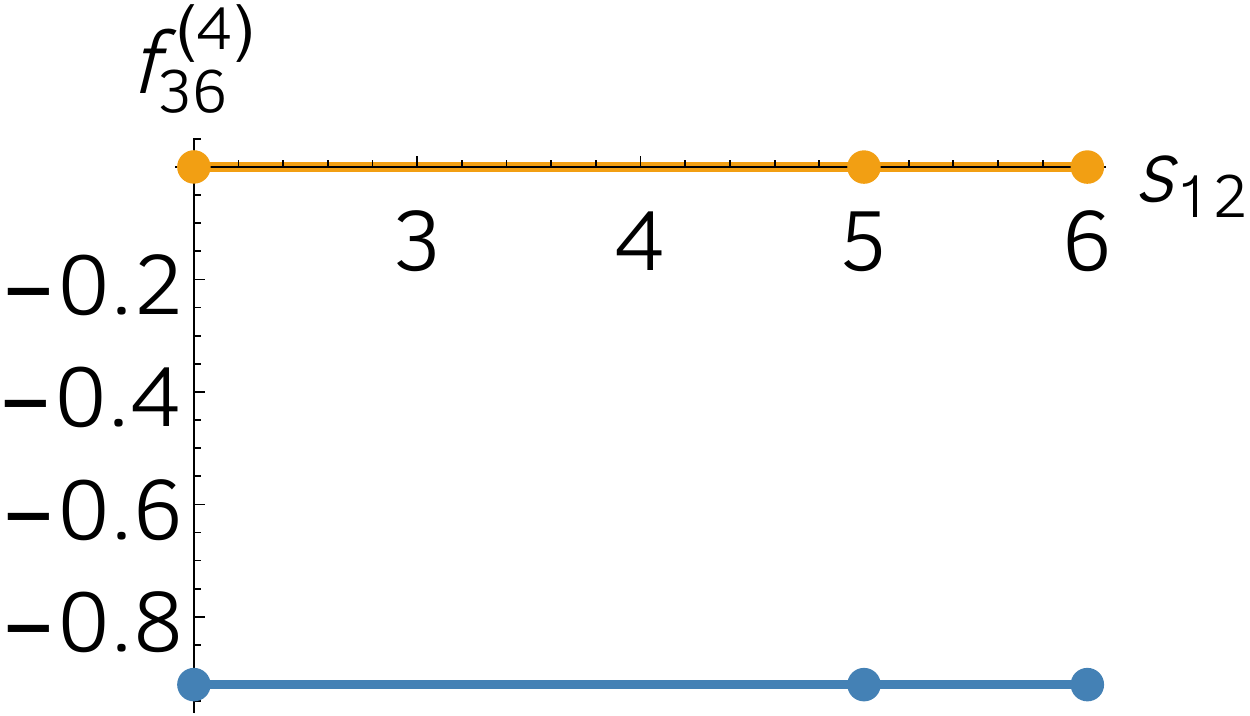} \\
\img{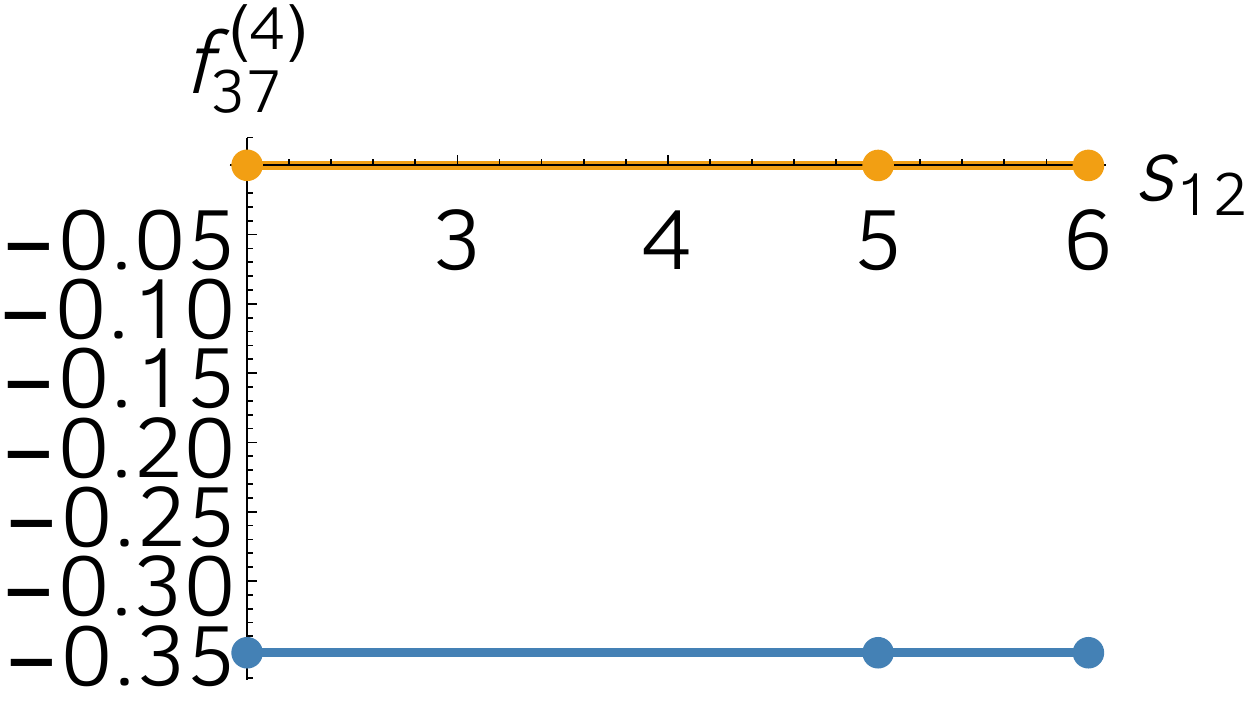} & \img{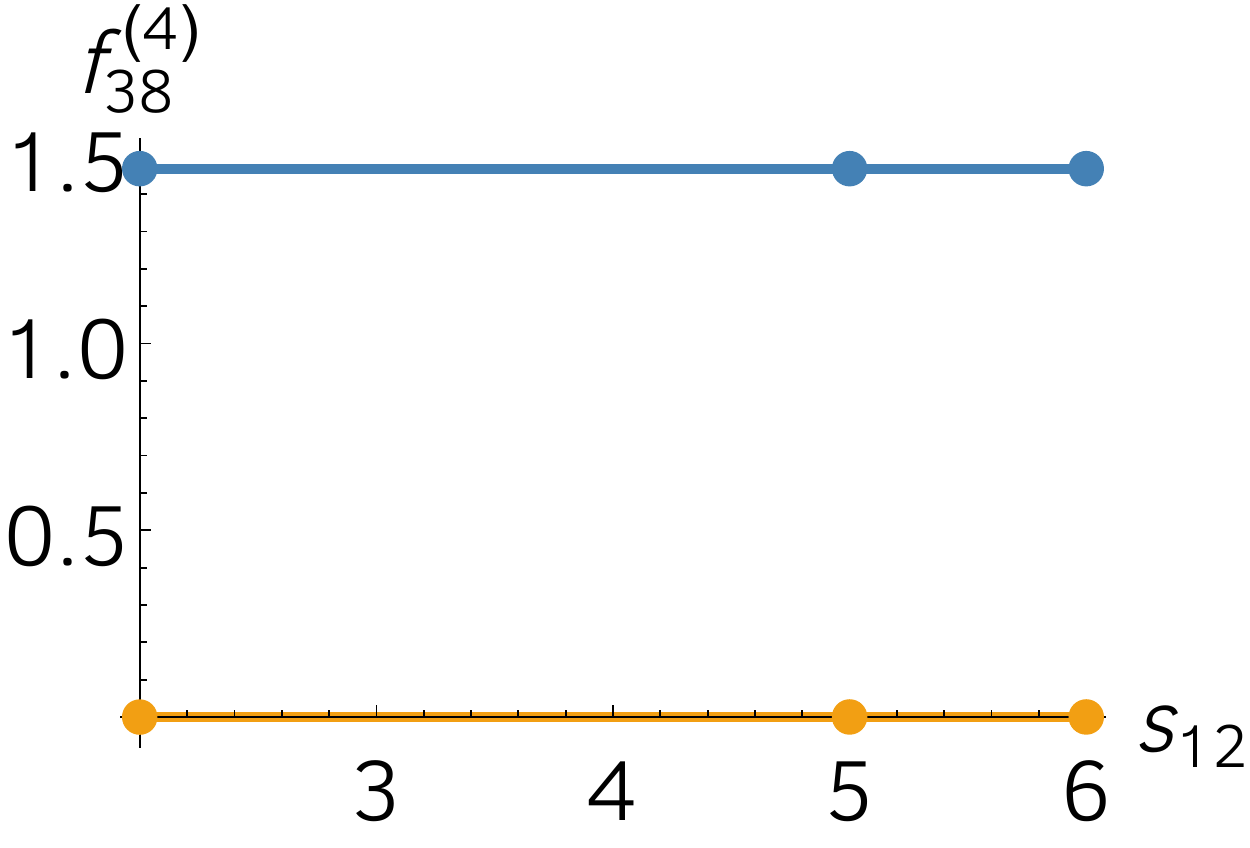} & \img{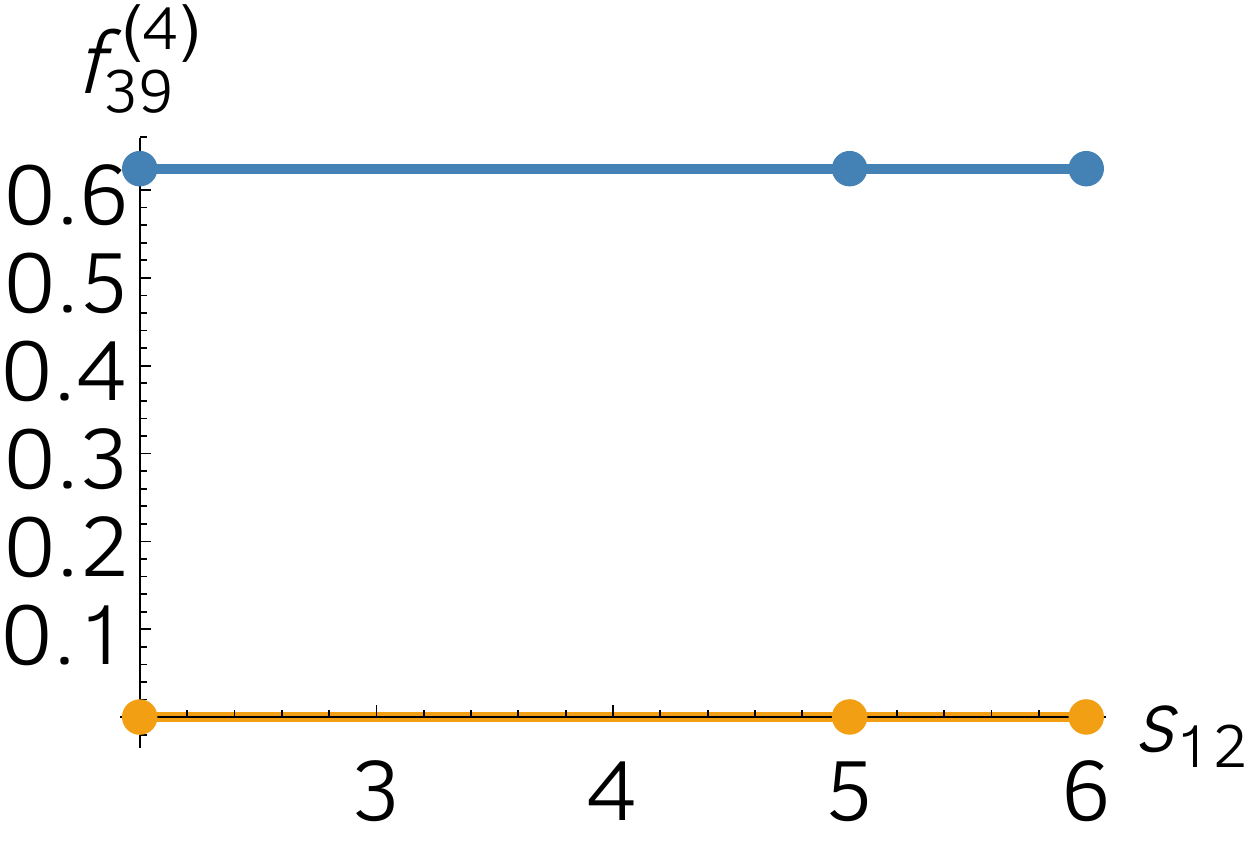} \\
\img{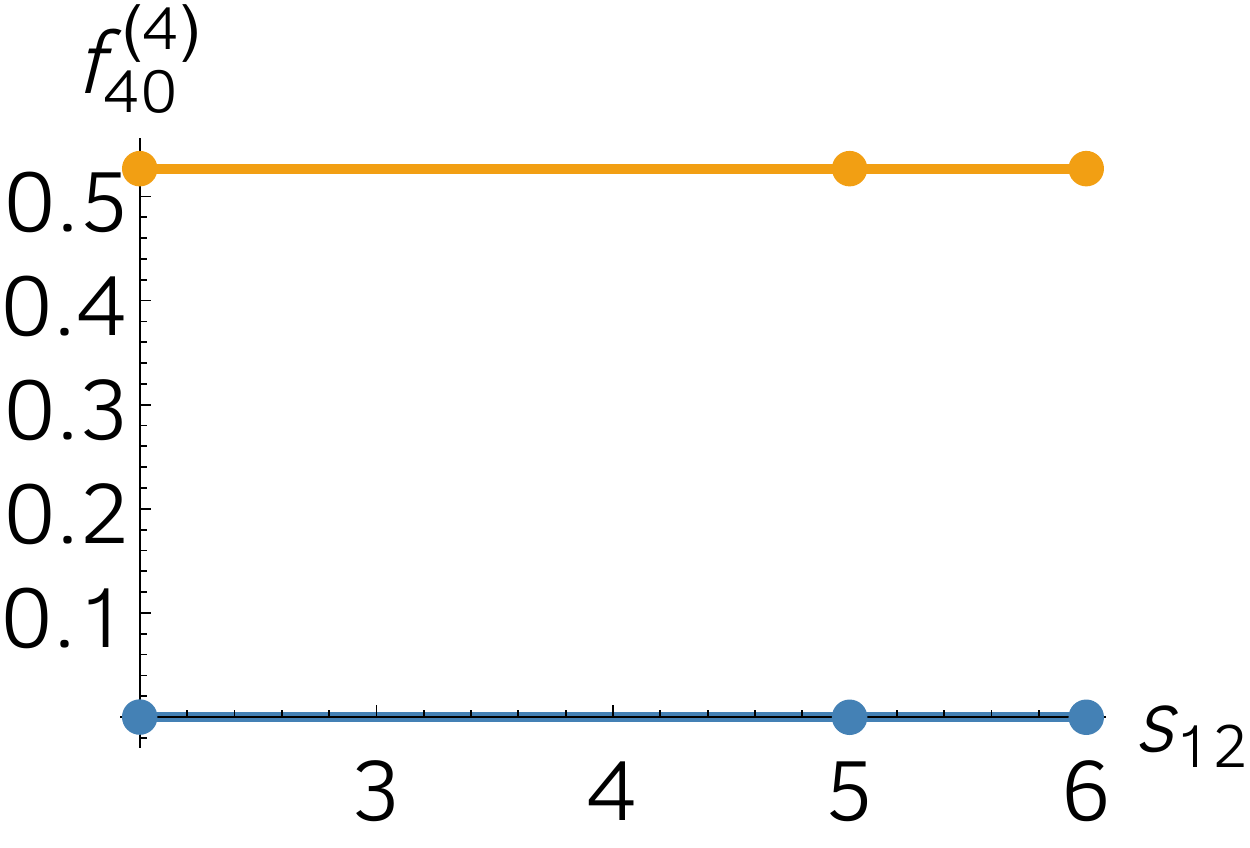} & \img{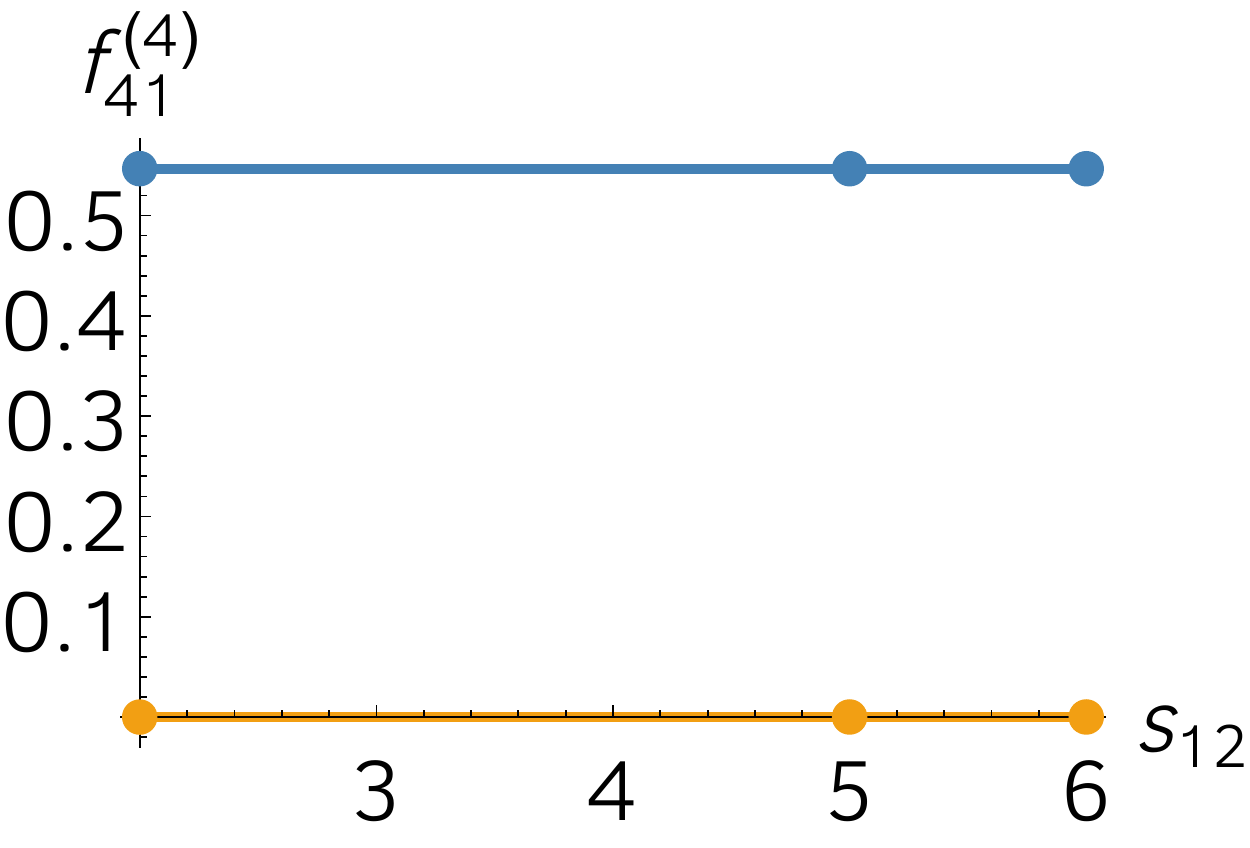} & \img{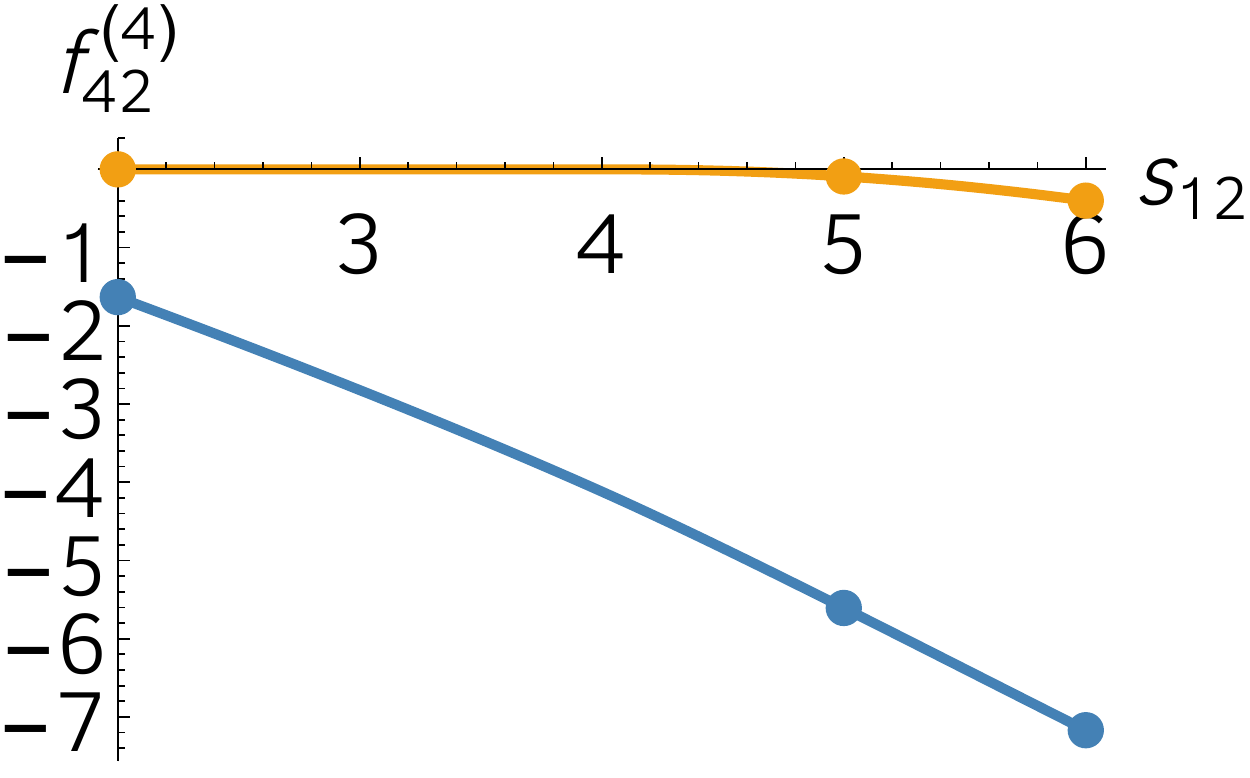} \\
\img{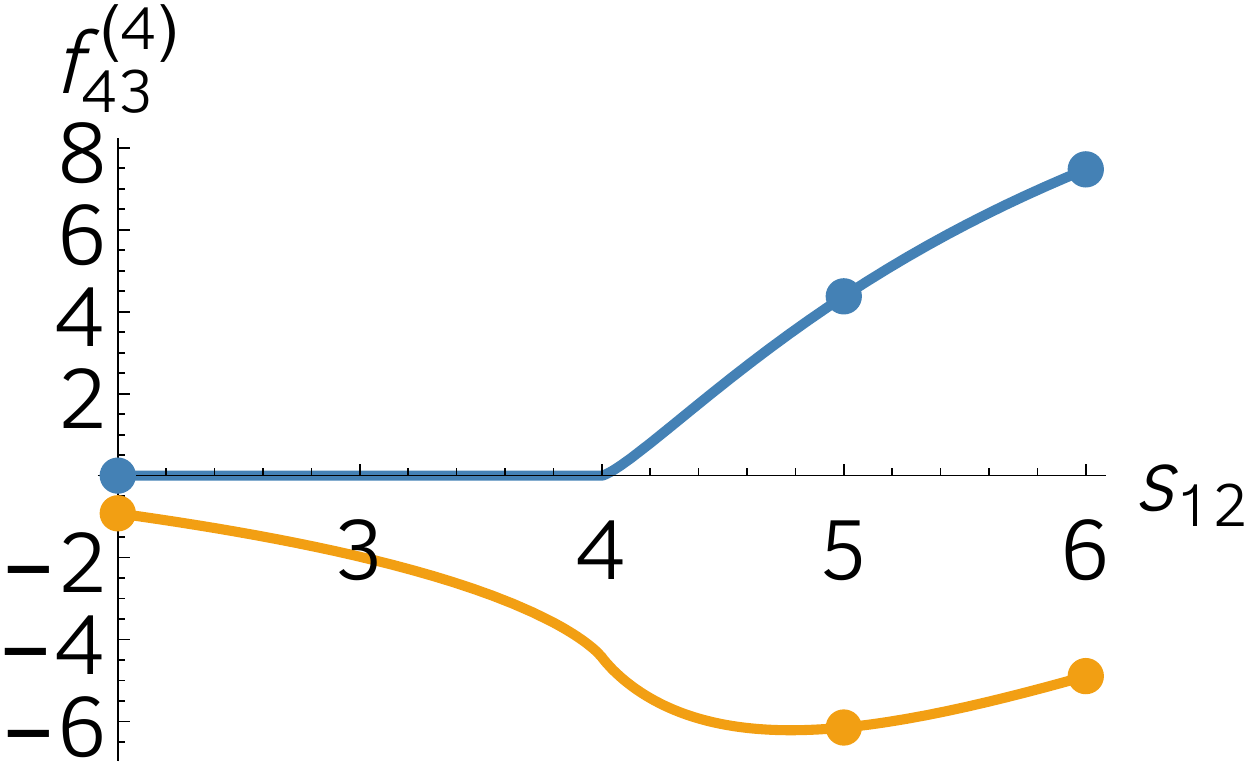} & \img{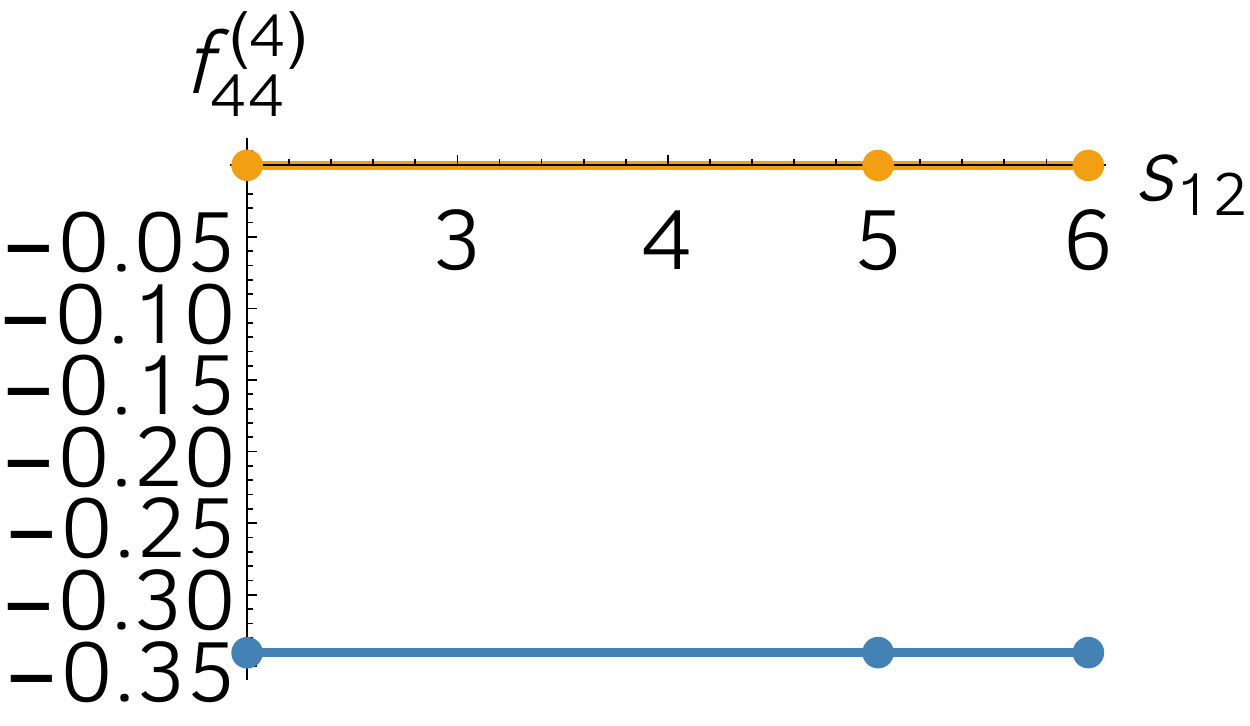} & \img{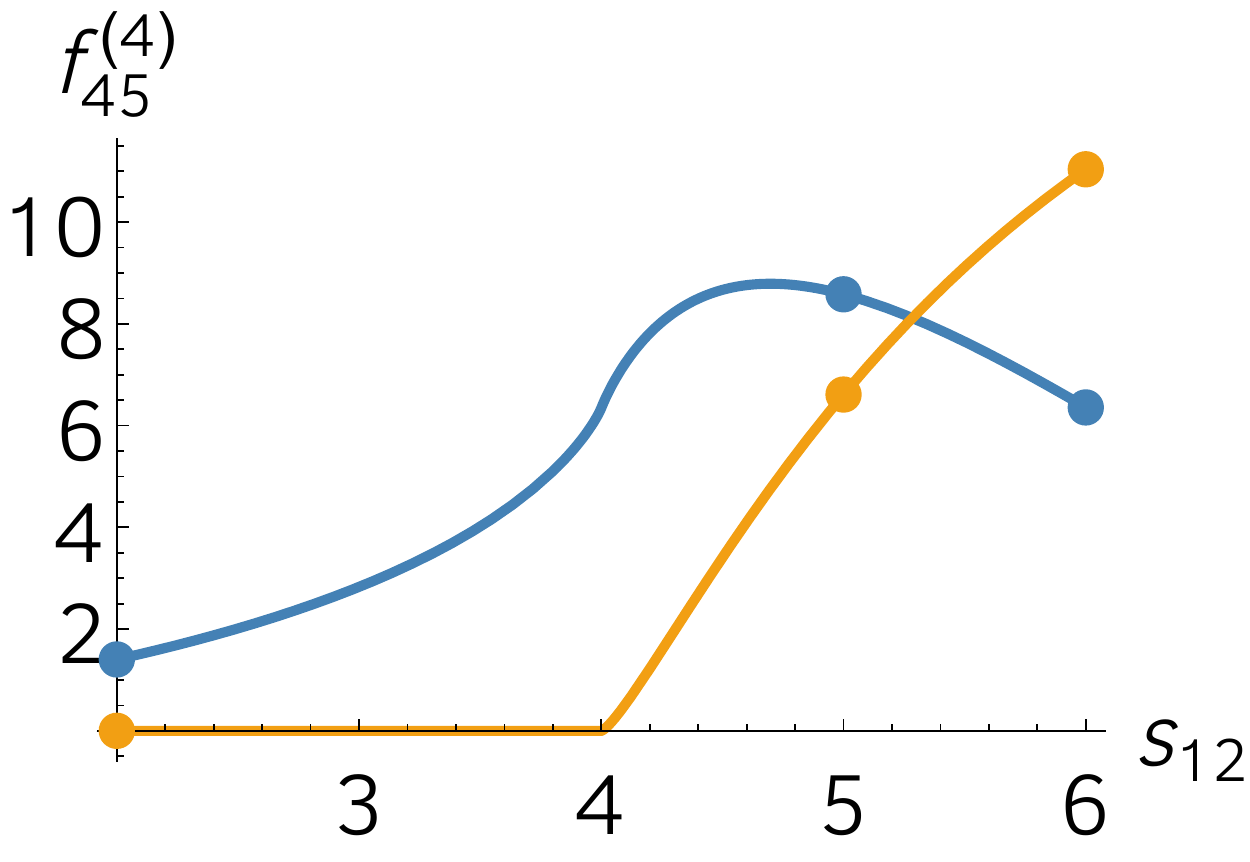} \\
\img{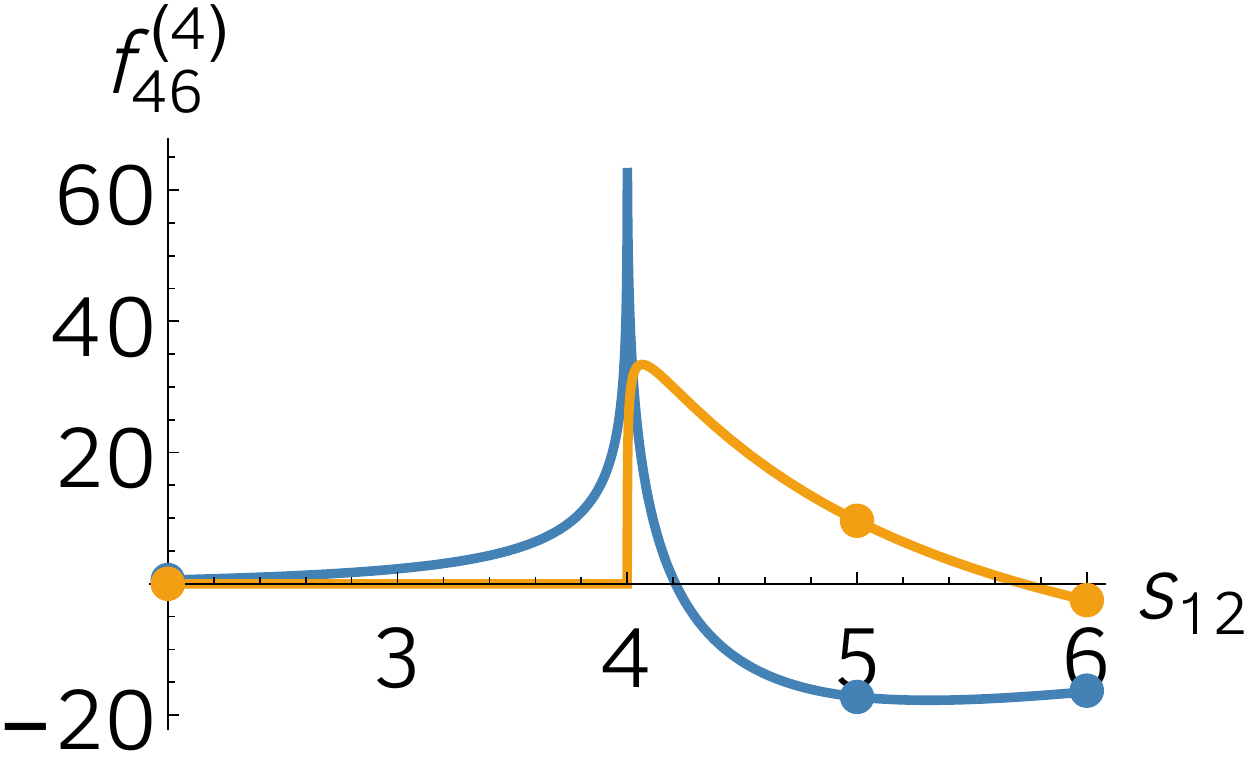} & \img{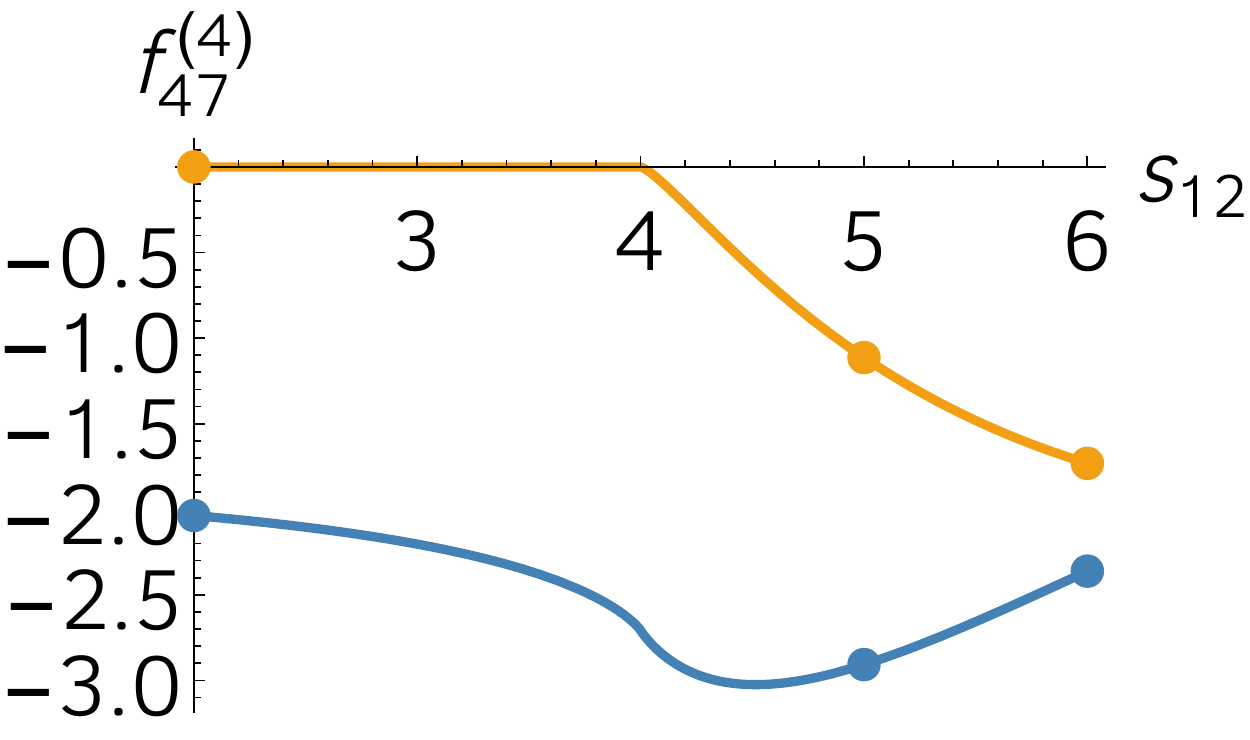} & \img{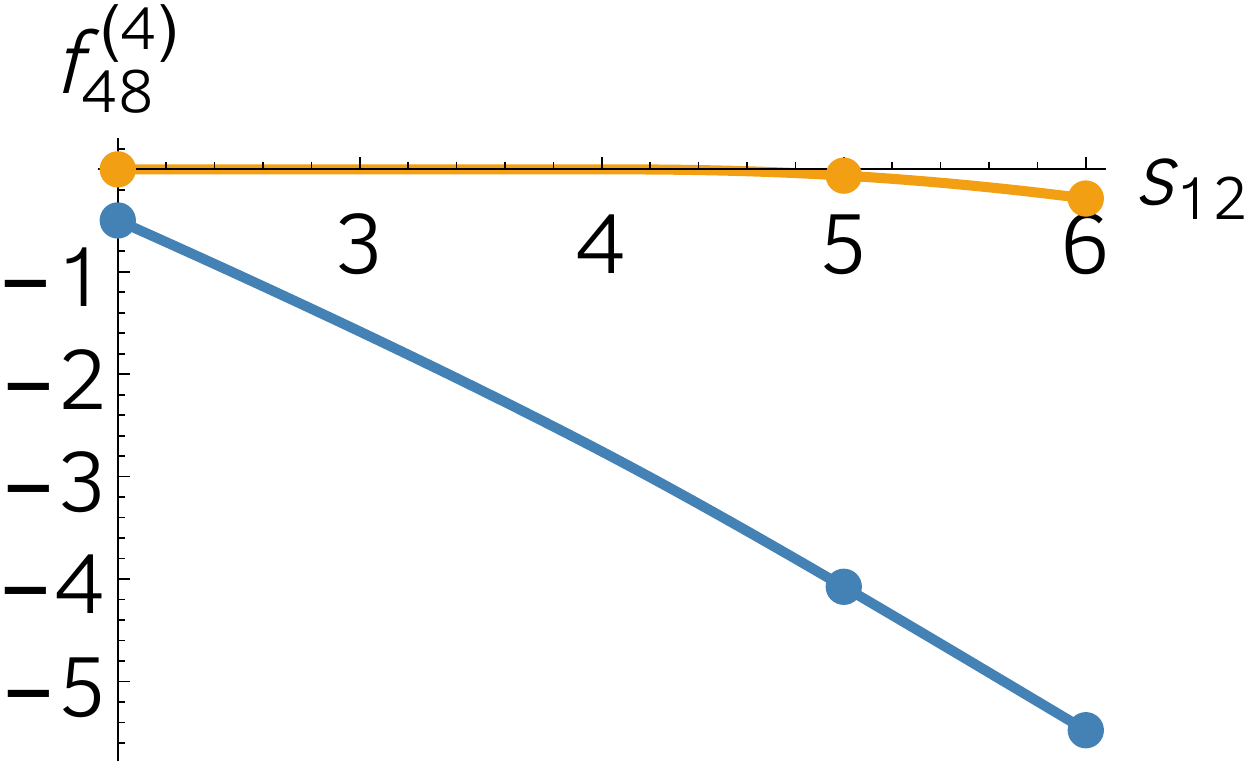} 
\end{tabular}
\end{table}
\begin{table}[ht]
	\begin{tabular}{ p{4.75cm}  p{4.75cm}  p{4.75cm} }
\img{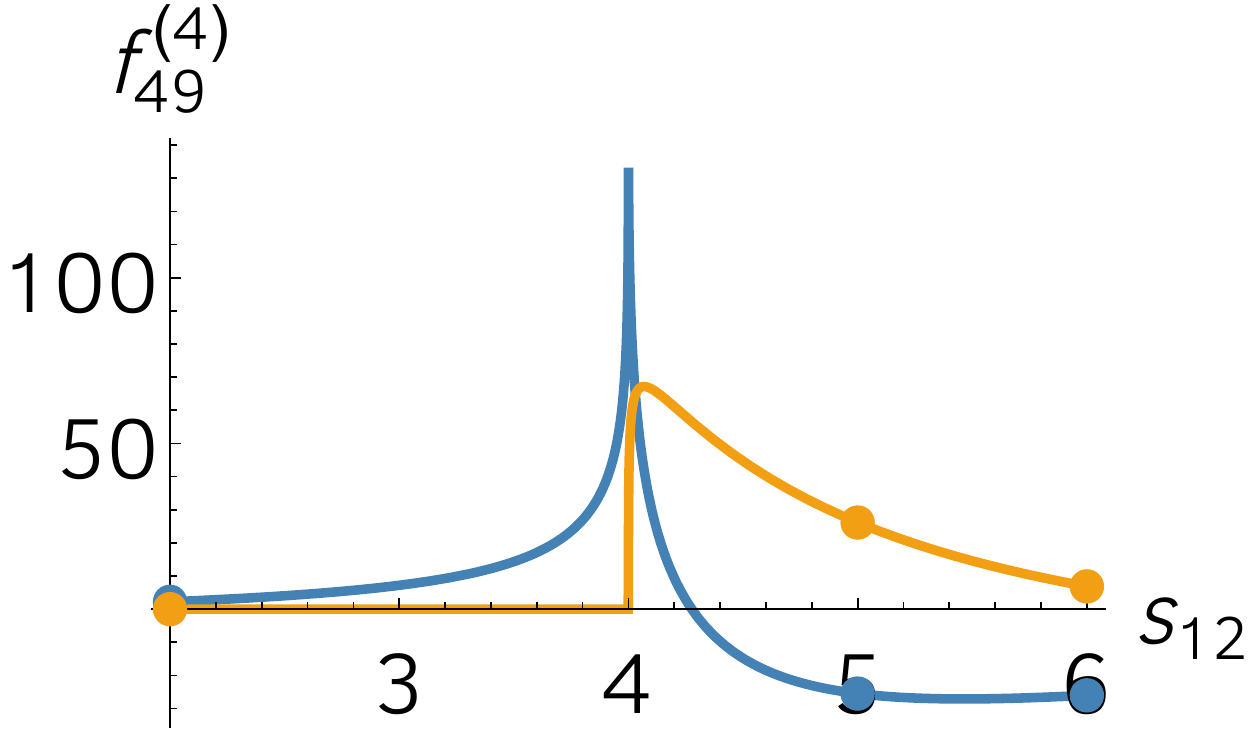} & \img{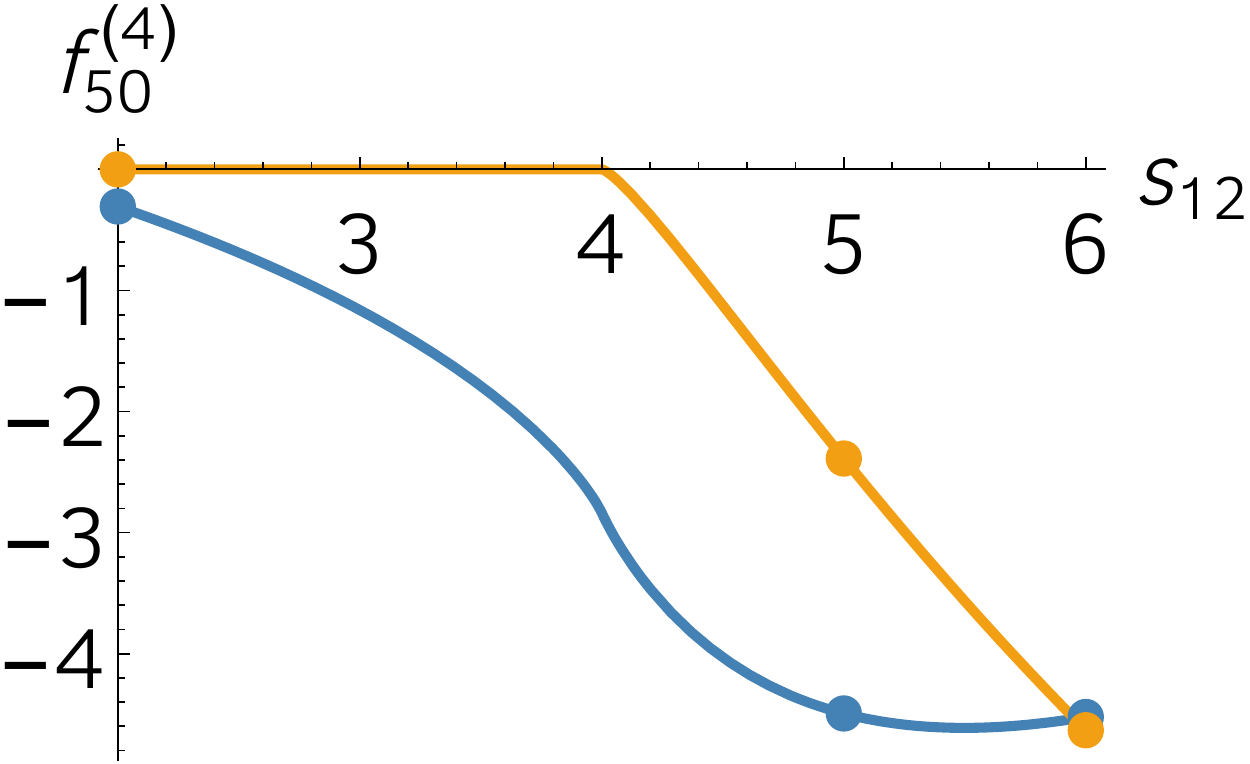} & \img{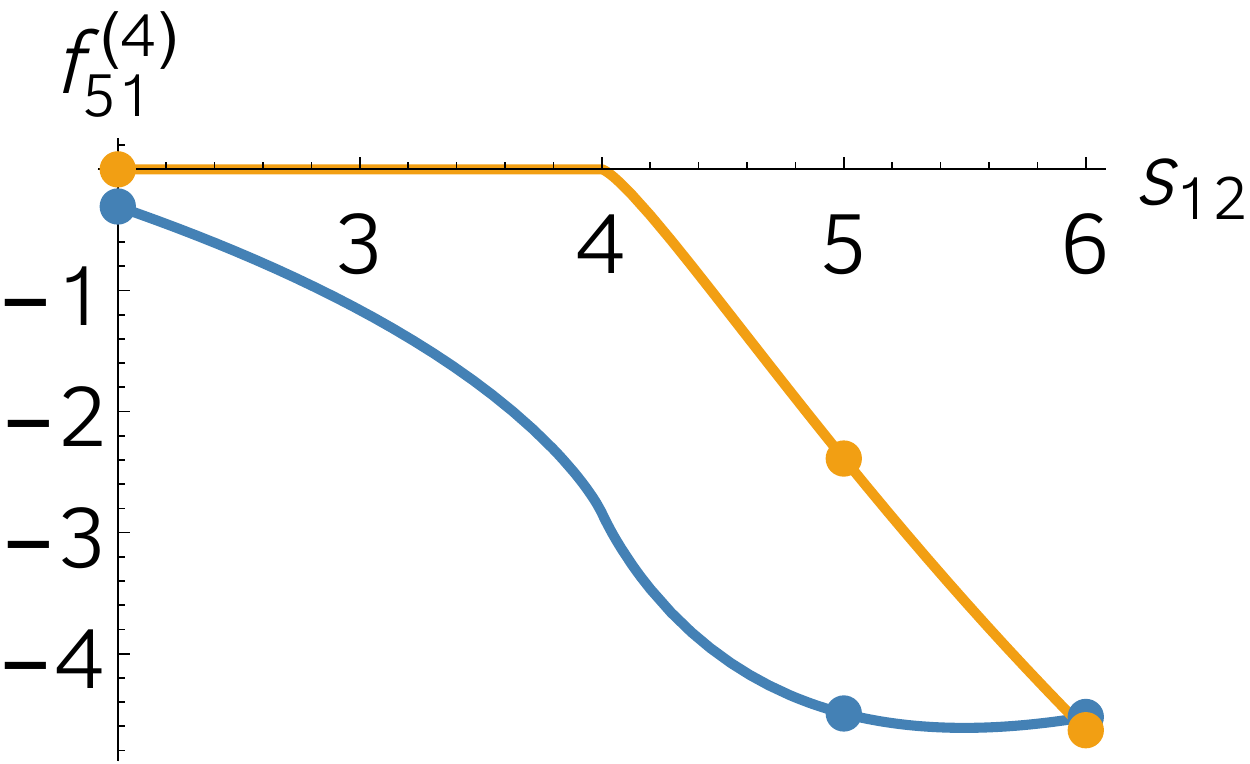} \\
\img{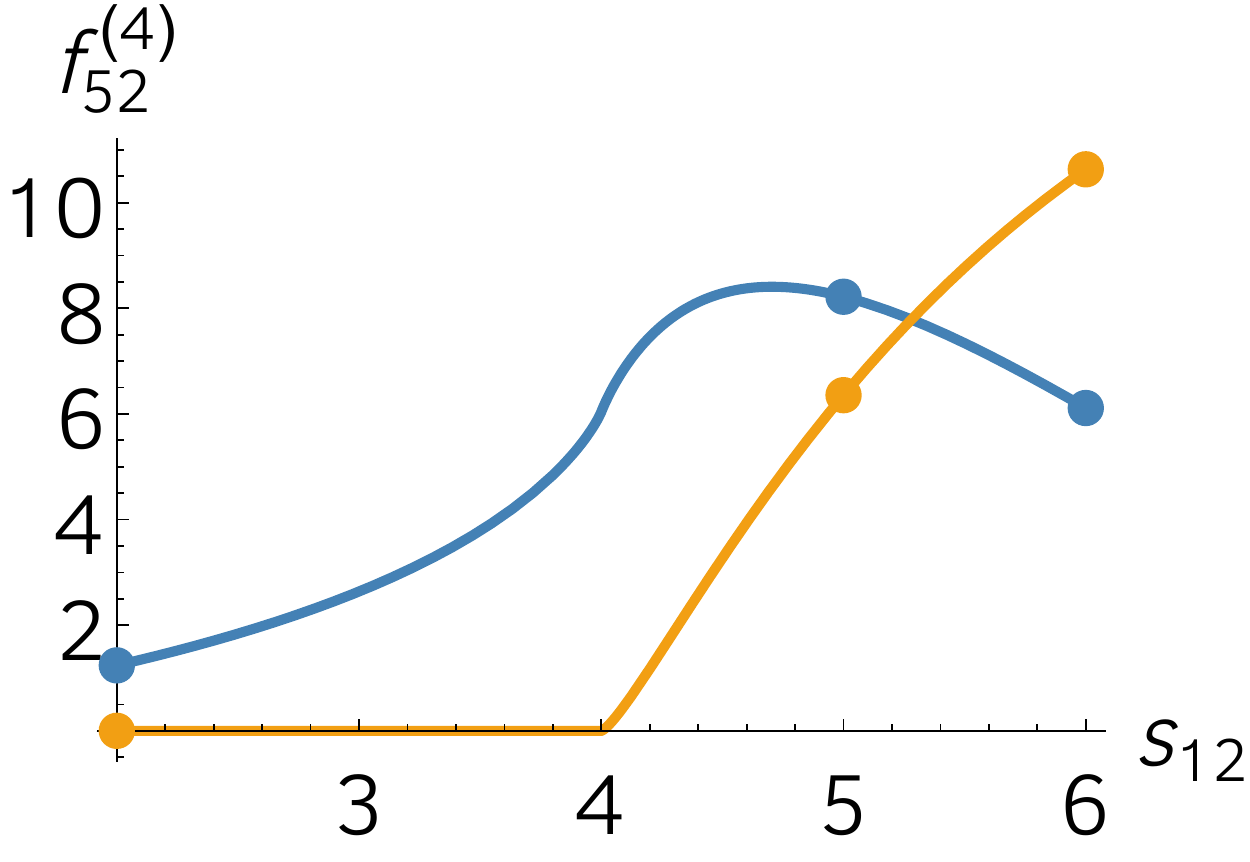} & \img{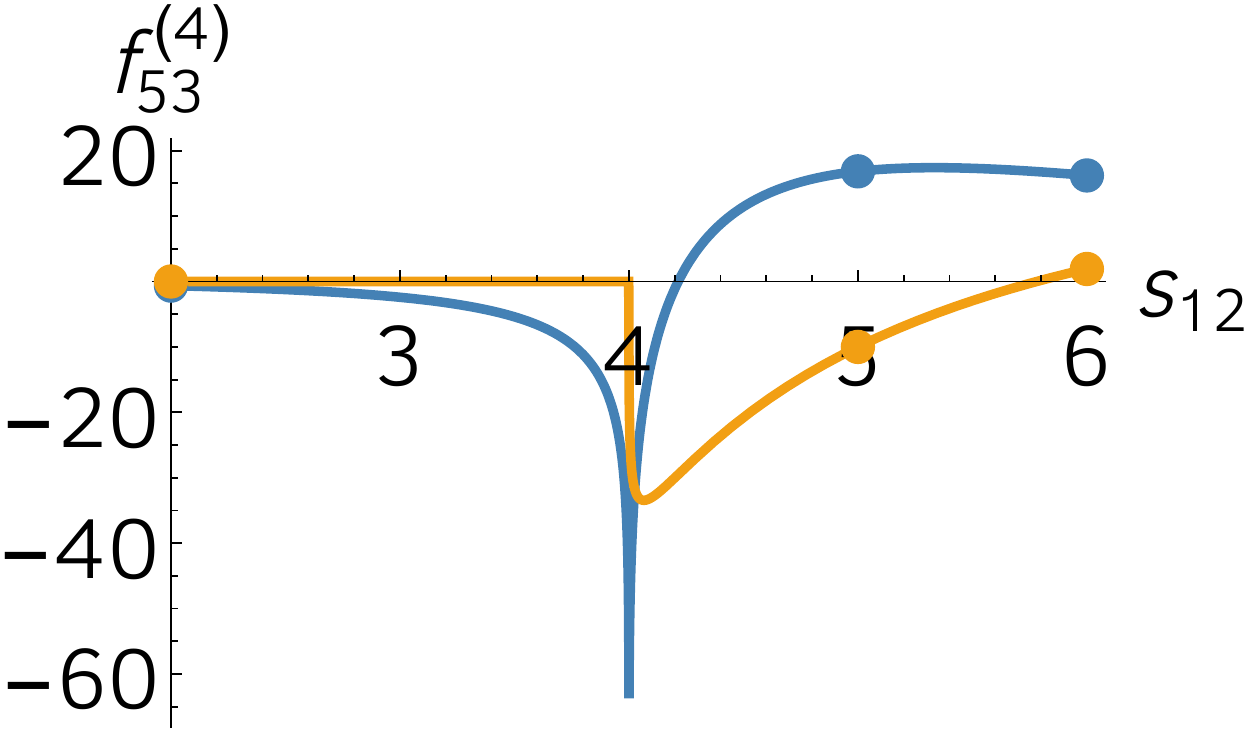} & \img{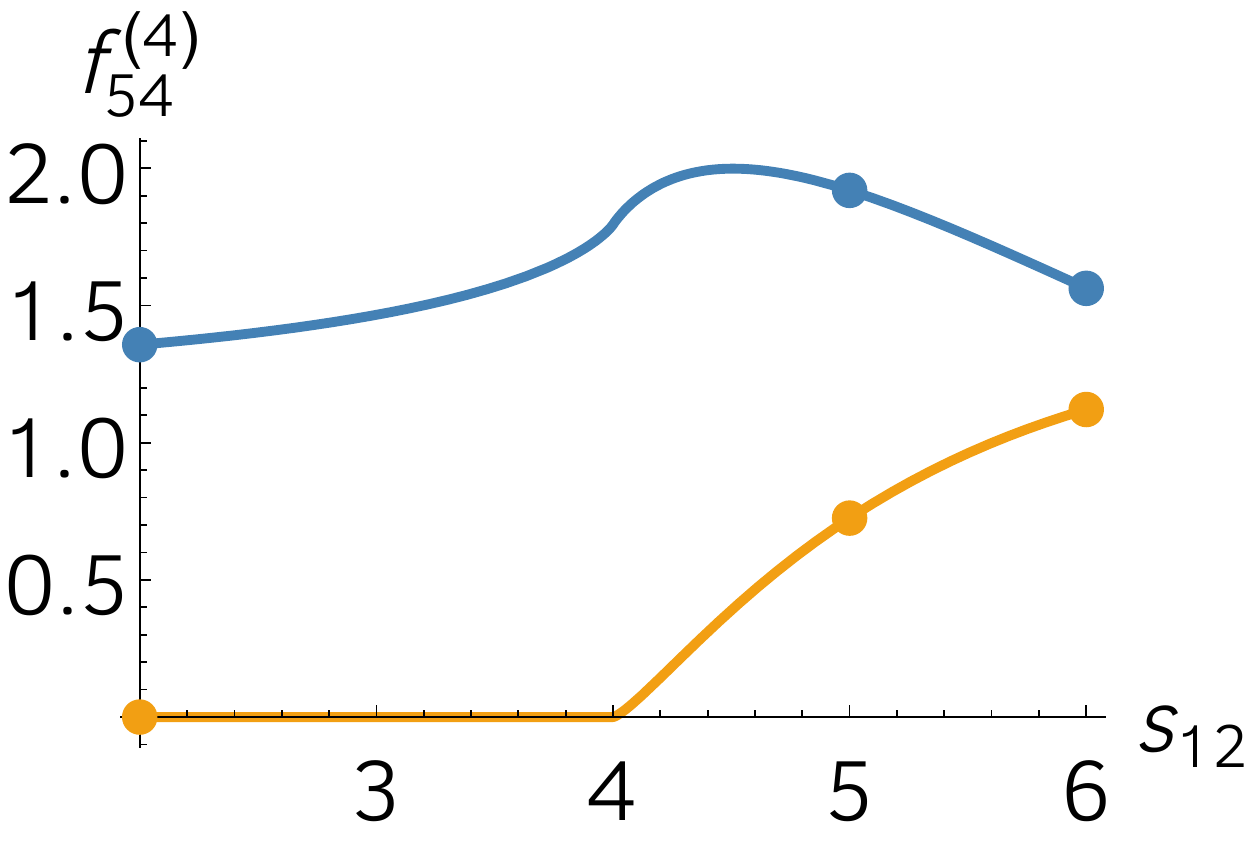} \\
\img{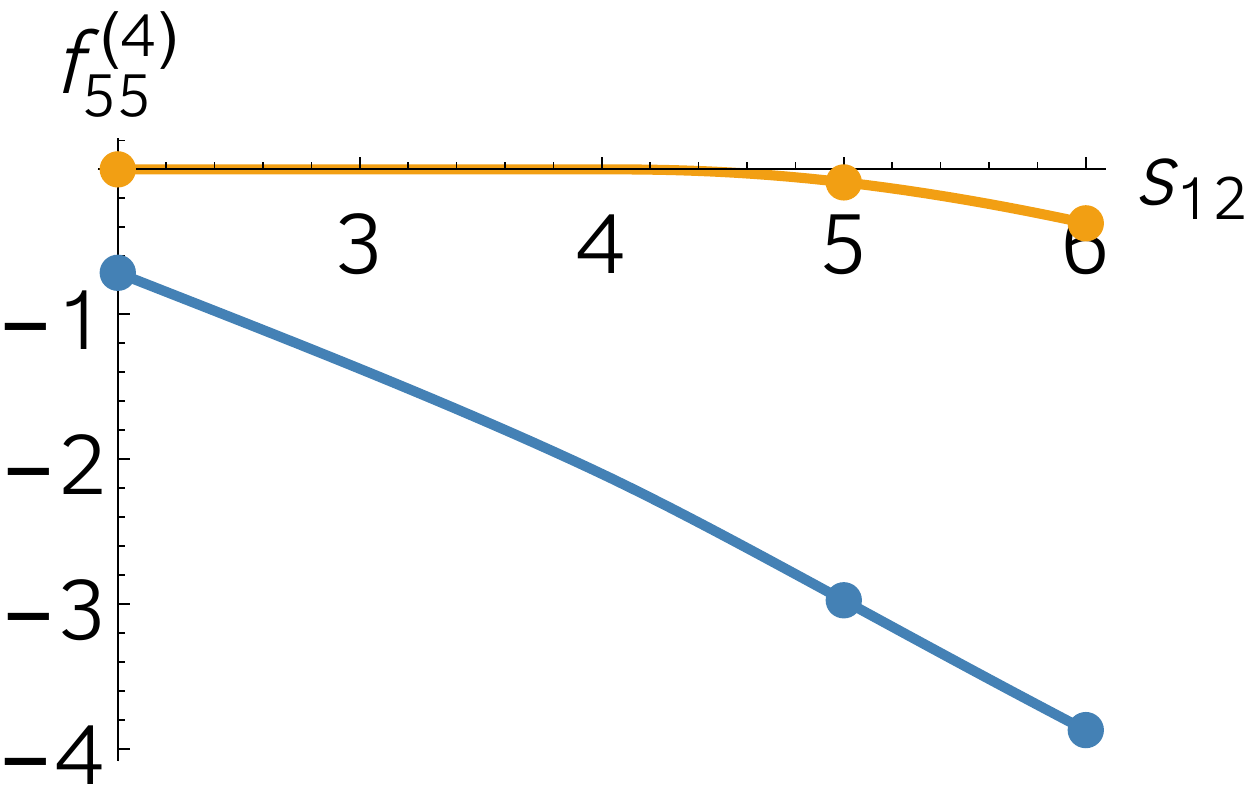} & \img{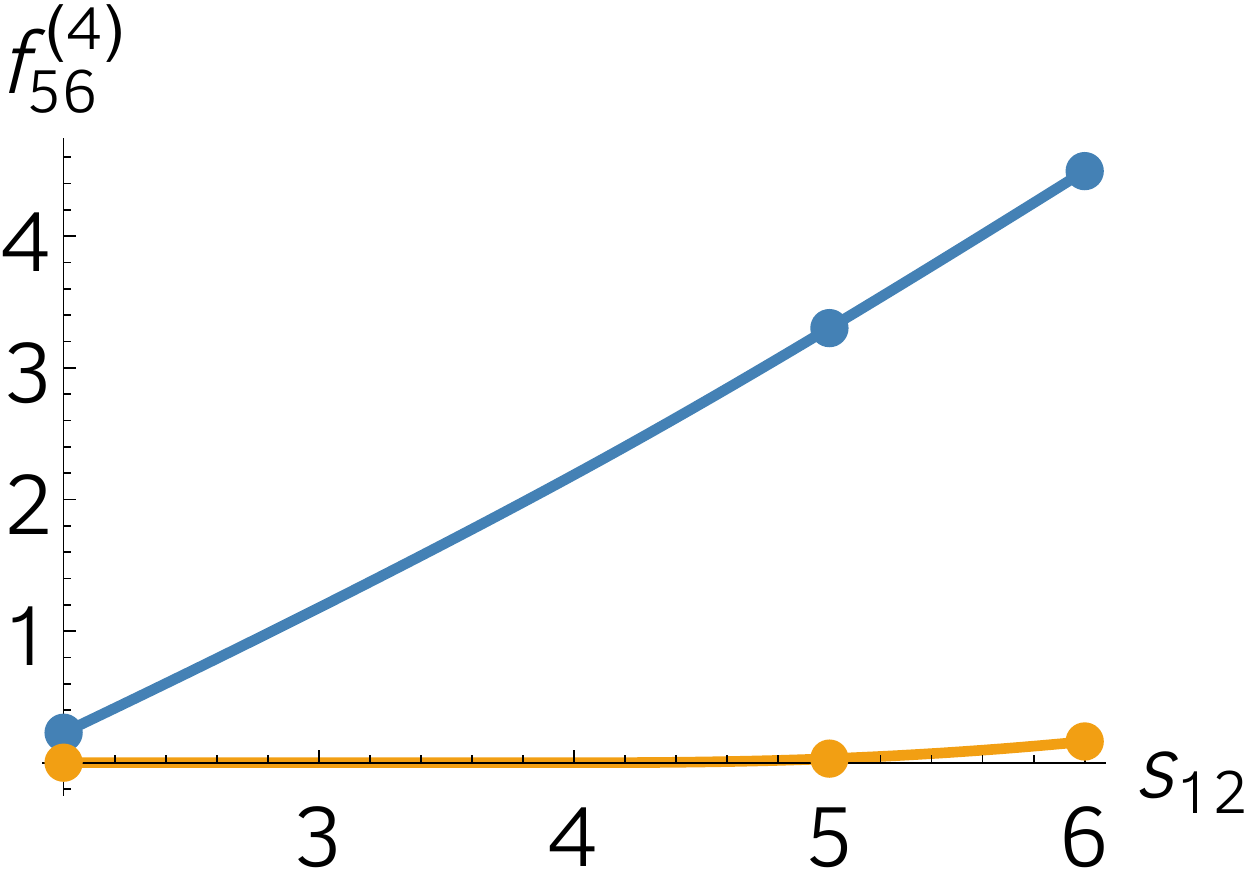} & \img{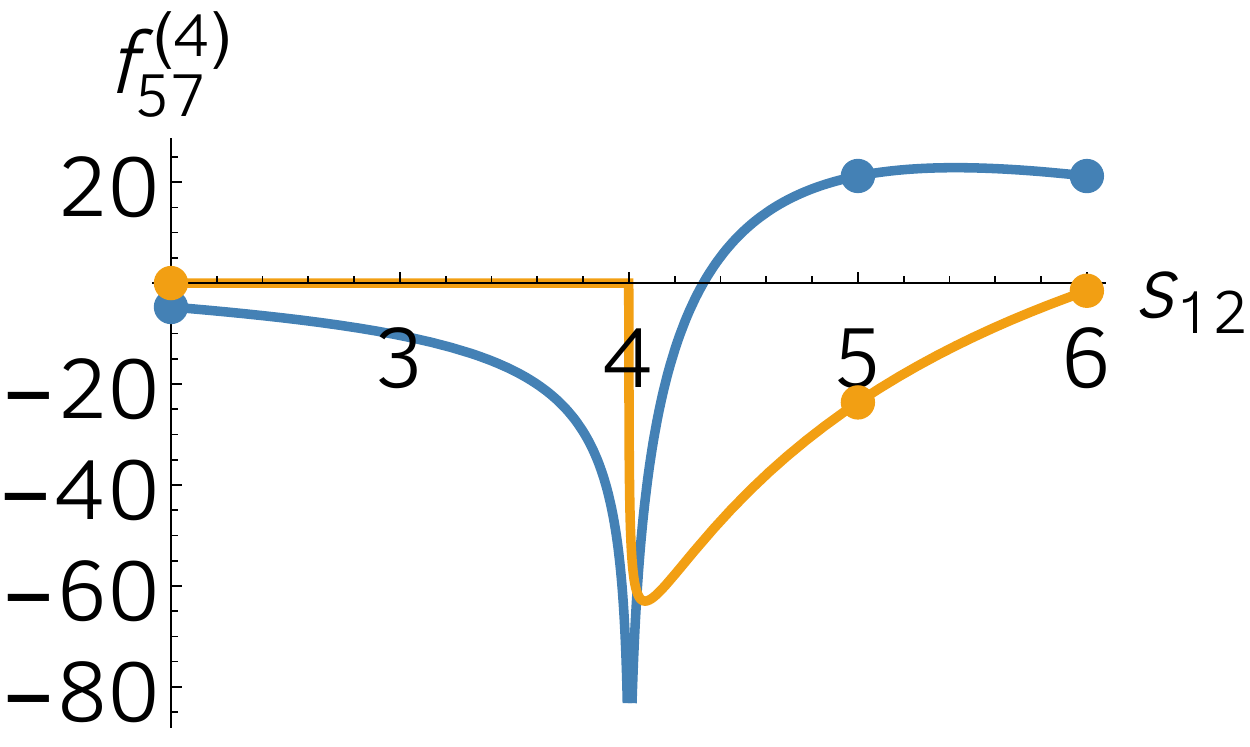} \\
\img{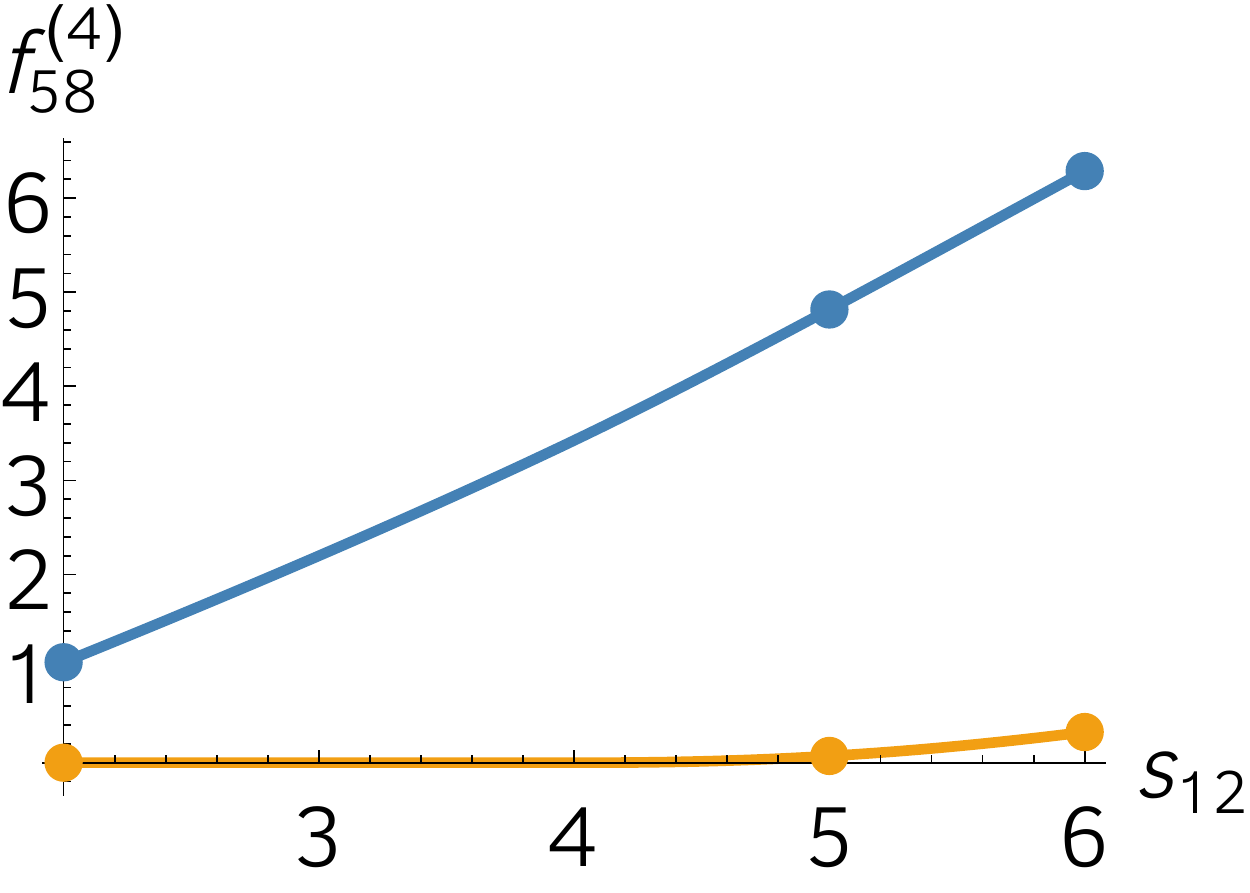} & \img{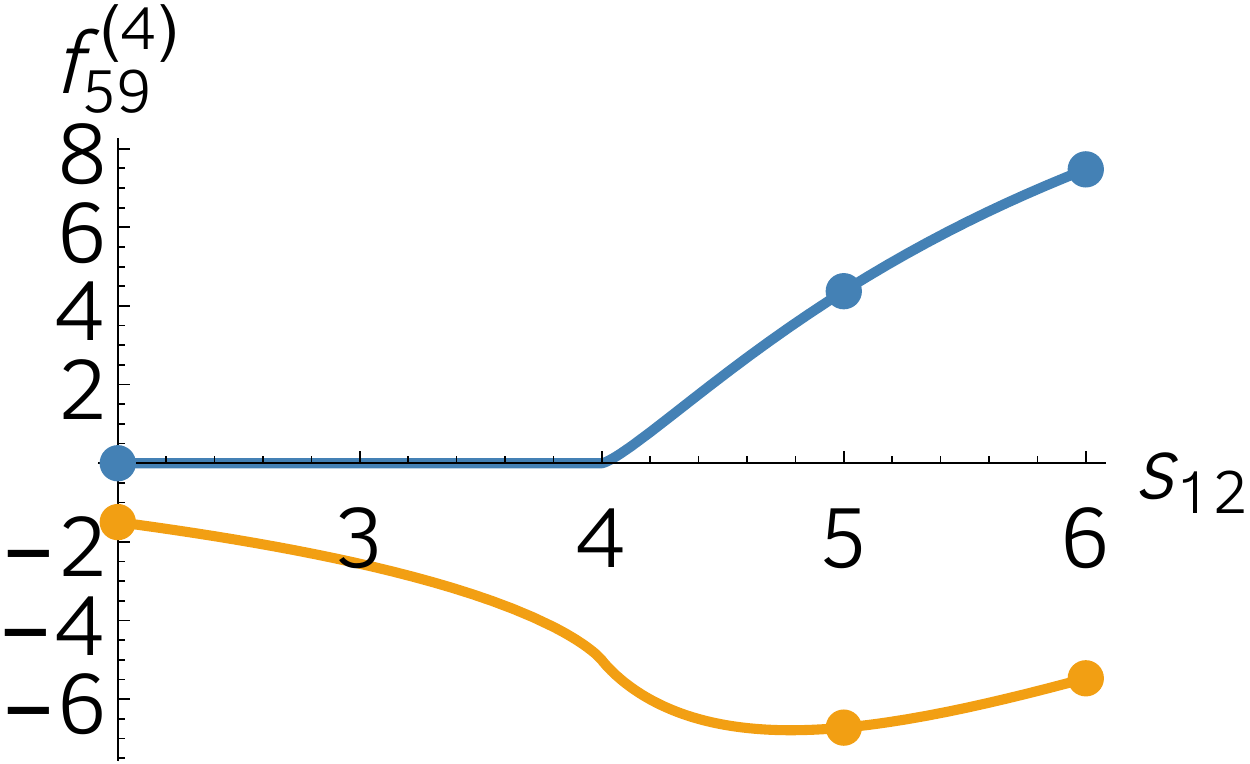} & \img{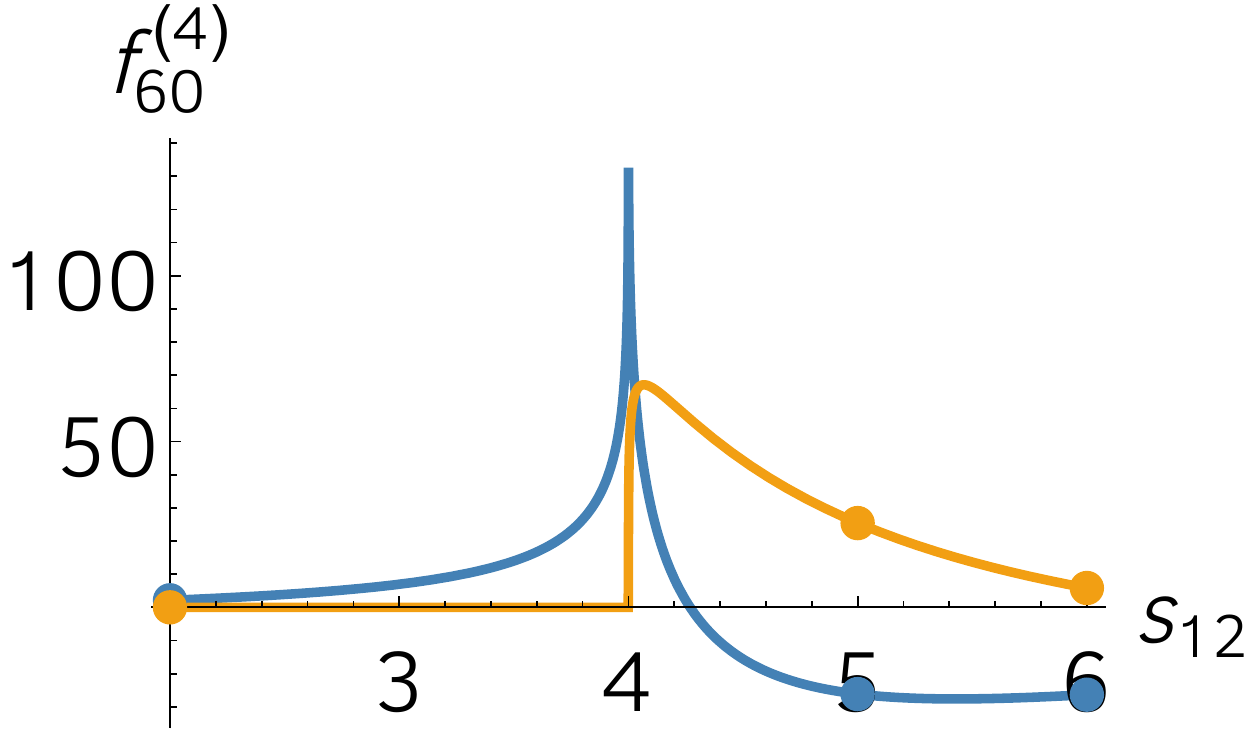} \\
\img{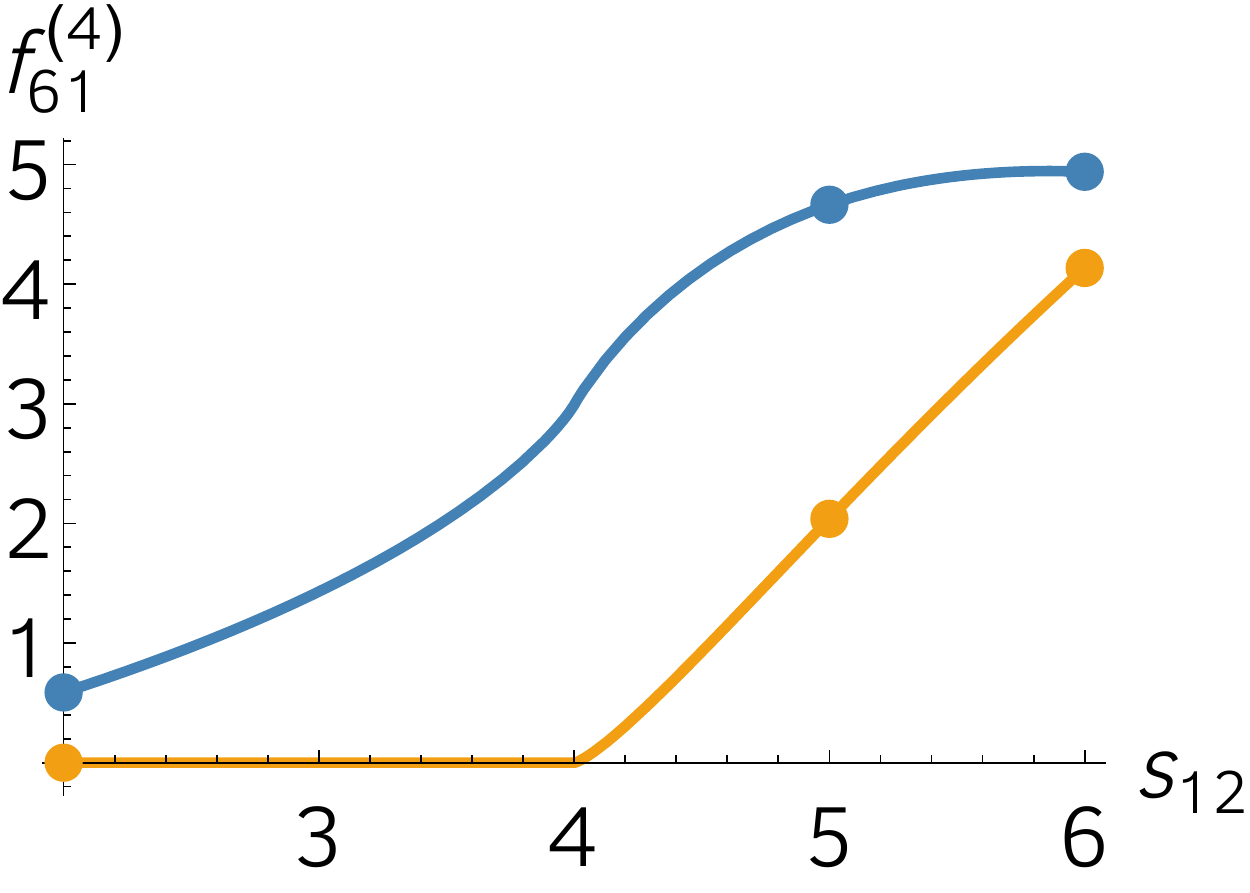} & \img{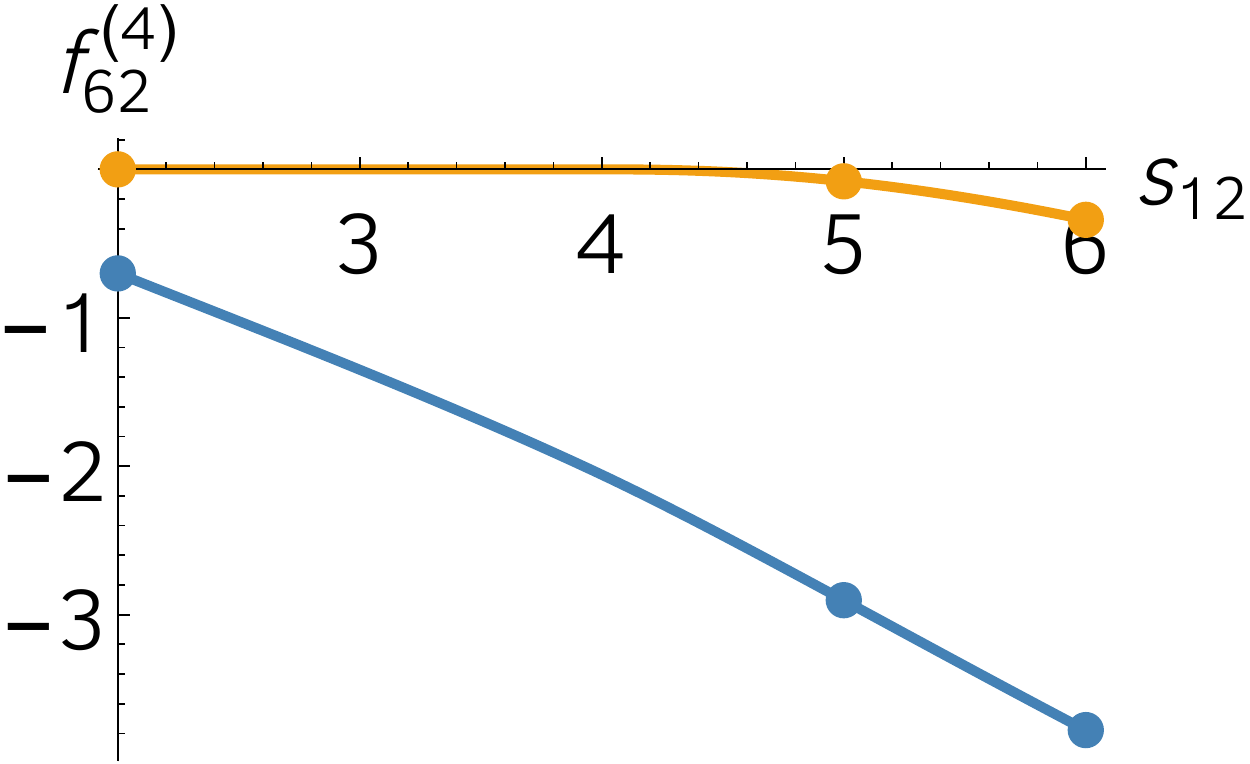} & \img{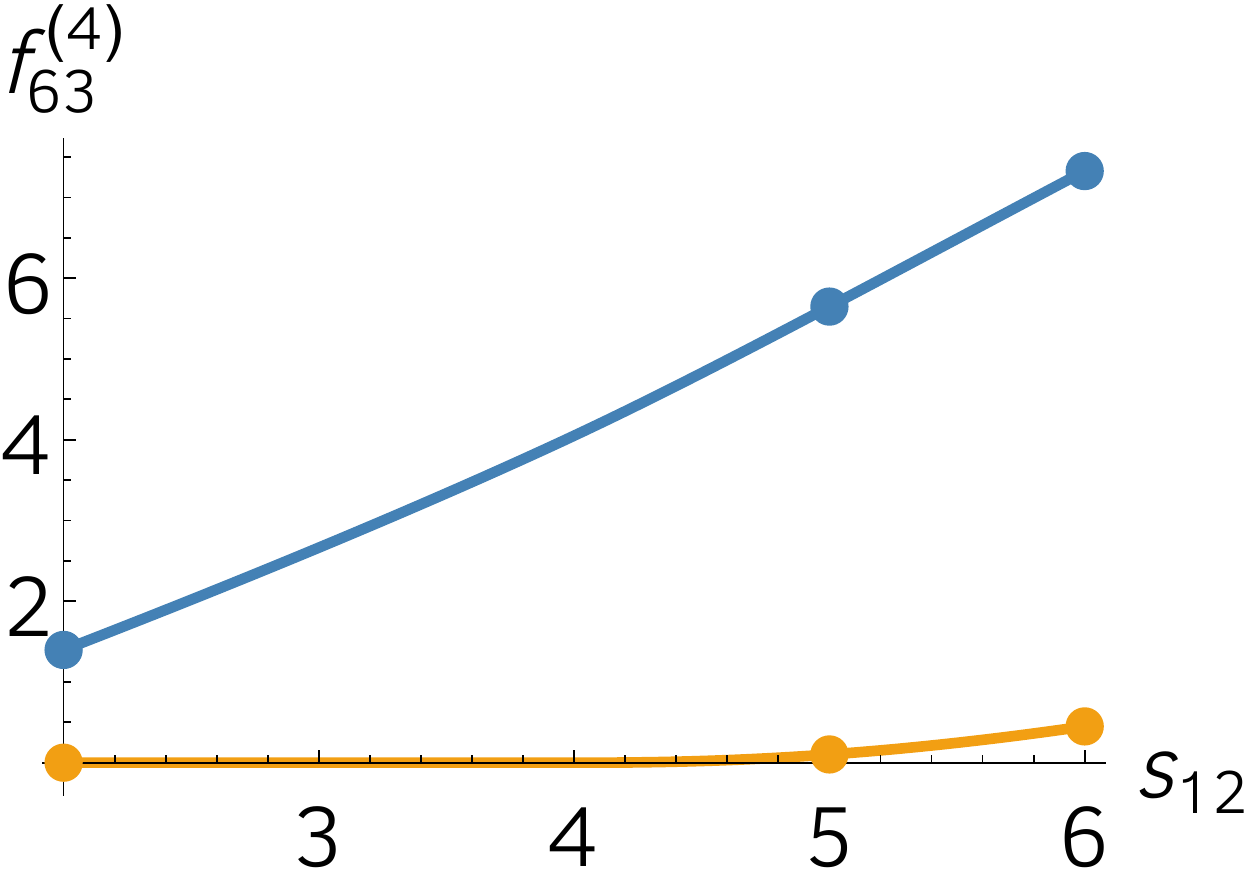} \\
\img{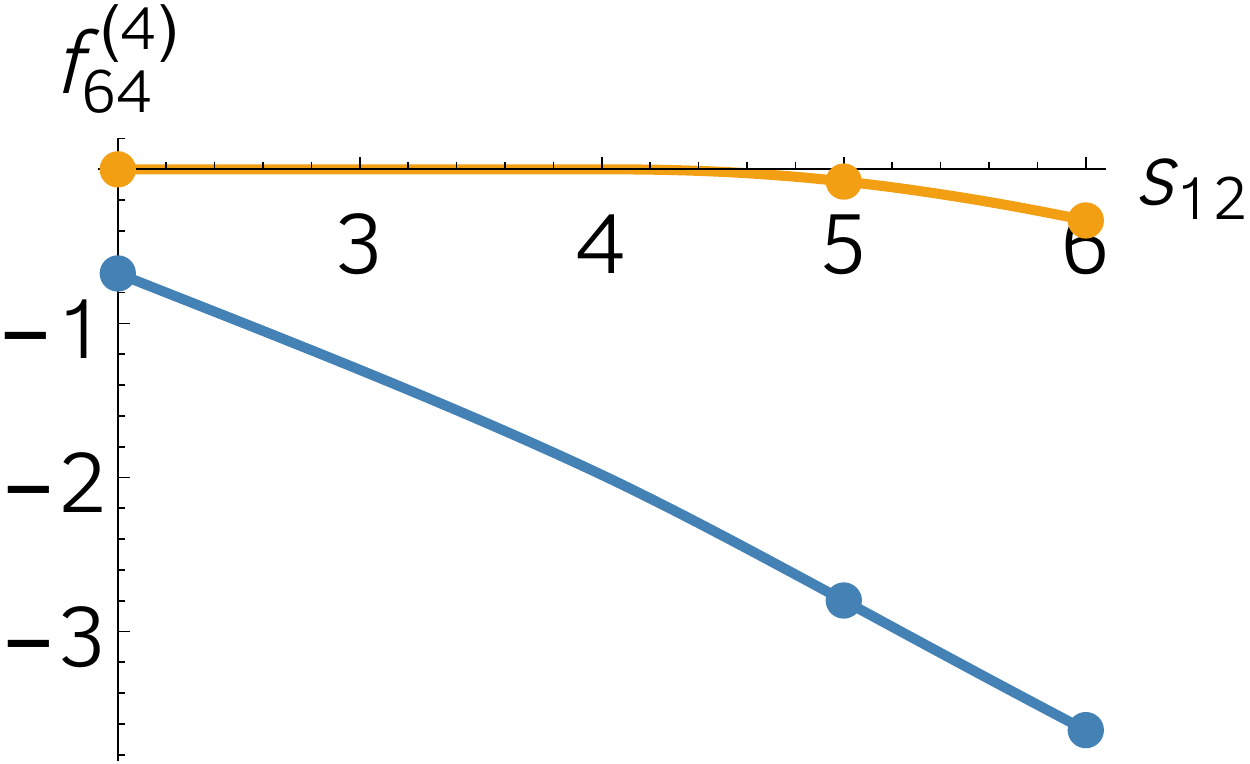} & \img{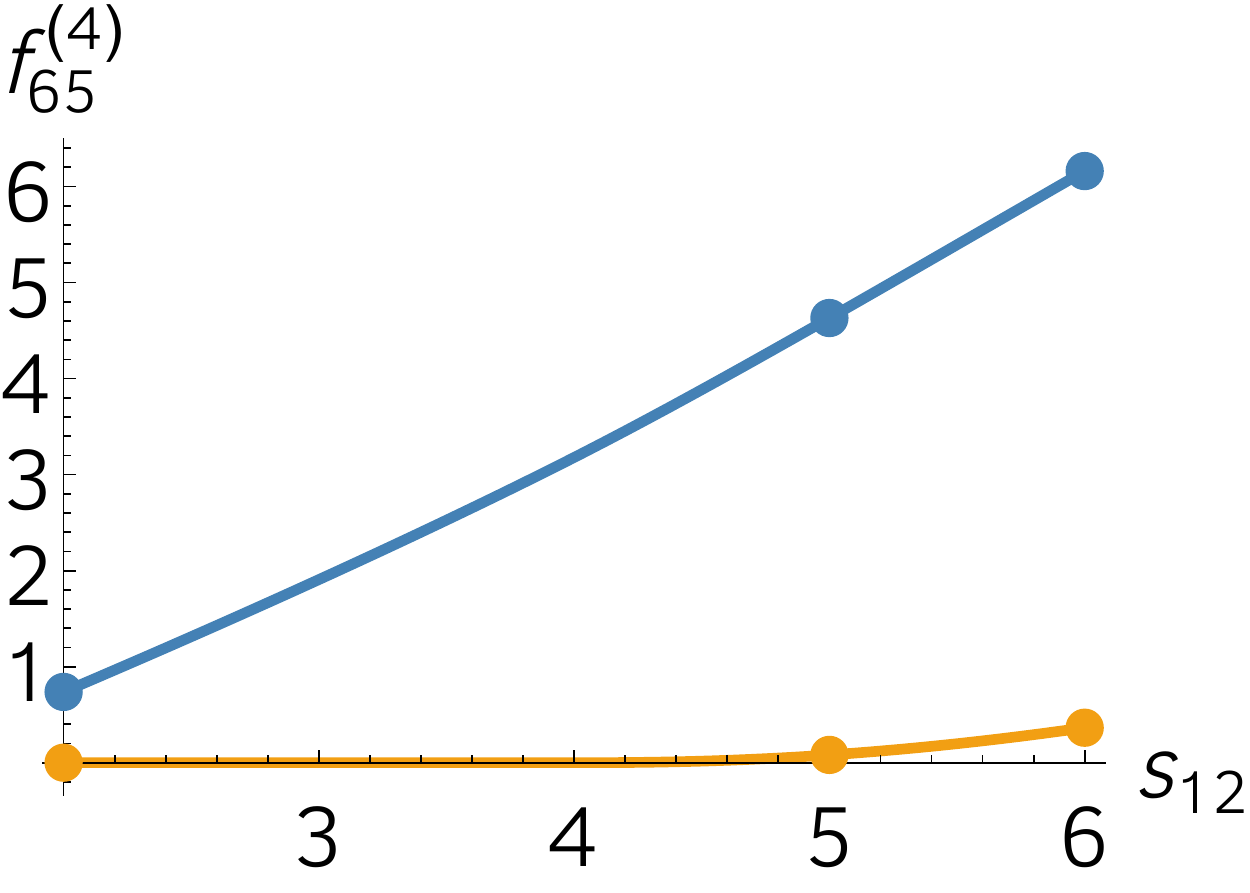} & \img{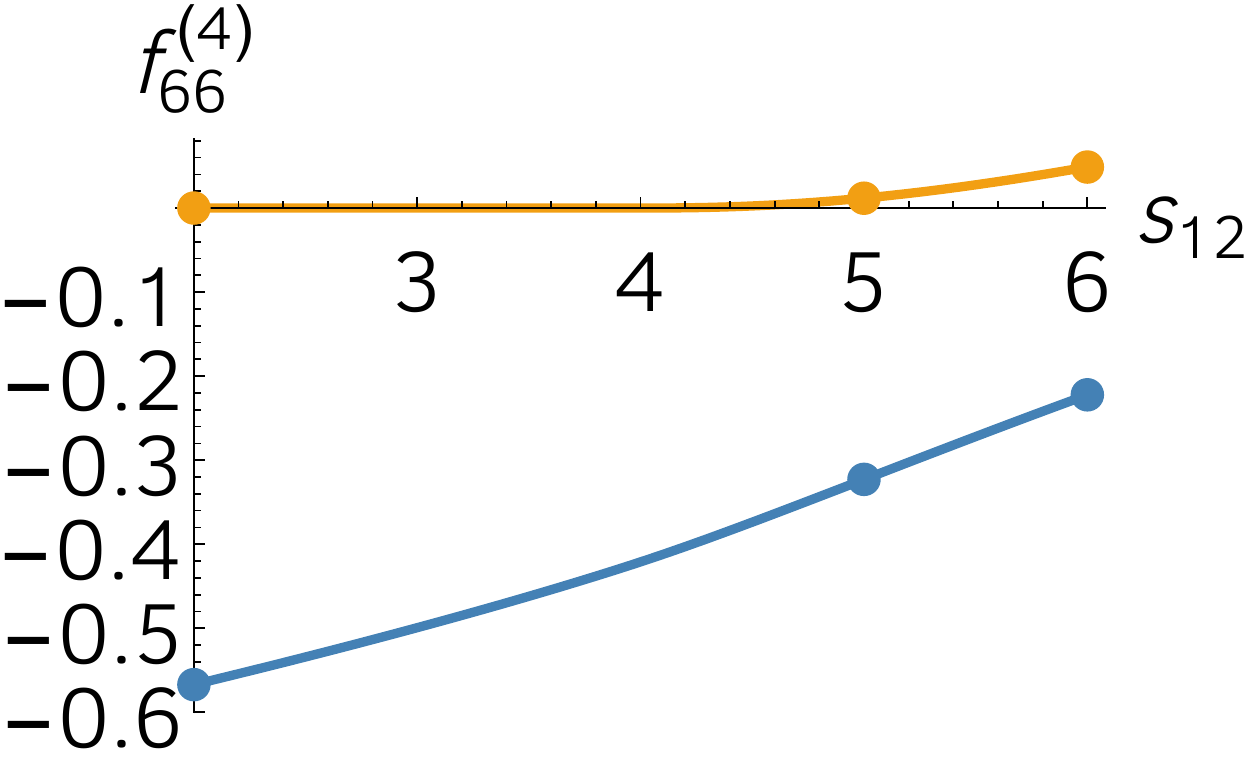} \\
\img{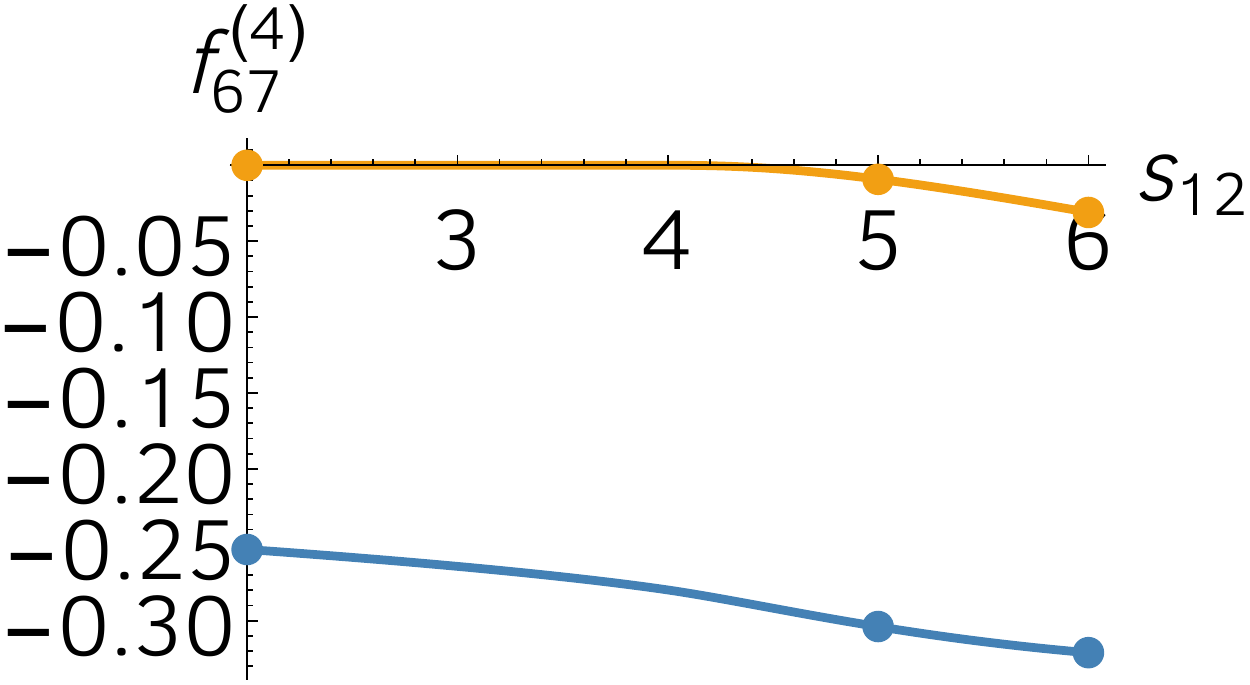} & \img{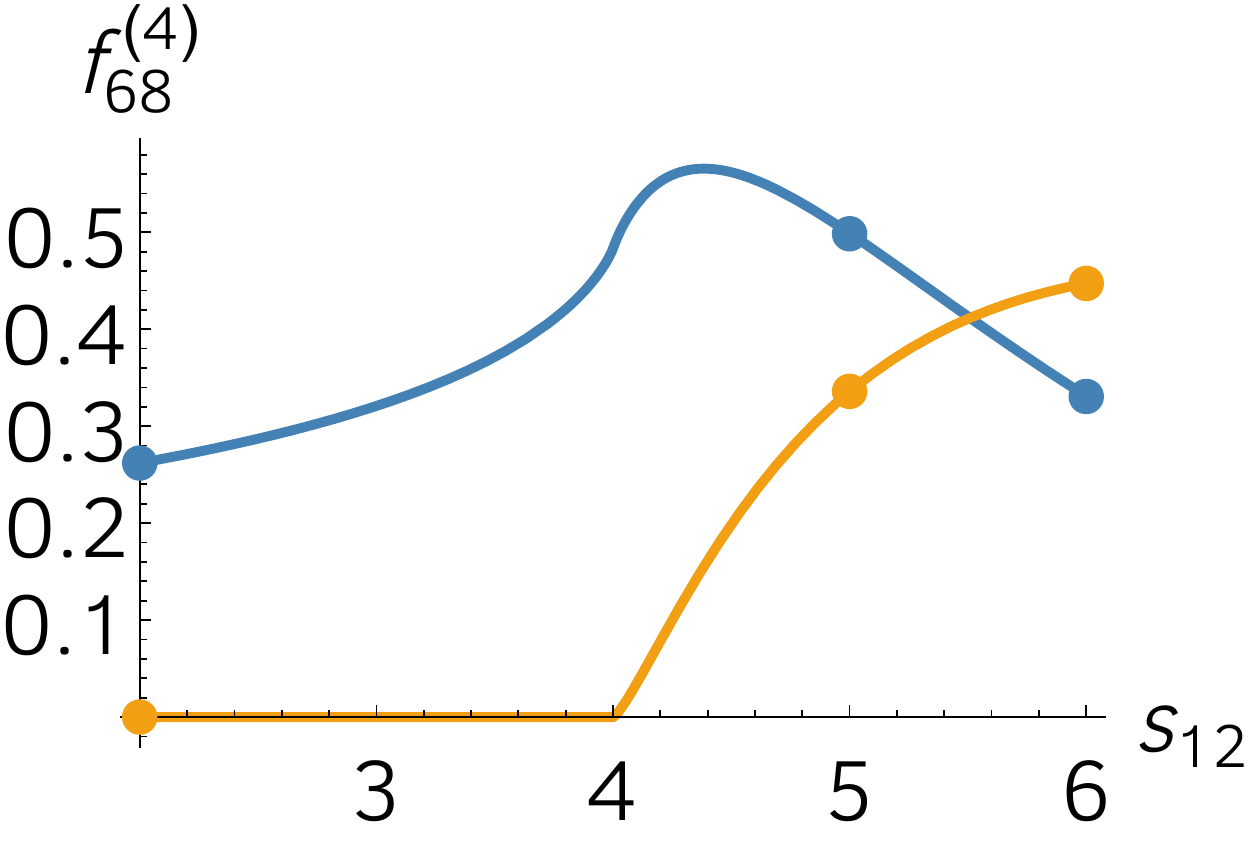} & \img{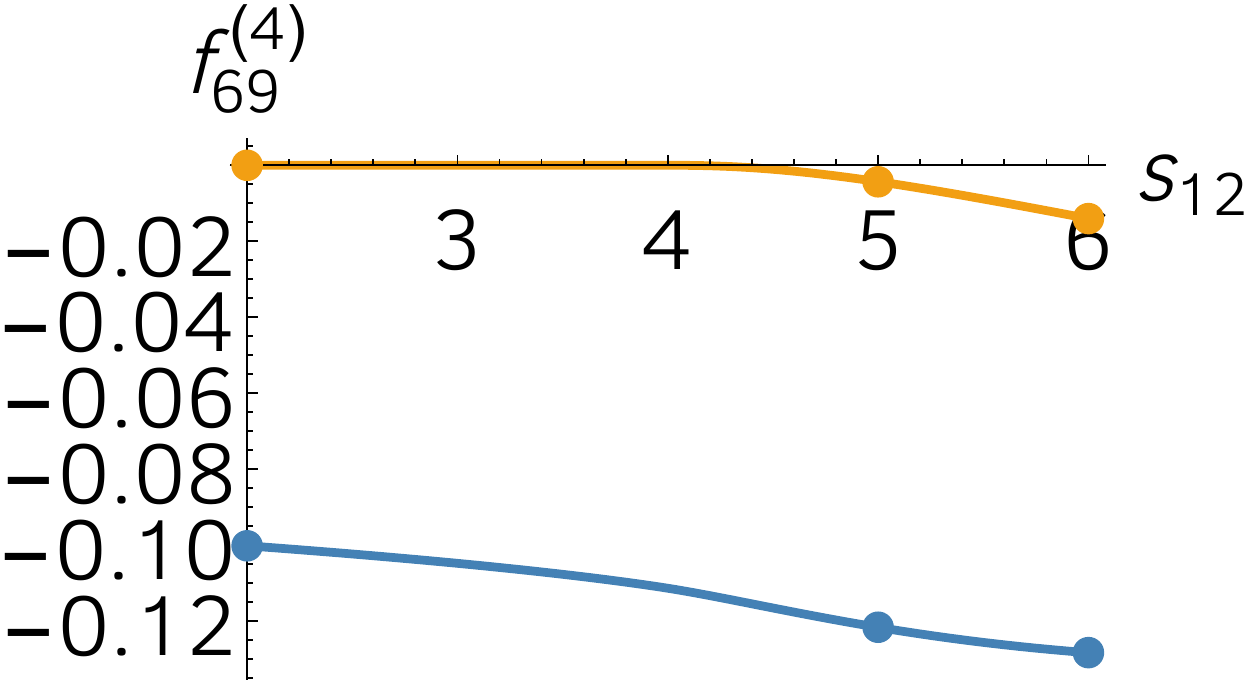} \\
\img{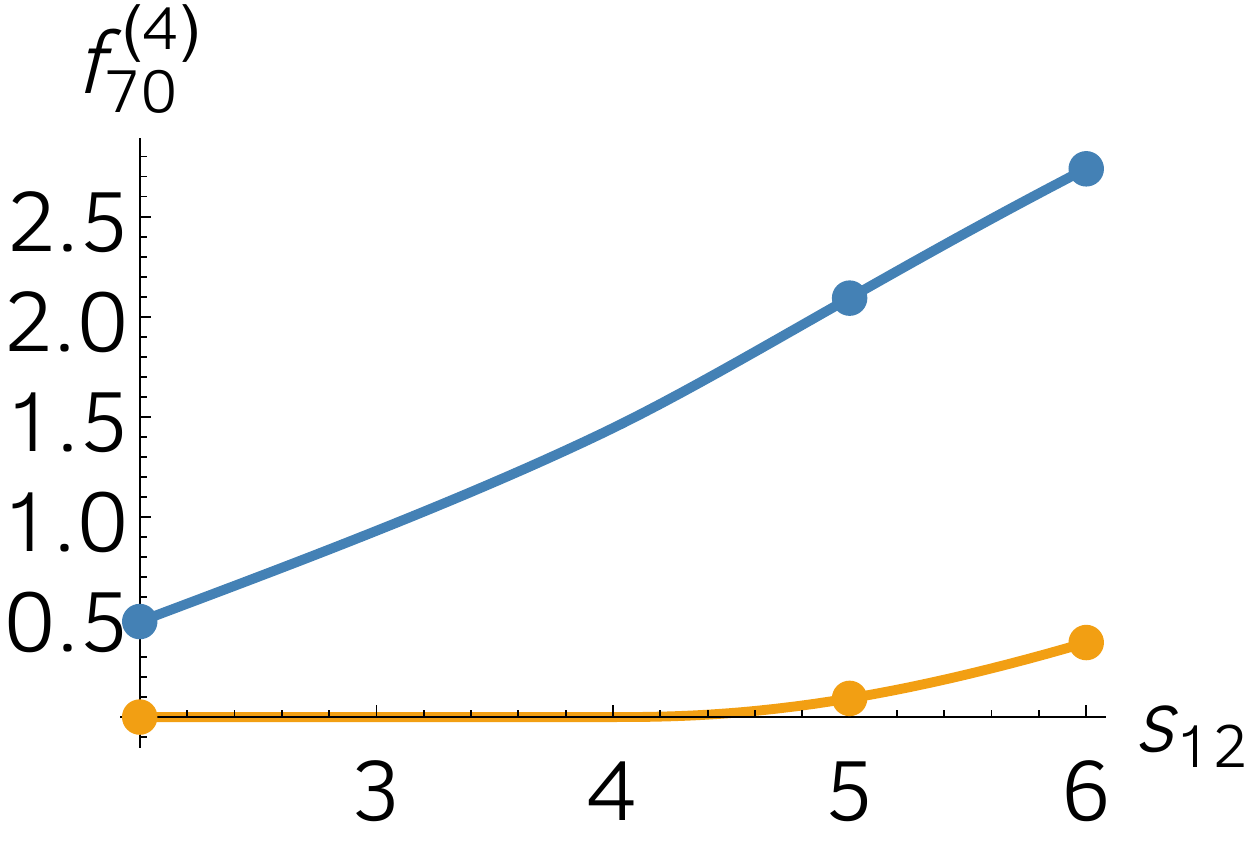} & \img{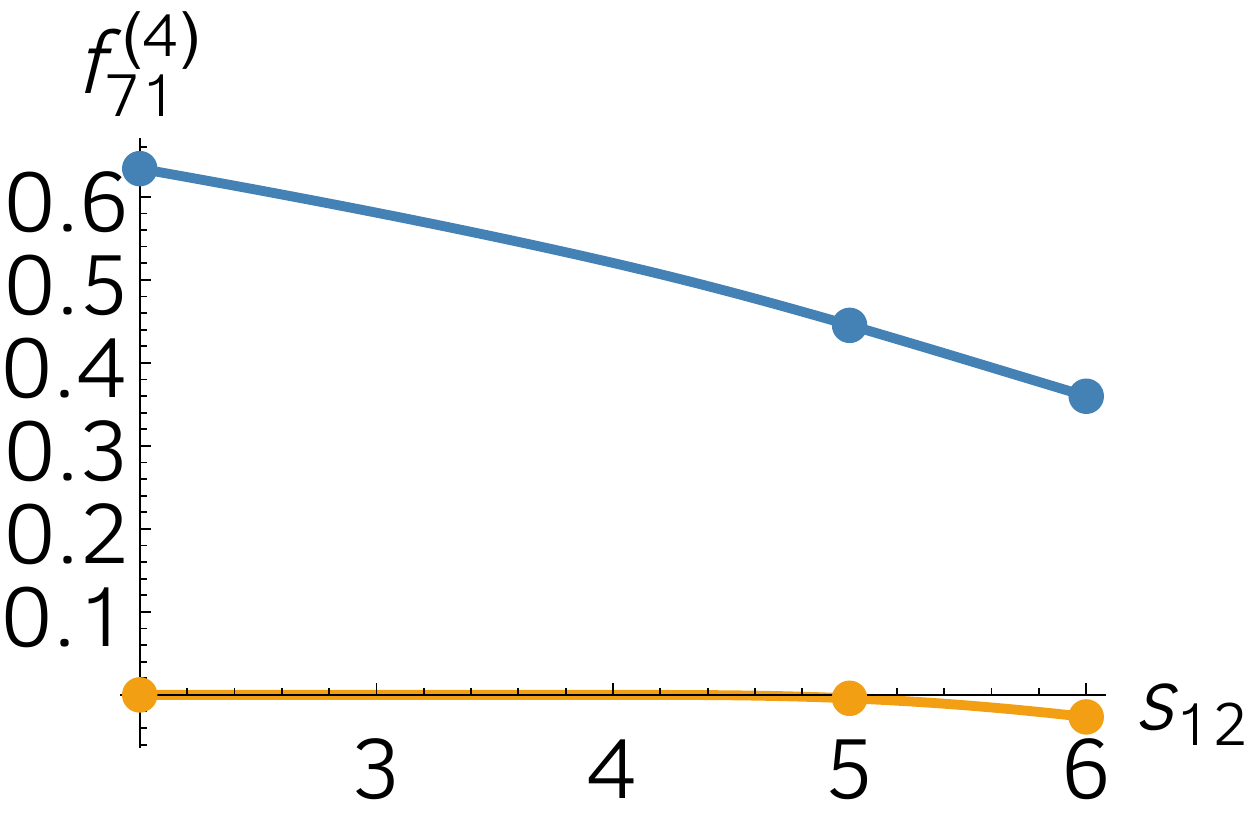} & \img{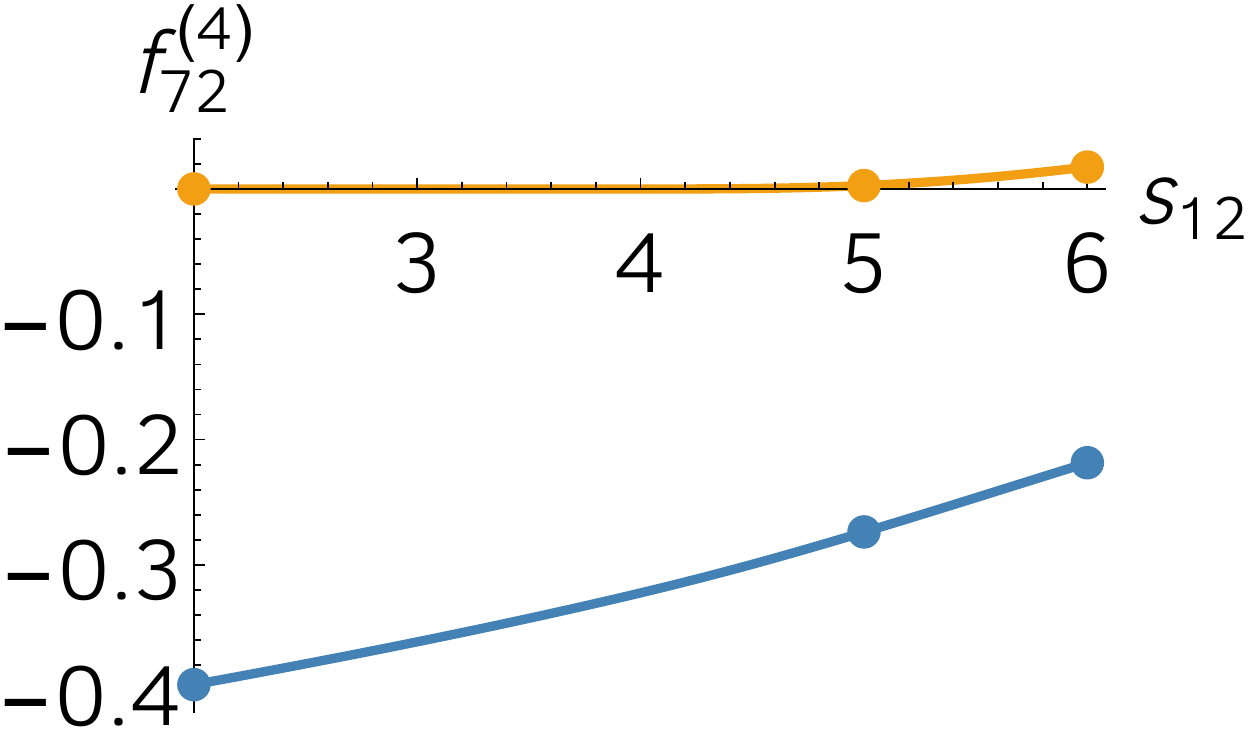} \\
\img{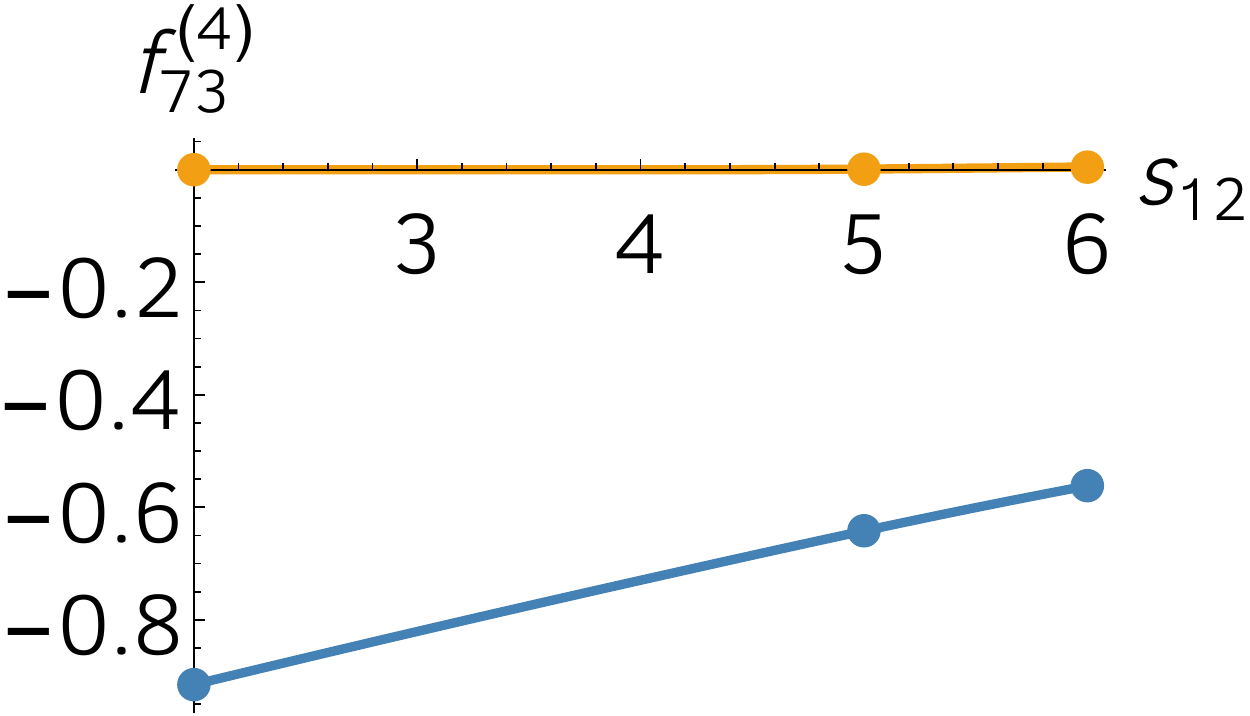} & &    
\end{tabular}
\end{table}

\clearpage

\bibliographystyle{JHEP}
\bibliography{refs}

\end{document}